\documentstyle[aps,prc,floats,epsfig]{revtex}
\begin{document}
\twocolumn
\setlength{\unitlength}{10mm}
\newcommand{\beq}{\begin{equation}}
\newcommand{\eps}{\epsilon}
\newcommand{\eeq}{\end{equation}}
\newcommand{\bea}{\begin{eqnarray}}
\newcommand{\eea}{\end{eqnarray}}
\renewcommand{\textfraction}{0.05}
\renewcommand{\bottomfraction}{0.95}
\renewcommand{\dbltopfraction}{0.99}
\renewcommand{\textfraction}{0.01}
\renewcommand{\bottomfraction}{0.99}
\setlength{\dbltextfloatsep}{5pt plus 1pt minus 1 pt}
\setlength{\dblfloatsep}{5pt plus 1pt minus 1 pt}
\setlength{\intextsep}{5pt plus 1pt minus 1 pt}
\setlength{\textfloatsep}{5pt plus 1pt minus 1 pt}
\def\nuc#1#2{\relax\ifmmode{}^{#1}{\protect\text{#2}}\else${}^{#1}$#2\fi}
\newcommand{\dsp}{\displaystyle}
\textwidth=16.51cm
\textheight=22.86cm
\topmargin=0pt
\headheight=-0.050cm
\headsep=0pt
\renewcommand \refname {\rm \normalsize REFERENCES}
\columnsep=0.65cm
\hoffset=6mm	
\topmargin=5mm
\title{\large 
\sf
COMPLEX PARTICLE AND LIGHT FRAGMENT EMISSION IN THE CASCADE-EXCITON MODEL
OF NUCLEAR REACTIONS
}
\vspace{1cm}
\author{ 
\sf 
Stepan G.\ Mashnik and Arnold J.\ Sierk\\
T-16, Theoretical Division\\
\sf
Los Alamos National Laboratory, Los Alamos, NM 87545, USA\\
\sf
Tel.: 505-667-9946, E-mail: mashnik@lanl.gov\\
\sf
Tel.: 505-667-6784, E-mail: t2ajs@lanl.gov\\
\sf
Konstantin K.\ Gudima\\
\sf
Nuclear Physics Laboratory, Institute of Applied Physics,\\
\sf
Academy of Science of Moldova, Kishinev, MD-2028, Moldova\\
\sf
Tel: (373)-276-90-81, E-mail: gudima@cc.acad.md
} 
\begin{sf}
\maketitle
\noindent  {\bf SUMMARY}\\
A brief description of our improvements and refinements that led from
the CEM95 version of the Cascade-Exciton Model (CEM) code to CEM97 and to
CEM2k is given.
The increased accuracy and predictive power of the code CEM2k 
are shown by several examples.
To describe fission and light-fragment (heavier than $^4$He) production, 
the CEM2k code has been merged with the GEM2 code of 
Furihata. We present some results on proton-induced
fragmentation and fission reactions predicted by this extended version of 
CEM2k.
We show that merging CEM2k with GEM2 allows us to
describe many fission and fragmentation reactions in addition to the 
spallation reactions which are already relatively well described.
Nevertheless, the current version of GEM2 does not
provide a completely satisfactory description of complex particle spectra, 
heavy-fragment emission, and spallation yields.
We have initiated another approach to describe fission, complex particles and
fragment emission by developing further our CEM2k code addressing
specifically these problems. In this effort, we have developed our
own universal approximation for inverse cross sections. 
We have also developed new routines to calculate Coulomb barriers and
widths of
emitted particles and to simulate their kinetic energy using arbitrary
approximations for the inverse cross sections. To describe fission-fragment
production,  we have
incorporated into CEM2k a thermodynamical model of fission
by Stepanov. This extended version of CEM2k allows us to describe 
much better complex particle emission and many fission fragments, but
it is still incomplete and needs further work.

\vspace{0.4cm}
\noindent {\bf I. INTRODUCTION}\\
The design of a hybrid reactor system driven with a high-current accelerator 
requires information about residual nuclides that are produced by high 
energy protons and secondary neutrons interacting in the target and in 
structural materials.
It is both physically and economically impossible to measure all necessary
data, which is why reliable models and codes are needed. A model with a 
good predictive power both for the spectra of emitted particles and 
residual nuclide yields is the Cascade-Exciton Model (CEM) of nuclear
reactions$^1$. An improved version of the CEM is contained in the code 
CEM95$^2$, which is available free from the NEA/OECD, Paris. 
Following an increased interest in intermediate-energy nuclear data in
relation to such projects as Accelerator Transmutation of nuclear Wastes (ATW),
Accelerator Production of Tritium (APT), the Spallation Neutron Source (SNS),
the Rare Isotope Accelerator (RIA), and others,
we developed a new version of the cascade-exciton model, CEM97$^{3,4}$.
CEM97 has a number of improved physics features, uses better 
elementary-particle cross sections for the cascade model, and due to some
significant algorithmic improvements is several
times faster than CEM95. A preliminary version has been incorporated
into the transport code system MCNPX, although unfortunately the
MCNPX team did not include the efficiency improvements$^5$.

The recent GSI measurements performed using inverse kinematics
for interactions of $^{208}$Pb$^{6,7}$ and $^{238}$U$^8$ at 1 GeV/nucleon and
$^{197}$Au at 800 MeV/nucleon$^9$ with liquid $^1$H provide
a very rich set of cross sections for production of practically
all possible isotopes from these reactions in a ``pure" form,
{\it i.e.}, individual cross sections from a specific given bombarding isotope
(or target isotope, when considering reactions in the usual kinematics,
p + A). Such cross sections are much easier to compare to models than the 
``camouflaged" data from $\gamma$-spectrometry measurements. These 
are often obtained only for a natural composition of isotopes in a target
and are mainly for cumulative production, where measured cross sections
contain contributions nor only from the direct production of
a given isotope, but also from all its decay-chain precursors. 
Analysis of these new data has motivated us to further
improve the CEM and to develop a preliminary version of a new code, 
CEM2k$^{10,11}$, still under development.

In our original motivation, different versions of the CEM codes were 
developed to reliably describe the yields of spallation products
and spectra of secondary nucleons, without a special emphasis
on complex-particle and light-fragment emission or on fission
fragments in reactions with heavy targets. In fact, CEM95, CEM97, 
and the initial version of the CEM2k code simulate spallation 
only and do not calculate the process of fission, and do
not provide fission fragments and a further possible evaporation of
particles from them. Thus, in simulating the compound stage of a 
reaction, when these codes encounter a fission, they simply tabulate 
this event (that permits calculation of fission cross sections and 
fissility) and finish the calculation of this event without a subsequent 
treatment of fission fragments.  To be able to describe 
nuclide production in the fission region, these codes have to be extended
by incorporating a model of high energy fission
({\it e.g.}, in the transport code MCNPX$^5$, where CEM97 and
CEM2k are used, 
they are supplemented by Atchison's fission model$^{12,13}$,
often cited in the literature as the RAL model of fission).

Since many nuclear and astrophysical applications require reliable
data also on complex particles (gas production) and light and/or heavy
fragment production, we here address for the first time these questions 
with our improved cascade-exciton model and present our
preliminary results on the study. We try two different ways
of solving these problems and the results from both
approaches are presented below after a short background on the
development of the CEM.

\vspace{0.4cm}
\noindent {\bf II. FROM CEM95 TO CEM97 TO CEM2K}\\
First, we recall the fundamental ingredients of the CEM and the main 
differences between 
the improved cascade-exciton model code CEM97$^{3,4}$ and its precursor,
CEM95$^2$.  The CEM assumes that the reactions occur in three stages. 
The first stage is the IntraNuclear Cascade (INC),
in which primary particles can be re-scattered and produce secondary
particles several times prior to absorption by, or escape from the nucleus.
The excited residual nucleus remaining after the emission of the
cascade particles determines the particle-hole configuration that is
the starting point for the second, pre-equilibrium stage of the
reaction. The subsequent relaxation of the nuclear excitation is
treated in terms of the modified exciton model of pre-equilibrium decay 
followed by the equilibrium fission/evaporation stage of the reaction.
Generally, all three components may contribute to experimentally measured 
particle spectra and distributions. 

An important ingredient of the CEM is the criterion for transition 
from the intranuclear cascade to the pre-equilibrium model. In 
conventional cascade-evaporation models
(like ISABEL and Bertini's INC used in LAHET$^{14}$,
fast particles are traced down to some minimal energy, the cutoff energy
$T_{cut}$ (or one compares the duration of the cascade stage of a reaction 
with a cutoff time, in ``time-like" INC models, such as the Liege INC$^{15}$).
This cutoff is usually about 7--10 MeV above the Fermi energy,
below which particles are considered to be absorbed by the
nucleus. The CEM uses a different criterion to decide when a primary
particle is considered to have left the cascade.

An effective local optical absorptive potential $W_{opt. mod.}(r)$ is 
defined from the local interaction cross section of the particle,
including Pauli-blocking effects. This imaginary potential is compared
to one defined by a phenomenological global optical model
$W_{opt. exp.}(r)$. We characterize the degree of similarity or difference
of these imaginary potentials by the parameter 
$${\cal P} =\mid (W_{opt. mod.}-W_{opt. exp.}) / W_{opt. exp.} \mid . $$

When $\cal P$ increases above an empirically chosen value, the particle
leaves the cascade, and is then considered to be an exciton.
Both CEM95 and CEM97 use the fixed value $\cal P$ = 0.3.
With this value, we find the cascade stage of the CEM is generally shorter 
than that in other cascade models.

The transition from the preequilibrium stage of a reaction to the
third (evaporation) stage occurs when the probability of nuclear
transitions changing the number of excitons $n$ with
$\Delta n = + 2$ becomes equal to the probability of
transitions in the opposite direction, with $\Delta n = - 2$,
{\it i.e.}, when the exciton model predicts an equilibration has been
established in the nucleus.

The improved cascade-exciton model
in the code CEM97 differs from the CEM95 version by incorporating new 
approximations for the elementary cross sections used in the cascade,
using more precise values for nuclear masses, $Q$-values, binding and 
pairing energies, using corrected systematics for the level-density
parameters, 
adjusting the cross sections of pion absorption on quasi-deuteron 
pairs inside a nucleus, 
allowing for nuclear transparency of pions, including the Pauli principle 
in the preequilibrium calculation, and implementing significant refinements 
and improvements in the algorithms of many subroutines,
decreasing the computing time by up to a
factor of 6 for heavy nuclei, which is very important when performing
practical simulations with transport codes like MCNPX.
On the whole, this set of improvements leads to a better description
of the particle spectra and yields of residual nuclei and a better 
agreement with available data for a variety of nuclear reactions.
Details and examples with some results from this work may be found 
in$^{3,16}$.

We also make a number of refinements in the calculation of the
fission channel, as described in$^{4,17}$.
Besides these modifications of the CEM95 code introduced especially 
for a better description of fission cross sections, we have been further 
improving the CEM, striving for a 
model capable of predicting different characteristics of nuclear reactions 
for arbitrary targets with a wide range of incident energies. 
Modifications made for a better description of the
preequilibrium, evaporative, and cascade stages
will also affect the fission channel.  We have incorporated into the CEM the
updated experimental atomic mass table by Audi and Wapstra$^{18}$,
the nuclear ground-state masses (where data does not exist)
and shell corrections by M\"oller {\it et al.}$^{19}$,
and the pairing energy shifts from M\"oller, Nix, and Kratz$^{20}$
into the level-density formula. In addition, we have derived a corrected
systematics for the level-density parameters using the Ignatyuk
expression$^{21}$,  
with coefficients fitted to the data analyzed by Iljinov {\it et al.}$^{22}$
(we discovered that Iljinov {\it et al.}\ used $11 / \sqrt{A}$ for the pairing 
energies $\Delta$ (see Eq.\ (3) in$^4$) 
in deriving their level-density systematics instead of the 
value of $12 / \sqrt{A}$ stated in$^{22}$).
We also derived additional semiempirical level-density-parameter systematics 
using the M\"oller {\it et al.}$^{19}$
ground-state microscopic 
corrections, both with and without the M\"oller, Nix, and Kratz$^{20}$ 
pairing gaps, and
introduced into the CEM a new empirical relation to take into account the 
excitation-energy dependence of the ground-state shell correction 
in the calculation of fission barriers (see$^4$).

As mentioned in the Introduction, analysis of the recent GSI measurements$^{6-9}$
has motivated us to further improve the CEM.
The authors of the GSI measurements performed a comparison of their data
to several codes, including LAHET$^{14}$,
YIELDX$^{23}$, ISABLA (ISABEL INC code from LAHET followed by the ABLA$^{24}$ 
evaporation code), CASCADO$^{25}$, and the Liege INC by Cugnon$^{15}$, 
and encountered
serious problems; none of these codes were able to accurately describe 
all their measurements. Most of the calculated distributions of isotopes 
produced as a function of neutron number were shifted toward
larger masses as compared to the data.
While in some disagreement with the measurements, the Liege INC and the CASCADO
codes provide a better agreement with the data than LAHET, ISABLA, and
YIELDX do. Being aware of this situation with the GSI data, 
we decided to consider them ourselves, leading to the development of CEM2k.

First, we calculated the $^{208}$Pb GSI reaction$^6$ 
with the standard versions of CEM95 and CEM97 and determined$^{10}$ 
that though CEM95 describes quite well production of 
several heavy isotopes near the target (we calculate p + $^{208}$Pb; 
therefore $^{208}$Pb is a target, not a projectile as in the GSI measurements),
it does not reproduce correctly the cross sections for lighter
isotopes in the deep spallation region. The disagreement
increases with increasing distance from the target, and all calculated
curves are shifted to the heavy-mass direction, just as was obtained
by the authors of the GSI measurements with the codes they tried.
The results of the CEM97 code are very similar to those of CEM95 
(see a figure with CEM97 results in$^{26}$). 
Later on, we performed an extensive set of calculations of the same data
with several more codes (HETC$^{27}$,
LAHET$^{14}$ 
with both ISABEL and Bertini options, CASCADE$^{28}$,
CASCADE/INPE$^{29}$,
INUCL$^{30}$,
and YIELDX$^{23}$)
and got very similar results:$^{26}$
all codes disagree with the data in the deep spallation region, the 
disagreement increases as one moves away from the target, and all 
calculated curves are shifted in the heavy-mass direction.
  
This means that for a given final element (Z), all
models predict emission of too few neutrons. Most of the neutrons are
emitted at the final (evaporation) stage of a reaction. One way to 
increase the number of emitted neutrons would be to increase the 
evaporative part of a reaction. In the CEM, this might be done 
in two different ways: the first would be to have a shorter 
preequilibrium stage, so that more excitation energy remains available
for the following evaporation; the second would be to have a longer
cascade stage, so that after the cascade, less exciton
energy is available for the preequilibrium stage, so fewer energetic
preequilibrium particles are emitted, leaving more excitation energy
for the evaporative stage.

One easy way to shorten the preequilibrium stage of a reaction
in CEM is to arbitrarily allow only transitions that increase the 
number of excitons, $\Delta  n = + 2$, {\it i.e.},
only allow the evolution of a nucleus toward the compound nucleus.
In this case, the time of the equilibration will be shorter and fewer
preequilibrium particles will be emitted, leaving more excitation
energy for the evaporation. Such an approach is used by some other exciton
models, for instance, by the Multistage Preequilibrium Model (MPM)
used in LAHET$^{14}$.
Calculations using this modification to CEM97 (see Fig.\ 2d in$^{10}$)
provide a shift of the calculated curves in the right direction, but 
only very slightly improve agreement with the GSI data. 

A second method of increasing evaporation is to enlarge the cascade 
part of a reaction; we may either enlarge the parameter $\cal P$ or 
remove it completely and resort to a cutoff energy $T_{cut}$,
as is done in other INC models. Calculations have shown that a reasonable 
increase of $\cal P$ doesn't solve the problem. However, if we do not 
use $\cal P$, but instead use a cutoff energy of $T_{cut} = 1$ MeV 
for incident energies above the pion production threshold, the code 
agrees with the GSI data significantly better (see Fig.\ 2e in$^{10}$).
Using both these conditions leads to results that 
describe the p + $^{208}$Pb GSI data very well.
We call this approach CEM2k. An example of CEM2k results for the yield of
Tm, Ir, and Tl isotopes from  p + $^{208}$Pb interactions compared to
the GSI$^6$ and ITEP$^{31}$ data and
with predictions by CEM95, LAHET-ISABEL, LAHET-Bertini, CASCADE, CASCADE/INPE,
INUCL, and YIELDX is shown in {\bf Fig.\ 1}. Similar comparisons for more
isotopes may be found in$^{10,26,31}$. 
We find that CEM2k agrees best with these GSI (and ITEP)
data of the codes tested here and in$^{10,26,31}$.
\begin{center}
\begin{figure} 
\hspace*{-2.0mm}
\includegraphics[angle=-90,width=8.0cm]{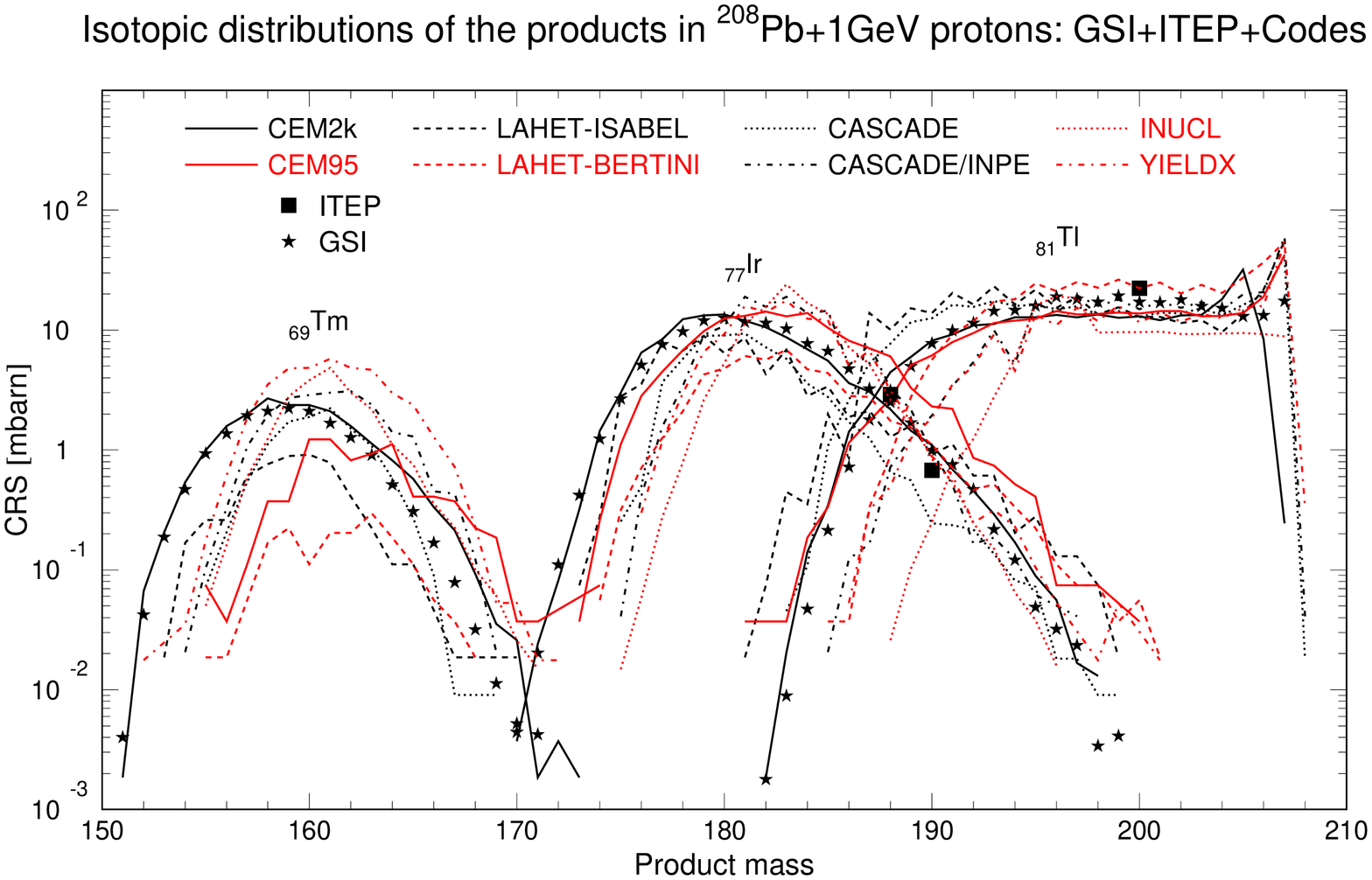}

\vspace*{2mm}
{\noindent  Figure 1. Mass distributions for independent production
of Tm, Ir, and Tl isotopes from 1 GeV protons colliding with $^{208}$Pb.
Stars are GSI measurements$^6$, while squares show ITEP  data$^{31}$. 
The calculations are identified as:
CEM2k---our results, CEM95$^{2}$, LAHET-ISABEL$^{14}$, 
LAHET-Bertini$^{14}$, CASCADE/INPE$^{29}$, 
CASCADE$^{28}$,
INUCL$^{30}$, and YIELDX$^{23}$.
}
\label{fig1}
\end{figure}
\end{center}

\vspace{-9.5mm}
Finding a good agreement of CEM2k with the isotope production, we wish 
to see how well it describes spectra of secondary particles in comparison 
with CEM97. {\bf Figure 2} shows examples of neutron spectra from interactions
of protons with the same target, $^{208}$Pb at 0.8 and 1.5 GeV
(we do not know of measurements of spectra at 1 GeV, 
the energy of the isotope-production data). We see that
CEM2k describes spectra of secondary neutrons comparably to 
CEM97, even possibly a little better at larger angles. 
\begin{center}
\begin{figure}
\vspace{-29mm} 
\hspace*{-2.5cm}
\includegraphics[angle=-0,width=13.cm]{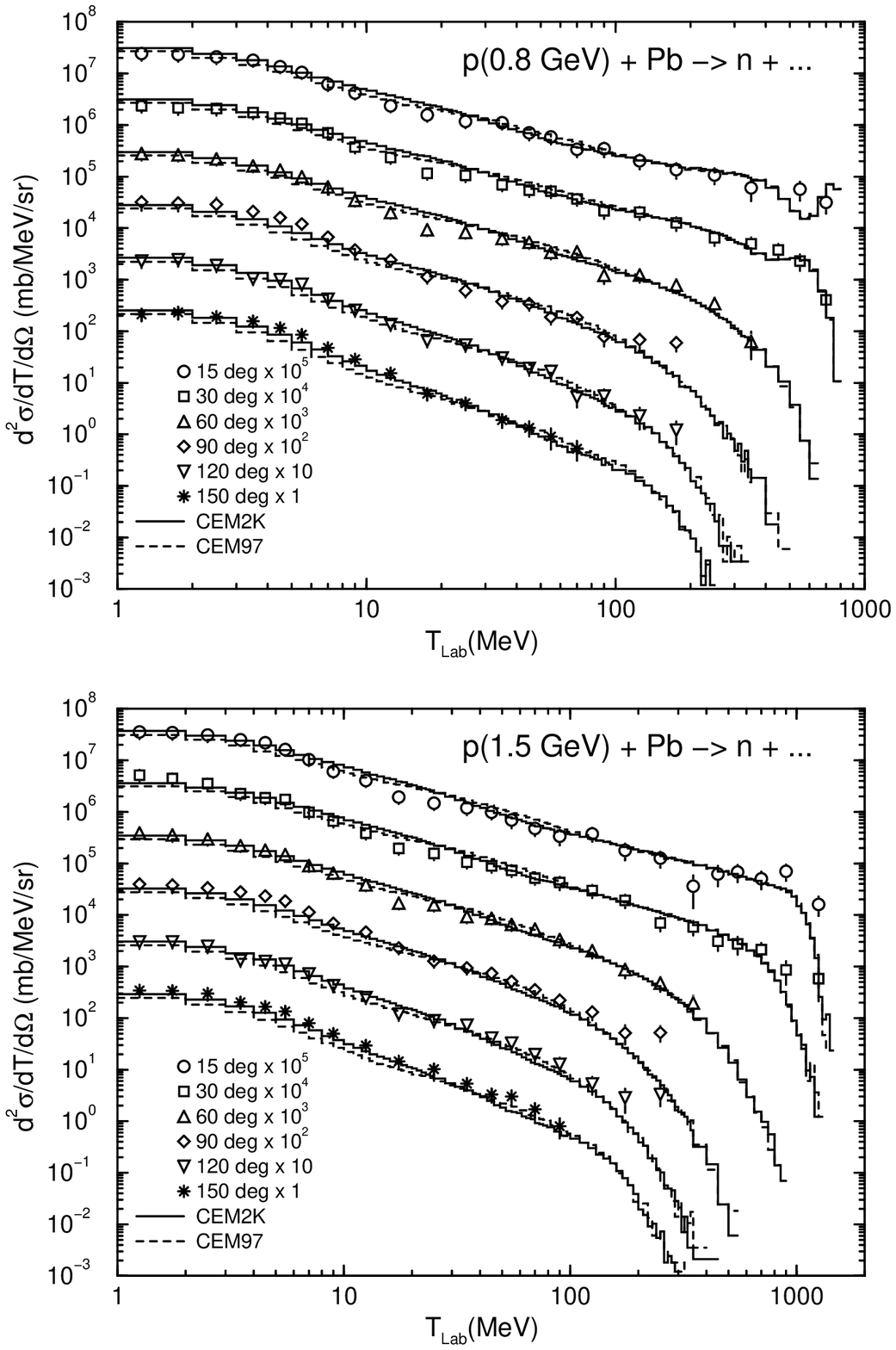}

\vspace*{-20mm}
{\noindent  Figure 2. Comparison of measured$^{32}$ double differential cross
sections of neutrons from 0.8 and 1.5 GeV protons on Pb to CEM2k and CEM97 
calculations. 
}
\label{fig2}
\end{figure}
\end{center}

\vspace{-9mm}

So this preliminary version of CEM2k
describes both the GSI data from $^{208}$Pb interactions with p
at 1 GeV/nucleon and the spectra of emitted neutrons from p+$^{208}$Pb
at 0.8 and 1.5 GeV better than its precursor CEM97.

We use CEM2k as fixed from our analysis of the
$^{208}$Pb data$^{6,7}$  without further modifications
to calculate the $^{197}$Au$^9$ and $^{238}$U$^8$ GSI measurements.
An example of the yield of several isotopes from $^{197}$Au calculated by
CEM2k is shown in {\bf Fig.\ 3} together with standard
CEM97 predictions and calculations by the LAHET-Bertini and YIELDX
codes from$^9$. We see that just as in the case of the $^{208}$Pb data,
CEM2k agrees best with the $^{197}$Au data in the spallation region
compared to the other codes tested here. 
Several more results for $^{197}$Au and $^{238}$U and their detailed discussion
may be found in$^{10}$.
\begin{center}
\begin{figure} 
\vspace*{-22mm}
\hspace*{-1.6cm}
\includegraphics[angle=-0,width=11.0cm]{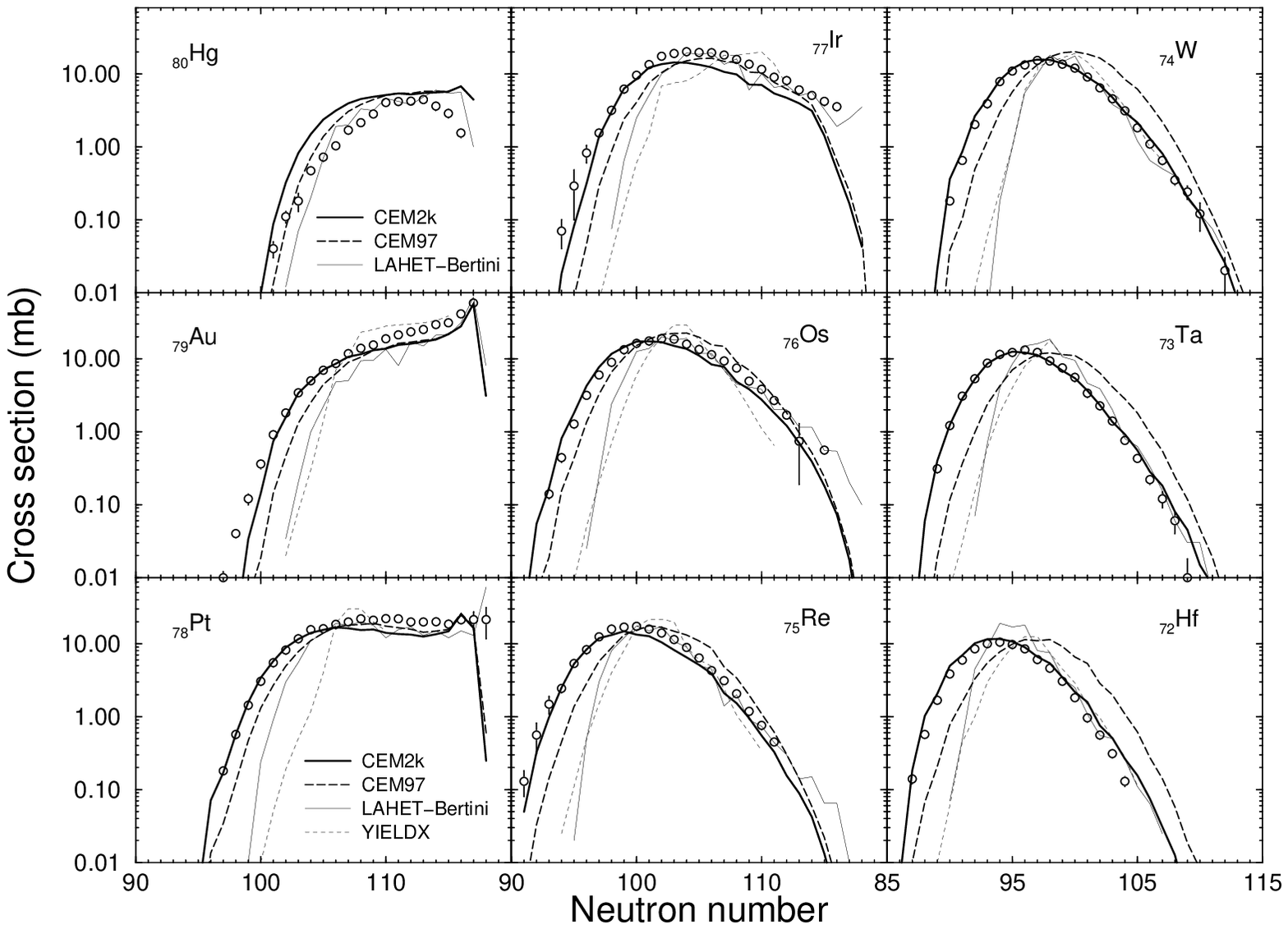}

\vspace*{-67mm}
{\noindent  Figure 3. Isotopic distributions of spallation products from 
the reaction $^{197}$Au + p at 800 $A$ MeV from mercury to hafnium.
Open circles are the GSI data$^9$, CEM2k (thick solid curves)
and CEM97 (thick dashed curves) are our present calculations,
LAHET-Bertini (thin solid curves) and YIELDX (thin dashed curves)
are calculations from$^9$.
}
\label{fig3}
\end{figure}
\end{center}
\vspace*{-4mm}

Besides the changes to CEM97 mentioned above, we also made a 
number of other improvements and refinements, such as 
imposing momentum-energy conservation for each simulated event
(the Monte Carlo algorithm previously used in CEM 
provides momentum-energy conservation only 
statistically, on the average, but not exactly for the cascade stage 
of each event);
using real binding energies for nucleons at the cascade 
stage instead of the approximation of a constant
separation energy of 7 MeV used in previous versions of the CEM; 
and using reduced masses of particles in the calculation of their
emission widths instead of using the approximation
of no recoil used previously.
In Ref$^{11}$.  we have shown that these refinements (which involve no 
further parameter fitting), while only improving slightly the agreement
with the GSI measurements (and other data on medium and heavy targets),
are crucial for light targets, especially when calculating 
$^4$He and other fragment emission from light nuclei. This is
especially important for applications where gas production is 
calculated.

Another improvement important for applications is
better representing the total reaction cross section. Previous
versions of CEM (just like many other INC-type models) calculate the 
total reaction cross section, $\sigma_{in}$ 
using the geometrical cross section, $\sigma_{geom}$, and the
number of inelastic, $N_{in}$, and elastic, $N_{el}$, simulated
events, namely: 
$\sigma_{in} = \sigma_{geom} N_{in} / (N_{in} + N_{el})$.
This approach provides a good agreement with available data at incident
energies above about 100 MeV, but is not reliable at lower
bombarding energies. To address this problem, we have incorporated into
CEM2k the NASA systematics by Tripathi {\it et al.}$^{33}$ for
all incident protons and neutrons with energies above the maximum in the NASA
reaction cross sections, and the Kalbach systematics$^{34}$ 
for neutrons of lower energy.  As shown in {\bf Fig.\ 4}, we can describe 
much better with CEM2k the total reaction cross sections 
(and correspondingly any other partial cross sections) for
n- and p-induced reactions, especially at energies below about 100 MeV.

\vspace*{-10mm}
\begin{center}
\begin{figure} 
\vspace*{-40mm}
\hspace*{-24.mm}
\includegraphics[angle=-0,width=12.5cm]{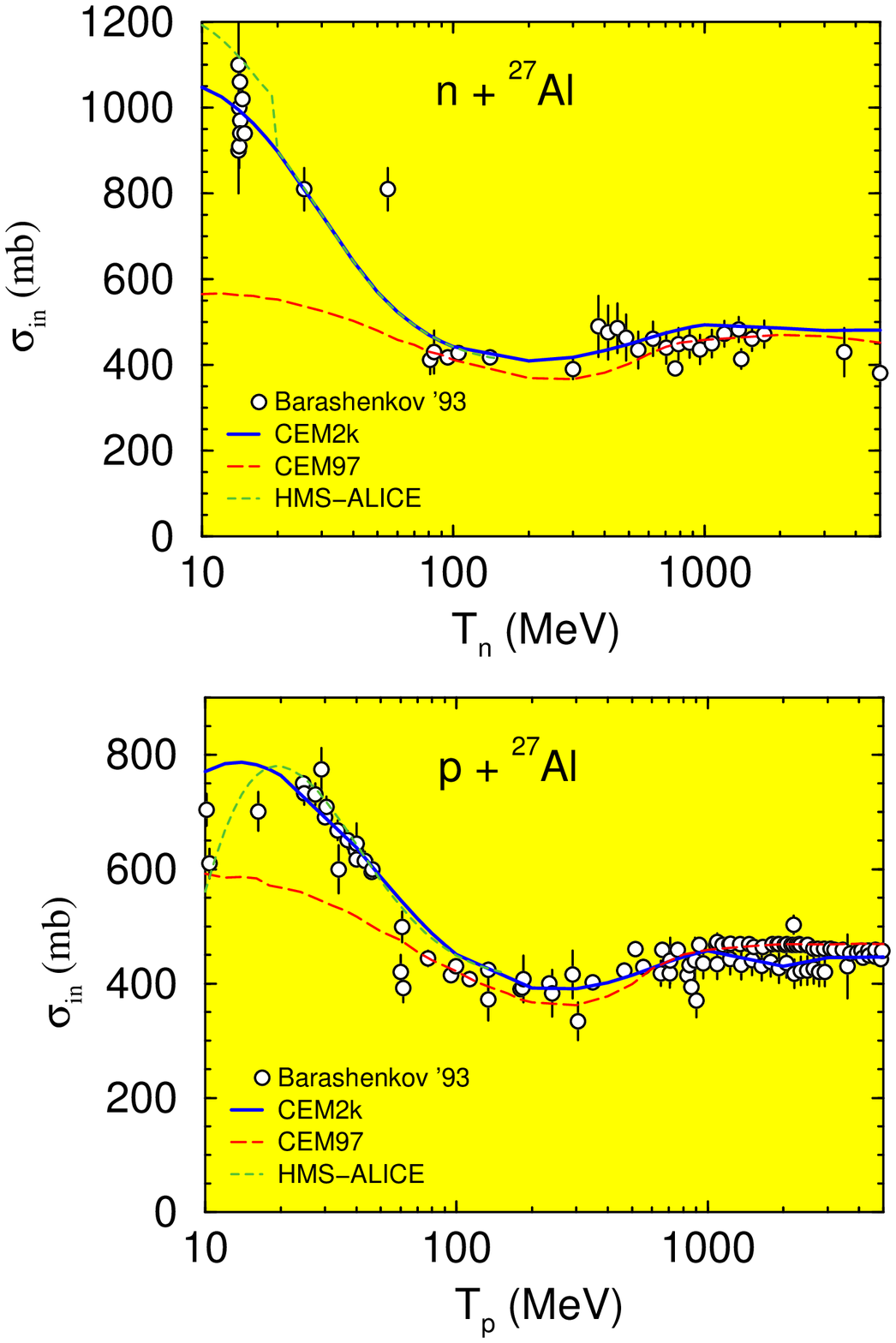}

\vspace*{-29mm}
{\noindent Figure 4. Total reaction cross sections for n- and p-induced 
reactions on Al calculated by CEM2k and CEM97 compared with experimental 
data compiled by Barashenkov$^{35}$ and calculations
from the HMS-ALICE code$^{36}$.
}
\label{fig4}
\end{figure}
\end{center}

\vspace{0.4cm}
\noindent {\bf III. MERGING CEM2K WITH GEM2}\\
As a first attempt to describe with CEM2k both emission of 
intermediate-mass 
fragments heavier than $^4$He and production of heavy fragments
from fission, we merged CEM2k with the Generalized Evaporation Model
(GEM) code by Furihata$^{37,38}$. GEM is an extension by Furihata
of the Dostrovsky evaporation model$^{39}$ as implemented in
LAHET$^{14}$ to include up to 66 types of particles and fragments that
can be evaporated from an excited compound nucleus plus a modification
of the version of Atchison's fission model$^{12,13}$ used in LAHET. Many 
of the parameters were adjusted for a better description of 
fission reactions when using it in conjunction with the extended 
evaporation model. We merged GEM2 (the last update of the GEM code)
with CEM2k as follows: we calculate the cascade and preequilibrium
stages of a reaction with our CEM2k, then we describe the subsequent 
evaporation of particles and fragments and fission from the remaining 
excited compound nuclei using GEM2. To understand the role of 
preequilibrium particle emission, we performed calculations of all the 
reactions we tested both with emission of preequilibrium particles and 
without them, {\it i.e.}, going directly to GEM2 after the intranuclear 
cascade stage of a reaction described by CEM2k.

A very detailed description of the GEM, together with a large amount
of results obtained for many reactions
using GEM coupled either with the Bertini or ISABEL INC models
in LAHET may be found in$^{37,38}$. Therefore, we present here
only the main features of GEM, following mainly$^{38}$ and using as
well useful information obtained in private communications from
Dr.\ Furihata.

\vspace{0.4cm}
\noindent {\bf A. Evaporation Model in GEM}\\
Furihata did not change in GEM the general algorithms used in LAHET
to simulate evaporation and fission. The decay widths of evaporated 
particles and 
fragments are estimated using the classical Weisskopf-Ewing statistical
model$^{40}$. In this approach, the decay probability $P_j$ for the 
emission of a particle $j$ from a parent compound nucleus $i$ with 
the total kinetic energy in the center-of-mass system between $\eps$ 
and $\eps + d \eps$ is
\beq
P_j(\eps) d \eps = g_j \sigma_{inv} (\eps )
{{\rho_d ( E -Q -\eps)}\over{\rho_i (E)}} \eps d \eps ,
\eeq
where $E$ [MeV] is the excitation energy of the parent nucleus $i$ with mass
$A_i$ and charge $Z_i$, and $d$ denotes a daughter nucleus with mass $A_d$ 
and charge $Z_d$ produced after the emission of ejectile $j$ with mass $A_j$ 
and charge $Z_j$ in its ground state. $\sigma_{inv}$ is the cross section for 
the inverse reaction, $\rho_i$ and $\rho_d$ are the level densities [MeV]$^{-1}$
of the parent and the daughter nucleus, respectively.
$g_j = (2S_j+1) m_j/\pi^2 \hbar^2$, where $S_j$ is the spin and $m_j$ is the mass
of the emitted particle $j$. The $Q$-value is calculated using the excess mass
$M(A,Z)$ as $Q=M(A_j,Z_j)+M(A_d,Z_d) - M(A_i,Z_i)$. In GEM, four mass tables 
are used to calculate $Q$-values, according to the following priority:
(1) the Audi-Wapstra mass table$^{18}$,
(2) theoretical masses calculated by M\"oller {\it et al.}$^{19}$,
(3) theoretical masses calculated by Comay {\it et al.}$^{41}$,
(4) the mass excess calculated using the old Cameron formula$^{42}$.
As does LAHET, GEM uses Dostrovsky's formula$^{39}$ to calculate
the inverse cross section $\sigma_{inv}$ for all emitted particles and 
fragments
\begin{equation}
\sigma_{inv} (\eps) = \sigma_{g} \alpha  \left(
1 + {\beta \over \eps} \right) \mbox{ ,}
\label{a2}
\end{equation}
which is often written as
$$
\sigma_{inv} (\eps) =
\cases{\sigma_g c_n (1 + b/ \eps)& for neutrons \cr
\sigma_g c_j (1- V/ \eps)& for charged particles ,\cr}
$$
where $\sigma_g = \pi R^2_b$ [fm$^2$] is the geometrical cross section, and
\beq
V = k_j Z_j Z_d e^2 / R_c
\eeq
is the Coulomb barrier in MeV. 

The new ingredient in GEM in comparison with LAHET which
considers evaporation of only 6 particles (n, p, d, t, $^3$He, and $^4$He)
is that Furihata included the possibility of evaporation of up to 66 types 
of particles and fragments and incorporated into GEM several sets of parameters
$b$, $c_j$, $k_j$, $R_b$, and $R_c$ for each particle.

The 66 ejectiles considered by GEM for evaporation 
are selected to satisfy the following criteria:
(1) isotopes with $Z_j \leq 12$;
(2) naturally existing isotopes or isotopes near the stability line;
(3) isotopes with half-lives longer than 1 ms. All the 66 ejectiles
considered by GEM are shown in Table 1.

\begin{center}
Table 1. The ejectiles considered by GEM

\vspace{2mm}
\begin{tabular}{rlllllll}
\hline\hline 
 $Z_j$\hspace{2mm} & \multicolumn{7}{l} {Ejectiles} \\
\hline
0\hspace{2mm}  & n       &         &         &         &         &         &         \\
1\hspace{2mm}  & p       &\hspace{1mm}   d     &\hspace{1mm}   t     &         &         &         &         \\
2\hspace{2mm}  &$^{3 }$He&\hspace{1mm}$^{4 }$He&\hspace{1mm}$^{6 }$He&\hspace{1mm}$^{8 }$He&         &         &         \\
3\hspace{2mm}  &$^{6 }$Li&\hspace{1mm}$^{7 }$Li&\hspace{1mm}$^{8 }$Li&\hspace{1mm}$^{9 }$Li&         &         &         \\
4\hspace{2mm}  &$^{7 }$Be&\hspace{1mm}$^{9 }$Be&\hspace{1mm}$^{10}$Be&\hspace{1mm}$^{11}$Be&\hspace{1mm}$^{12}$Be&         &         \\
5\hspace{2mm}  &$^{8 }$B &\hspace{1mm}$^{10}$B &\hspace{1mm}$^{11}$B &\hspace{1mm}$^{12}$B &$\hspace{1mm}^{13}$B &         &         \\
6\hspace{2mm}  &$^{10}$C &\hspace{1mm}$^{11}$C &\hspace{1mm}$^{12}$C &\hspace{1mm}$^{13}$C &\hspace{1mm}$^{14}$C &\hspace{1mm}$^{15}$C &\hspace{1mm}$^{16}$C \\
7\hspace{2mm}  &$^{12}$N &\hspace{1mm}$^{13}$N &\hspace{1mm}$^{14}$N &\hspace{1mm}$^{15}$N &\hspace{1mm}$^{16}$N &\hspace{1mm}$^{17}$N &         \\
8\hspace{2mm}  &$^{14}$O &\hspace{1mm}$^{15}$O &\hspace{1mm}$^{16}$O &\hspace{1mm}$^{17}$O &\hspace{1mm}$^{18}$O &\hspace{1mm}$^{19}$O &\hspace{1mm}$^{20}$O \\
9\hspace{2mm}  &$^{17}$F &\hspace{1mm}$^{18}$F &\hspace{1mm}$^{19}$F &\hspace{1mm}$^{20}$F &\hspace{1mm}$^{21}$F &         &         \\
10\hspace{2mm} &$^{18}$Ne&\hspace{1mm}$^{19}$Ne&\hspace{1mm}$^{20}$Ne&\hspace{1mm}$^{21}$Ne&\hspace{1mm}$^{22}$Ne&\hspace{1mm}$^{23}$Ne&\hspace{1mm}$^{24}$Ne\\
11\hspace{2mm} &$^{21}$Na&\hspace{1mm}$^{22}$Na&\hspace{1mm}$^{23}$Na&\hspace{1mm}$^{24}$Na&\hspace{1mm}$^{25}$Na&         &         \\
12\hspace{2mm} &$^{22}$Mg&\hspace{1mm}$^{23}$Mg&\hspace{1mm}$^{24}$Mg&
\hspace{1mm}$^{25}$Mg&\hspace{1mm}$^{26}$Mg&\hspace{1mm}$^{27}$Mg&
\hspace{1mm}$^{28}$Mg\\
\hline\hline 
\end{tabular}
\end{center}

GEM includes several options for the parameter set in expressions (2,3):

1) The ``simple'' parameter set is given as $c_n = c_j =k_j = 1$
$b=0$, and $R_b  = R_c = r_0 (A^{1/3}_j + A^{1/3}_d)$ [fm];
users need to input $r_0$.

2) The ``precise'' parameter set is used in GEM as default,
and we use this set in our present work.

A) For all light ejectiles
up to $\alpha$ ($A_j \leq 4$), the
parameters determined by Dostrovsky 
{\it et al.}$^{39}$ are used in GEM, namely:
$c_n = 0.76 + c_a A_d^{-1/3}$, $b=(b_a A^{-2/3}_d - 0.050)/
(0.76 + c_a A^{-1/3}_d)$
(and $b=0$ for $A_d \geq 192$), where $c_a = 1.93$ and $b_a = 1.66$,
$c_p = 1 + c$, $c_d = 1 + c/2$, $c_t = 1 + c/3$, $c_{^3He} = c_\alpha = 0$,
$k_p = k$, $k_d = k + 0.06$, $k_t = k + 0.12$,
$k_{^3He} = k_\alpha - 0.06$, where $c$, $k$, and $k_\alpha$ are listed
in Table 2 for a set of $Z_d$. Between the $Z_d$ values listed in Table 2,
$c$, $k$, and $k_\alpha$ are interpolated linearly.The nuclear distances 
are given by $R_b = 1.5 A^{1/3}$ for neutrons and protons, and
$1.5 (A^{1/3}_d + A^{1/3}_j)$ for d, t, $^3$He, and $\alpha$.

\begin{center}
Table 2. $k$, $k_\alpha$, and $c$ parameters used in GEM

\vspace{2mm}
\begin{tabular}{|c|c|c|c|}
\hline \hline
\hspace{5mm} $Z_d$\hspace{5mm} &\hspace{5mm} $k$\hspace{5mm} &\hspace{5mm}
 $ k_\alpha$\hspace{5mm} &\hspace{5mm} $c$\hspace{5mm}  \\
\hline
$\leq 20$ & 0.51 & 0.81 & 0.0 \\
30 & 0.60 & 0.85 & -0.06 \\
40 & 0.66 & 0.89 & -0.10 \\
$\geq 50$ & 0.68 & 0.93 & -0.10\\
\hline \hline
\end{tabular}
\end{center}

The nuclear distance for the Coulomb barrier is expressed as
$R_c = R_d + R_j$, where $R_d = r^c_0 A^{1/3}$, $r^c_0 = 1.7$,
and $R_j=0$ for neutrons and protons, and $R_j = 1.2$
for d, t, $^3$He, and $^4$He.  We note that several of these parameters 
are similar to the original values published by
Dostrovsky {\it et al.}$^{39}$ (and used, for example in our CEM codes)
but not exactly the same. Dostrovsky {\it et al.}$^{39}$ had $c_a = 2.2$, 
$b_a = 2.12$, and $r^c_0 = 1.5$. Also, for the $k$, $k_\alpha$, and $c$
parameters shown in Table 2, they had slightly different values, shown 
in Table 3. 

\begin{center}
Table 3. $k_p$, $c_p$, $k_\alpha$, and $c_\alpha$ parameters from Ref.$^{39}$

\vspace{2mm}
\begin{tabular}{|c|c|c|c|c|}
\hline \hline
\hspace{5mm} $Z_d$\hspace{5mm} &\hspace{5mm} $k_p$\hspace{5mm} &\hspace{5mm}
 $c_p$\hspace{5mm} & $ k_\alpha$\hspace{5mm} &\hspace{5mm} $c_\alpha$\hspace{5mm}  \\
\hline
10        & 0.42 & 0.50 & 0.68 & 0.10\\
20        & 0.58 & 0.28 & 0.82 & 0.10\\
30        & 0.68 & 0.20 & 0.91 & 0.10 \\
50        & 0.77 & 0.15 & 0.97 & 0.08 \\
$\geq 70$ & 0.80 & 0.10 & 0.98 & 0.06\\
\hline \hline
\end{tabular}
\end{center}

B) For fragments heavier that $\alpha$ ($A_j \geq 4$), ``the precise'' parameters 
of GEM use values by Matsuse {\it et al.}$^{43}$, namely:
$c_j = k = 1$,
$R_b = R_0(A_j) + R_0(A_d) + 2.85$ [fm],
$R_c = R_0(A_j) + R_0(A_d) + 3.75$ [fm], where
$R_0(A) = 1.12A^{1/3} - 0.86 A^{-1/3}$.

3) The upgraded version of GEM realized in the code GEM2 
contains two other options for the parameters of the inverse cross sections.

A) A set of parameters due to Furihata for light ejectiles in combination
with Matsuse's parameters for fragments heavier than $\alpha$. Furihata 
and Nakamura 
determined $k_j$ for p, d, t, $^3$He, and $\alpha$ as follows:$^{44}$
$$k_j = c_1 \log (Z_d) + c_2 \log (A_d) + c_3 .$$
The coefficients $c_1$, $c_2$, and $c_3$ for each ejectile are shown 
in Table 4.

\begin{center}
Table 4. $c_1$, $c_2$, and $c_3$ for p, d, t, $^3$He, and $\alpha$ from$^{44}$

\vspace{2mm}
\begin{tabular}{|c|c|c|c|}
\hline \hline
\hspace{5mm} Ejectile\hspace{5mm} &\hspace{5mm} $c_1$\hspace{5mm} &\hspace{5mm}
 $ c_2$\hspace{5mm} &\hspace{5mm} $c_3$\hspace{5mm}  \\
\hline
p        & 0.0615 & 0.0167 & 0.3227\\
d        & 0.0556 & 0.0135 & 0.4067\\
t        & 0.0530 & 0.0134 & 0.4374 \\
$^3$He   & 0.0484 & 0.0122 & 0.4938 \\
$\alpha$ & 0.0468 & 0.0122 & 0.5120\\
\hline \hline
\end{tabular}
\end{center}

When these parameters are chosen in GEM2, the following nuclear
radius $R$ is used in the calculation of $V$ and $\sigma_g$:

$$
R =\cases{0&for  $A=1$ ,\cr
   1.2&for $2\leq A \leq 4$ ,\cr
   2.02&for $5\leq A \leq 6$ ,\cr
   2.42&for $A=7$ ,\cr
   2.83&for $A=8$ ,\cr
   3.25&for $A=9$ ,\cr
   1.414A^{1/3}_d + 1&for $A \geq 10$ .\cr}
$$

B) The second new option in GEM2 is to use 
Furihata's parameters for light ejectiles up to $\alpha$ and the
Botvina {\it et al.}$^{45}$ parameterization for inverse cross sections
for heavier ejectiles. Botvina {\it et al.}$^{45}$ found that
$\sigma_{inv}$ can be expressed as
\beq
\sigma_{inv} = \sigma_g \cases{(1 - V / \eps)&for  $\eps \geq V + 1$ [MeV],\cr
   {e^{\alpha (\eps - V - 1)} \over {V+1}}&for $\eps < V + 1$ [MeV],\cr}
\eeq
where
$$ \alpha = 0.869 + 9.91 / Z_j ,$$
$$ V = {Z_j Z_d \over {r_0^b (A^{1/3}_j + A^{1/3}_d)}} ,$$
$$r_0^b = 2.173 {{1+6.103 \times 10^{-3}Z_jZ_d}
\over{1 + 9.443 \times 10^{-3}Z_j Z_d}} \mbox{ [fm].}$$

The expression of $\sigma_{inv}$ for $\eps < V + 1$ shows the fusion reaction
in the sub-barrier region. When using Eq.\ (4) instead of Eq.\ (2), the total
decay width for a fragment emission can not be calculated analytically.
Therefore, the total decay width must be calculated numerically and takes 
much CPU time.

The total decay width $\Gamma_j$ is calculated by integrating Eq.\ (1) with 
respect to the total kinetic energy $\eps$ from the Coulomb barrier $V$ up to the
maximum possible value, $(E-Q)$. The good feature of Dostrovsky's 
approximation for the inverse cross sections, Eq.\ (2),  
is its simple energy dependence that allows the analytic integration of Eq.\ (1).
By using Eq.\ (2) for $\sigma_{inv}$, the total decay width for the particle 
emission is
\beq
\Gamma_j = {{g_j\sigma_g \alpha}\over{\rho_i(E)}}
\int^{E-Q}_V \eps \Bigl(1 + {\beta\over\eps} \Bigr)
\rho_d (E -Q-\eps) d \eps .
\eeq
The level density $\rho(E)$ is calculated in GEM according to the Fermi-gas
model using the expression$^{46}$
\beq
\rho(E) = { \pi \over {12} } 
{ {\exp ({2\sqrt{a(E-\delta)}}) } \over {a^{1/4}(E-\delta)^{5/4}} } ,
\eeq
where $a$ is the level density parameter and $\delta$ is the pairing energy
in MeV. As does LAHET, GEM uses the $\delta$ values evaluated by Cook
{\it et al.}$^{47}$ For those values not evaluated by Cook {\it et al.},
$\delta$'s from Gilbert and Cameron$^{46}$ are used instead. The simplest
option for the level density parameter in GEM is $a = A_d /8$ [MeV$^{-1}$], but the
default is the Gilbert-Cameron-Cook-Ignatyuk (GCCI) parameterization from
LAHET:$^{14}$
\beq
a = \tilde a {{1 - e^{-u}} \over {u}} +
a_I\Biggl( 1 - {{1 - e^{-u}} \over {u}}\Biggr),
\eeq
where $u=0.05(E-\delta)$, and
$$a_I = (0.1375 -8.36 \times 10^{-5} A_d) \times A_d ,$$
$$
\tilde a = \cases{ A_d/8&for  $Z_d < 9$ or $N_d < 9$,\cr
 A_d(a' + 0.00917 S)&for others.\cr}
$$
For deformed nuclei with 
$54 \leq Z_d \leq 78$, 
$86 \leq Z_d \leq 98$,
$86 \leq N_d \leq 122$,
or $130\leq N_d \leq 150$,
$a' = 0.12$ while $a' = 0.142$ for other nuclei. The shell corrections $S$ 
is expressed as a sum of separate contributions from neutrons and protons,
{\it i.e.} $S = S(Z_d) + S(N_d)$ from$^{46,47}$ and are tabulated in$^{38}$.

The level density is calculated using Eq.\ (6) only for high excitation
energies, $E \geq E_x$, where $E_x = U_x + \delta$ and $U_x = 2.5 + 150/A_d$
(all energies are in MeV). At lower excitation energies,
the following$^{46}$ is used for the level density:
\beq
\rho(E) = {\pi \over {12} } {1 \over T} \exp((E-E_0)/T) ,
\eeq
where $T$ in the nuclear temperature defined as $1/T = \sqrt{a/U_x} - 
1.5/U_x$. To provide a smooth connection of Eqs.\ (6) and (8) at $E=E_x$,
$E_0$ is defined as 
$E_0 = E_x - T(\log T - 0.25 \log a - 1.25 \log U_x + 2 \sqrt{aU_x})$.

For $E-Q-V < E_x$, substituting Eq.\ (8) into Eq.\ (5) we can calculate the
integral analytically, if we neglect the dependence of the level density parameter
$a$ on $E$:
\beq
\Gamma_j = { {\pi g_j \sigma_g \alpha} \over {12 \rho_i (E)} } \nonumber \\ 
\{I_1(t,t) + (\beta + V) I_0(t)\} , \nonumber
\eeq
where $I_0(t)$ and $I_1(t,t_x)$ are expressed as
\bea
I_0(t) &=& e^{-E_0/T}  (e^t - 1) , \nonumber \\
I_1(t,t_x) &=& e^{-E_0/T}  T\{(t-t_x + 1) e^{t_x} -t -1\} ,  \nonumber 
\eea
where $t = (E-Q-V)/T$ and $t_x = E_x/T$.
For $E-Q-V \geq E_x$, the integral of Eq.\ (5) cannot be solved analytically 
because of the denominator in Eq.\ (6). However, it is approximated as
\bea
\Gamma_j &=& { {\pi g_j \sigma_g \alpha} \over {12 \rho_i (E)} }
[ I_1(t,t_x) + I_3(s,s_x)e^s + (\beta + V) \nonumber \\ 
&\times & \{ I_0(t_x) - I_2(s,s_x) e^s \} ] ,
\eea
where $I_2(s,s_x)$ and $I_3(s,s_x)$ are given by
\bea
I_2(s,s_x) &=& 2 \sqrt{2} \{s^{-3/2} + 1.5 s^{-5/2} + 3.75 s^{-7/2} \nonumber \\
&-& (s^{-3/2}_x + 1.5 s^{-5/2}_x + 3.75s^{-7/2}_x) e^{s_x -s} \} , \nonumber
\eea
\bea
I_3(s,s_x) 
&=& (\sqrt{2}a)^{-1} [ 2 s^{-1/2} + 4 s^{-3/2} + 13.5s^{-5/2} \nonumber \\
&+& 60.0s^{-7/2} + 325.125s^{-9/2}  \nonumber \\
&-& \{(s^2-s^2_x)s^{-3/2}_x + (1.5s^2 + 0.5 s^2_x)s^{-5/2}_x \nonumber \\
&+& (3.75s^2 + 0.25s^2_x) s^{-7/2}_x \nonumber \\
&+& (12.875s^2 + 0.625s^2_x)s^{-9/2}_x \nonumber \\
&+& (59.0625s^2 + 0.9375s^2_x) s^{-11/2}_x \nonumber \\
&+& (324.8s^2_x + 3.28s^2_x ) s^{-13/2}_x \} e^{s_x - s} ], \nonumber 
\eea
with $s=2\sqrt{a(E-Q-V-\delta)}$ and $s_x = 2 \sqrt{a(E_x-\delta)}$.   

The ejectile $j$ to be evaporated is selected in GEM by the Monte Carlo
method according to the probability distribution calculated as 
$P_j = \Gamma_j / \sum_j \Gamma_j$, where $\Gamma_j$ is given by
Eqs.\ (9) or (10). The total kinetic energy $\eps$ of the emitted
particle $j$ and the recoil energy of the daughter nucleus is chosen 
according to the probability distribution given by Eq.\ (1). The angular
distribution of ejectiles is simulated to be isotropic in the 
center-of-mass system.

According to Friedman and Lynch$^{48}$,
it is important to include excited states in the particle emitted via the 
evaporation process along with evaporation of particles in their ground states,
because it greatly enhances the yield of heavy particles.
Taking this into consideration, GEM includes evaporation of complex particles and
light fragments both in the ground states and excited states. An excited
state of a fragment is included in calculations if its half-lifetime $T_{1/2}(s)$
satisfies the following condition:
\beq
{ T_{1/2} \over {\ln 2} } > { \hbar \over {\Gamma_j^*} } ,
\eeq
where $\Gamma_j^*$ is the decay width of the excited particle (resonance).
GEM calculates  $\Gamma_j^*$ in the same manner as for a ground-state particle 
emission.
The $Q$-value for the resonance emission is expressed as $Q^* = Q + E^*_j$,
where $E^*_J$ is the excitation energy of the resonance. The spin state of
the resonance $S^*_j$ is used in the calculation of $g_j$, instead of the
spin of the ground state $S_j$. GEM uses the ground state masses $m_j$ for
excited states because the difference between the masses is negligible.

Instead of treating a resonance as an independent particle, GEM simply enhances the
decay width $\Gamma_j$ of the ground state particle emission as follows:
\beq
\Gamma_j = \Gamma^0_j + \sum_{n} \Gamma^n_j ,
\eeq
where $\Gamma^0_j$ is the decay width of the ground state particle
emission, and $\Gamma^n_j$ is that of the $n$th excited state 
of the particle $j$ emission which satisfies Eq.\ (11).

The total kinetic energy distribution of the excited particles is 
assumed to be the same as that of the ground state particle emission.
$S^*_j$, $E^*_j$, and $T_{1/2}$ used in GEM are extracted from the
Evaluated Nuclear Structure Data File (ENSDF) database 
maintained by the National Nuclear Data Center at Brookhaven National 
Laboratory$^{49}$. 

Note that when including evaporation of up to 66 particles in GEM,
its running time increases many times compared to the case when
evaporating only 6 particles, up to $^4$He. The major particles emitted from
an excited nucleus are n, p, d, t, $^3$He, and $^4$He. For most of the
cases, the total emission probability of particles heavier than $\alpha$
is negligible compared to those for the emission of light ejectiles. 
To keep the GEM running time to a reasonable level, calculations
of probability of emission of particles heavier than $\alpha$ are done
only for 5\% of simulated evaporations for $A_i > 40$,
 7\% of simulated evaporations for $30 < A_i \leq  40$, and
 30\% of simulated evaporations for $20 < A_i \leq 30$.
This empirical criterion was determined by Furihata from actual 
simulations$^{38}$.

\vspace{0.4cm}
\noindent {\bf B. Fission Model in GEM}\\
The fission model used in GEM is based on Atchison's model$^{12,13}$
as implemented in LAHET$^{14}$,
often referred in the literature as the Rutherford Appleton Laboratory (RAL)
model, which is where Atchison developed it. There are two choices of 
parameters for the fission model: one of them is the original parameter set by
Atchison$^{12,13}$ as implemented in LAHET$^{14}$, and the other is a
parameter set evaluated by Furihata$^{37,38}$.\\

\noindent{\bf B.1. Fission Probability.}
The Atchison fission model is designed to only describe fission of
nuclei with $Z \geq 70$. It assumes that fission competes only with
neutron emission, {\it i.e.}, from the widths $\Gamma_j$ of n, p, d, 
t, $^3$He, and $^4$He,
the RAL code calculates the probability of evaporation of any 
particle. When a charged particle is selected to be evaporated, 
no fission competition is taken into account. When a neutron is
selected to be evaporated, the code does not actually simulate its evaporation,
instead it considers that fission may compete,
and chooses either fission or evaporation of a neutron according to
the fission probability $P_f$. This quantity is treated by the RAL code differently
for the elements above and below $Z=89$. The reasons Atchison split the 
calculation of the fission probability $P_f$ are: (1) there is
very little experimental information on fission in the region $Z=85$ to 88,
(2) the marked rise in the fission barrier for nuclei with $Z^2/A$ below
about 34 (see Fig.\ 2 in$^{13}$) together with the disappearance of asymmetric 
mass splitting, indicates that a change in the character of the fission
process occurs. If experimental information were available, a split between
regions about $Z^2/A \approx 34$ would more sensible$^{13}$.

1) $70 \leq Z_j \leq 88$. For fissioning nuclei with $70 \leq Z_j \leq 88$,
GEM uses the original Atchison calculation of the neutron emission
width $\Gamma_n$ and fission width $\Gamma_f$ to estimate the fission
probability as
\beq
P_f = {\Gamma_f \over {\Gamma_f + \Gamma_n} } = {1 \over {1+ \Gamma_n/\Gamma_f} }.
\eeq
Atchison uses$^{12,13}$ the Weisskopf and Ewing statistical model$^{40}$
with an energy-independent pre-exponential factor for the level density 
(see Eq.\ (6)) and Dostrovsky's$^{39}$ inverse cross section for neutrons
and estimates the neutron width $\Gamma_n$ as
\bea
\Gamma_n = 0.352 \bigl(1.68 J_0 + 1.93A_i^{1/3}J_1 \nonumber \\
+ A_i^{2/3}(0.76J_1 - 0.05 J_0)\bigr),
\eea
where $J_0$ and $J_1$ are functions of the level density parameter $a_n$
and $s_n (=2\sqrt{a_n(E-Q_n-\delta)})$ as
$$J_0 = { {(s_n-1) e^{s_n} + 1} \over {2 a_n} },$$
$$J_1 = { {(2s_n^2 - 6s_n + 6) e^{s_n} + s_n^2 -6} \over {8a_n^2} }.$$
Note that the RAL model uses
a fixed value for the level density parameter $a_n$, namely
\beq
a_n = (A_i - 1) /8,
\eeq
and this approximation is kept in GEM when
calculating the fission probability according to Eq.\ (13), though it differs from
the GCCI parameterization (7) used in GEM to calculate particle
evaporation widths.
The fission width for nuclei with $70 \leq Z_j \leq 88$ is calculated in the RAL 
model and in GEM as
\beq
\Gamma_f = { {(s_f - 1)e^{s_f} + 1} \over a_f },
\eeq
where $s_f = 2 \sqrt{a_f (E-B_f - \delta)}$ and the level density parameter
in the fission mode $a_f$ is fitted by Atchison to describe the measured
$\Gamma_f / \Gamma_n$ as:$^{13}$
\beq
a_f = a_n \Bigl(1.08926 + 0.01098 ( \chi - 31.08551)^2\Bigr),
\eeq
and $\chi = Z^2/A$.
The fission barriers $B_f$ [MeV] are estimated as
\beq
B_f = Q_n + 321.2 - 16.7 { {Z^2_i} \over A} + 0.218 
\Biggl({ {Z^2_i} \over {A_i} }\Biggr)^2 .
\eeq
Note that neither the angular momentum nor the excitation energy of the 
nucleus are taken into account in the estimate of the fission barriers.

2) $Z_j \geq 89$. For heavy fissioning nuclei with $Z_j \geq 89$
GEM follows the RAL model$^{12,13}$ and does not calculate at all
the fission width $\Gamma_f$ and does not use Eq.\ (13) to estimate
the fission probability $P_f$. Instead, the following semi-empirical
expression obtained by Atchison$^{12,13}$ by approximating the experimental
 values of
$\Gamma_n / \Gamma_f$ published by
Vandenbosch and Huizenga$^{50}$ is used to calculate
the fission probability:
\beq
\log (\Gamma_n / \Gamma_f ) = C(Z_i) ( A_i - A_0(Z_i)),
\eeq
where $C(Z)$ and $A_0(Z)$ are constants dependent on the nuclear charge $Z$
only. The values of these constants are those
used in the current version of LAHET$^{14}$ and are tabulated in Table 5 
(note that some adjustments of these values have been done since
Atchison's papers$^{12,13}$ were published).  

\begin{center}
Table 5. $C(Z)$ and $A_0(Z)$ values used in GEM

\vspace{2mm}
\begin{tabular}{|c|c|c|}
\hline \hline
\hspace{5mm} Z\hspace{5mm} &\hspace{5mm} $C(Z)$\hspace{5mm} &\hspace{5mm}
 $ A_0(Z)$\hspace{5mm} \\
\hline
89 & 0.23000 & 219.40\\
90 & 0.23300 & 226.90\\
91 & 0.12225 & 229.75\\
92 & 0.14727 & 234.04\\
93 & 0.13559 & 238.88\\
94 & 0.15735 & 241.34\\
95 & 0.16597 & 243.04\\
96 & 0.17589 & 245.52\\
97 & 0.18018 & 246.84\\
98 & 0.19568 & 250.18\\
99 & 0.16313 & 254.00\\
100& 0.17123 & 257.80\\
101& 0.17123 & 261.30\\
102& 0.17123 & 264.80\\
103& 0.17123 & 268.30\\
104& 0.17123 & 271.80\\
105& 0.17123 & 275.30\\
106& 0.17123 & 278.80\\
\hline \hline
\end{tabular}
\end{center}

In this approach the fission probability $P_f$ is
independent of the excitation energy of the fissioning nucleus
and its angular momentum.\\

\noindent{\bf B.2. Mass Distribution.}
The selection of the mass of the fission fragments depends on whether the
fission is symmetric or asymmetric. For a pre-fission nucleus with
$Z^2_i/A_i \leq 35$, only symmetric fission is allowed. For $Z^2_i/A_i > 35$,
both symmetric and asymmetric fission are allowed, depending on the
excitation energy of the fissioning nucleus. No new parameters were determined
for asymmetric fission in GEM.

For nuclei with $Z^2_i/A_i > 35$, whether the fission is symmetric or not is 
determined by the asymmetric fission probability $P_{asy}$
\beq
P_{asy} = { {4870e^{-0.36E}} \over {1+4870e^{-0.36E} } } .
\eeq
{\bf B.2.a. Asymmetric fission.} For asymmetric fission, the mass of one of the 
post-fission fragments $A_1$ is selected from a Gaussian distribution of mean 
$A_f = 140$ and width $\sigma_M = 6.5$. The mass of the second fragment is
$A_2 = A_i -A_1$.

\noindent{\bf B.2.b. Symmetric fission.} For symmetric fission, $A_1$ is selected from
the Gaussian distribution of mean $A_f = A_i/2$ and two options for the width $\sigma_M$
as described below.

The first option for choosing $\sigma_M$ is the original Atchison approximation:
\beq
\sigma_M =
\cases{3.97+0.425(E-B_f)-0.00212(E-B_f)^2 ,& \cr 
25.27 ,&\cr}
\eeq
 for $(E-B_f)$ below or above 100 MeV, respectively. In this expression all
values are in MeV and the fission barriers $B_f$ are calculated according
to Eq.\ (18) for nuclei with $Z_i \leq 88$. For nuclei with $Z_i > 88$,
the expression by Neuzil and Fairhall$^{51}$ is used:
\beq
B_f = C - 0.36 (Z^2_i/A_i) ,
\eeq
where $C = 18.8, 18.1, 18.1,$ and $18.5$ [MeV] for odd-odd, even-odd, 
odd-even, and even-even nuclei, respectively.
 
The second option in GEM for $\sigma_M$ (used here) was found by 
Furihata$^{37,38}$ as:
\beq
\sigma_M = C_3 (Z^2_i/A_i)^2 + C_4 (Z^2_i/A_i) + C_5(E-B_f) + C_6 .
\eeq
The constants $C_3=0.122$, $C_4 =-7.77$, $C_5=3.32\times10^{-2}$, and
$C_6=134.0$ were obtained by fitting with GEM the recent Russian collection of
experimental fission-fragment mass distributions$^{52}$. In this expression, 
the fission barriers $B_f$ by Myers and Swiatecki$^{53}$ are used.
More details may be found in Ref.$^{38}$\\

\noindent{\bf B.3. Charge Distribution.}
The charge distribution of fission fragments is assumed to be a Gaussian distribution 
of mean $Z_f$ and width $\sigma_Z$. $Z_f$ is expressed as
\beq
Z_f = { {Z_i+Z'_1 -Z'_2} \over 2} ,
\eeq
where
\beq
Z_l' = {{65.5A_l} \over {131+A_l^{2/3}}}, \mbox{$l=1$ or 2}.
\eeq
The original Atchison model uses $\sigma_Z = 2.0$. An investigation
by Furihata$^{38}$ suggests that $\sigma_Z = 0.75$ provides a better 
agreement with data; therefore $\sigma_Z = 0.75$ is used in GEM
and in our calculations.\\

\noindent{\bf B.4. Kinetic Energy Distribution.}
The kinetic energy of fission fragments [MeV] is determined by a
Gaussian distribution with mean $\eps_f$ and width $\sigma_{\eps_f}$.

The original parameters in the Atchison model are:\\
$$ \eps_f = 0.133Z_i^2/A_i^{1/3} - 11.4 ,$$
$$\sigma_{\eps_f} = 0.084 \eps_f .$$\\

Furihata's parameters in GEM, which we also use, are:
\beq
\eps_f =
\cases{0.131 Z^2_i/A_i^{1/3} ,& \cr 
0.104Z^2_i/A_i^{1/3} + 24.3 ,&\cr}
\eeq
for $Z^2_i/A^{1/3}_i \leq 900$ and $900 < Z^2_i/A_i^{1/3} \leq 1800$,
respectively, according to Rusanov et al.$^{52}$
By fitting the experimental data by Itkis {et al.}$^{54}$, Furihata
found the following expression for $\sigma_{\eps_f}$
\beq
\sigma_{\eps_f} =
\cases{C_1 (Z^2_i/A^{1/3}_i - 1000) + C_2 ,& \cr 
C_2 ,&   \cr}
\eeq
for $Z^2_i/A_i^{1/3}$ above and below 1000, respectively, and the values
of the fitted constants are $C_1 = 5.70 \times 10^{-4}$ and
$C_2 = 86.5$
The experimental data used by Furihata for fitting are the extrapolated 
values to the nuclear temperature 1.5 MeV by Itkis {\it et al.}$^{54}$
More details may be found in$^{38}$.

We note that Atchison also has modified his original version using 
recent data and published$^{55}$ an improved (and more complicated)
parameterization for many quantities and distributions in his model,
but these modifications$^{55}$ are not yet included either in LAHET or 
in GEM.

\vspace{0.4cm}
\noindent {\bf C. Results from CEM2k+GEM2}\\
We have merged the GEM2 code with our CEM2k, initially keeping all
the default options in GEM2. We began by concentrating on an analysis of the 
recent GSI measurements in inverse kinematics$^{6-9}$ as the richest and
best data set for testing this kind of model. 
As mentioned above, to understand the role of preequilibrium
particle emission, we performed calculations of all the reactions we
tested both taking into account preequilibrium particle emission and
ignoring it, {\it i.e.}, going directly to GEM2 after the intranuclear 
cascade stage of a reaction described by CEM2k. 
The size of the present paper allows us to present only
a few results for one reaction measured at GSI, which we choose to be
p(800 MeV) + Au$^{9}$. Results for other reactions may be found in$^{56}$.

\vspace*{-6mm}
\begin{center}
\begin{figure} 
\vspace*{-40mm}
\hspace*{-24.mm}
\includegraphics[angle=-0,width=12.5cm]{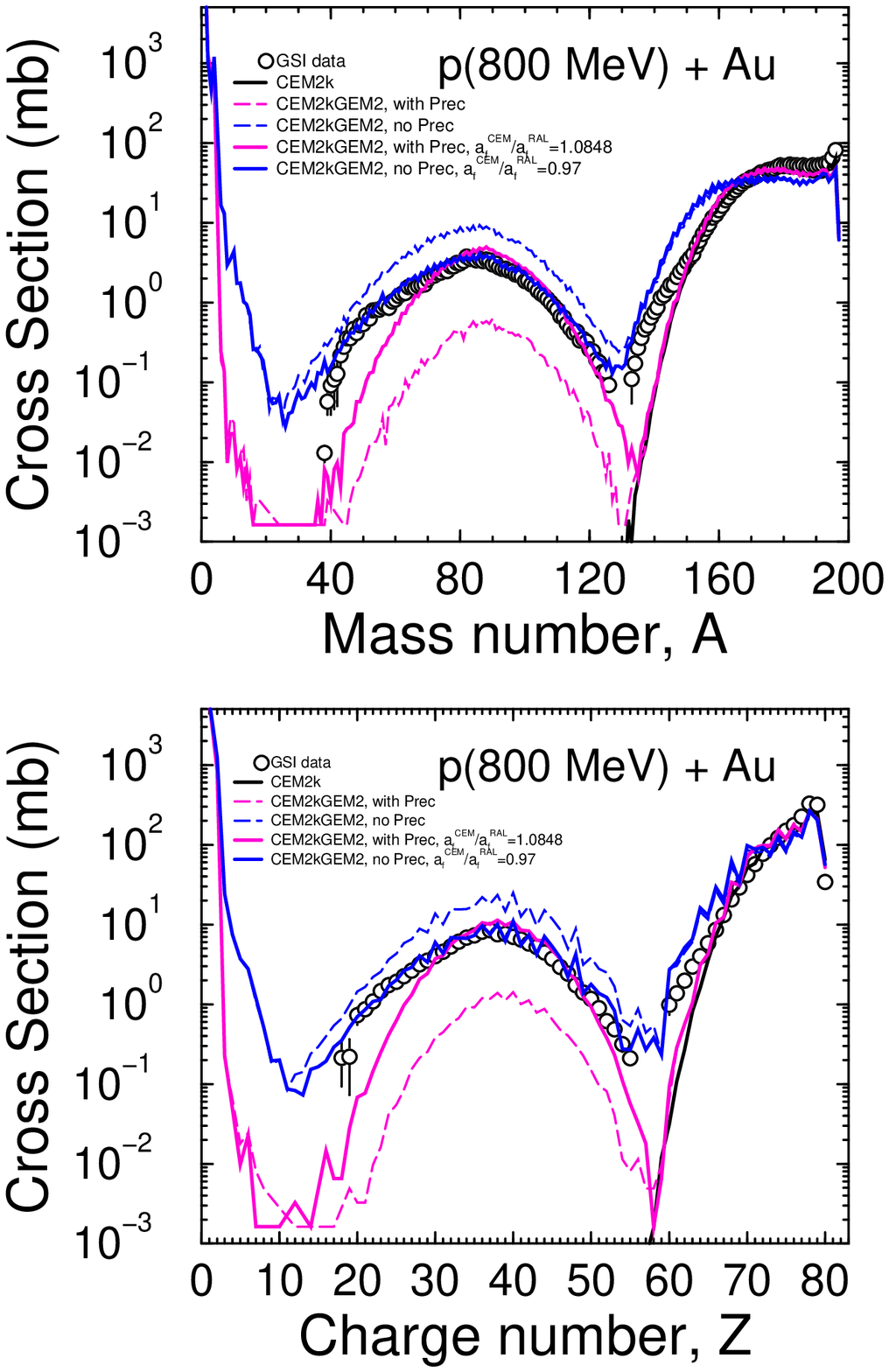}

\vspace*{-24mm}
{\noindent Figure 5. 
Comparison of the experimental$^{9}$ mass and charge distributions of the
nuclides produced in the reaction p(800 MeV) + Au (circles) with
different calculations. The black solid lines show our calculations
with CEM2k without contributions from fission. The dashed red and
blue lines show results found by merging CEM2k with GEM2 without any
modifications when preequilibrium emission is (red lines) or is not 
(blue lines) included. Solid color lines show results
from CEM2k+GEM2 with a modified $a_f$: red lines are
for the case with preequilibrium emission ($a_f^{CEM} / a_f^{RAL} = 1.0848)$
and blues lines show the results without preequilibrium emission
($a_f^{CEM} / a_f^{RAL} = 0.97)$.
}
\label{fig5}
\end{figure}
\end{center}

If we merge GEM2 with CEM2k without any modifications, the new code
does not describe correctly the fission cross section (and the yields of
fission fragments) whether we take into account preequilibrium emission
(see the dashed red line on Fig.\ 5) or not (see the dashed blue
line on Fig.\ 5). Such results were anticipated, as Atchison
fitted the parameters of his RAL fission model when it was coupled 
with the Bertini INC$^{57}$ which differs
from our INC. In addition, he did not model 
preequilibrium emission. Therefore, the distributions of fissioning
nuclei in $A$, $Z$, and excitation energy $E^*$ simulated by Atchison
differ significantly of the distributions we get; 
as a consequence, all the fission characteristics are also different.

Furihata used GEM2 coupled either with the Bertini INC$^{57}$ or with
the ISABEL$^{58}$ INC code, which also differs from our INC, and did 
not include preequilibrium particle emission. Therefore the
real fissioning nuclei simulated by Furihata differ from the ones in
our simulations, and the parameters adjusted by Furihata to work the best 
with her INC should not be the best for us. To get a good description 
of the fission cross section (and fission-fragment yields)
we need to modify at least one parameter in GEM2, namely to adjust
the level density parameter $a_f$ to get the correct fission cross 
section (see Eq.\ (17)), 
in the case of fissioning nuclei with $Z \leq 88$ (pre-actinides), and 
the parameter $C(Z)$ (see Eq.\ (19) and Tab.\ 5) for fissioning nuclei
with $Z > 88$ (actinides). From the dashed lines on Fig.\ 5 we see that we 
need to enlarge $a_f$ in our code to get a proper fission cross section
when we include preequilibrium emission
(the excitation energy of our fissioning nuclei and their $A$ and $Z$ are
smaller than provided by the Bertini or ISABEL INC without preequilibrium),
and we need to decrease $a_f$ in the case without preequilibrium.
By increasing $a_f$ by 1.0848 compared
with the original RAL and GEM2 value ($a_f^{CEM} /a_f^{RAL} = 1.0848$), 
we are able to reproduce correctly the fission cross section when we 
take into account preequilibrium emission (below, we label such 
results as ``with Prec'').
In the case with no preequilibrium emission, a proper
fission cross section is obtained for $a_f^{CEM} /a_f^{RAL} = 0.97$
(we label such results as ``no Prec''). We choose
these values for $a_f$ for all our further calculations of this reaction
and do not change any other parameters. 

The solid lines in Fig.\ 5 show results with
these values of $a_f$. One can see that the ``no Prec''
version provide a good description of both the mass and charge
distributions and agrees better with the data for these characteristics
than the ``with Prec'' version
(that is not true for isotopic distributions of individual elements,
as we show below).
The ``with Prec'' version reproduces correctly the 
position of the maximum in both $A$ and $Z$ distributions and the yields
of fission fragments not too far from these maximums, but the calculated
distributions are narrower than the experimental ones.
This is again because both Atchison and Furihata fitted
their $A$ and $Z$ distributions using models without preequilibrium
emission, which provide higher values for the excitation energy,
$A$, and $Z$ of fissioning nuclei. This means that to get a good
description of  $A$ and $Z$ distributions for fission fragments using
GEM2 in CEM2k ``with Prec'', we would need to modify the 
$A$ and $Z$ distributions of fission fragments in GEM2
(see Sec. B.2 and B.3), making them wider. This would take us beyond the
scope of the present work and here we do not vary any more parameters 
than we have already discussed.

Fig.\ 6 shows the GSI measurements$^{9}$ of the $A$ and
$Z$ distributions of the kinetic energy of products from the same reaction
compared with our CEM2k+GEM2 calculations both with and without
preequilibrium emission. Both versions of our calculations are in
reasonable agreement with the data, with slightly better agreement for the
version ``no Prec''. These results suggest to us that a small adjustment
of kinetic-energy distribution parameters in GEM (see Sec. B.4) may be
required for a better description of the data with our CEM2k+GEM2 code.

\begin{center}
\begin{figure} 
\vspace*{-40mm}
\hspace*{-24.mm}
\includegraphics[angle=-0,width=12.5cm]{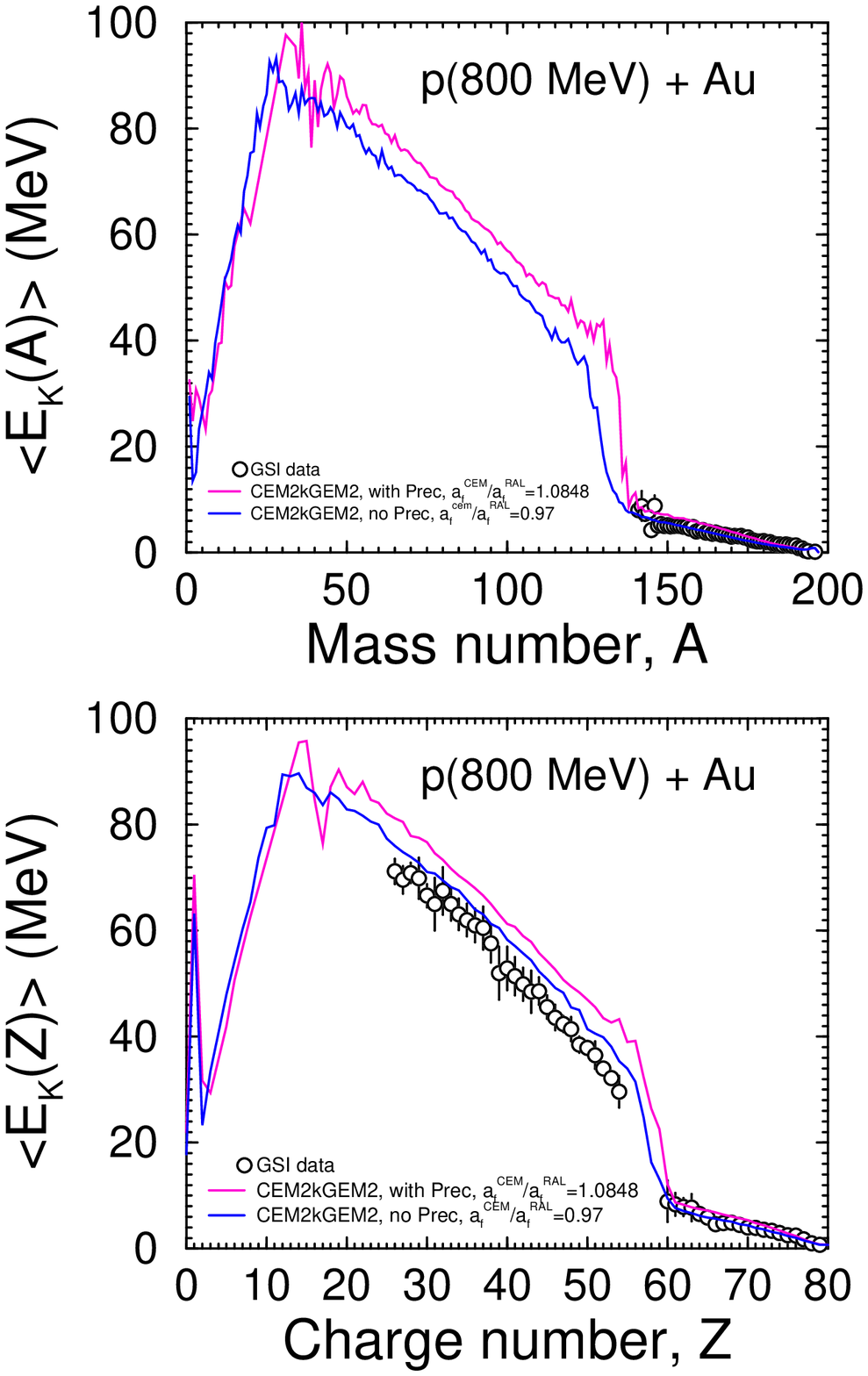}

\vspace*{-29mm}
{\noindent Figure 6. 
Comparison of the experimental$^{9}$ mass and charge distributions of the
kinetic energy of the 
nuclides produced in the reaction p(800 MeV) + Au (circles) with
out CEM2k+GEM2 calculations: ``with Prec'' results are shown
by red lines, ``no Prec'' results are shown by blue lines.
}
\label{fig6}
\end{figure}
\end{center}

\vspace*{-0.8cm}

Mass and charge distributions of the yields or kinetic energies of the
nuclides produced show only general trends and are not sensitive enough
to the details of a reaction. It is much more informative to study the
characteristics of individual nuclides and particles produced in a reaction.
Fig.\ 7 shows a comparison of the experimental data on production yields 
of twelve separate isotopes with $Z$ lying from 20 to 80 from the same 
reaction measured at GSI$^{9}$ with our calculations using both the
``with Prec'' (upper plot) and ``no Prec'' (lower plot) versions. 

\begin{center}
\begin{figure} 
\vspace*{-34mm}
\hspace*{-11.mm}
\includegraphics[angle=-0,width=10.0cm]{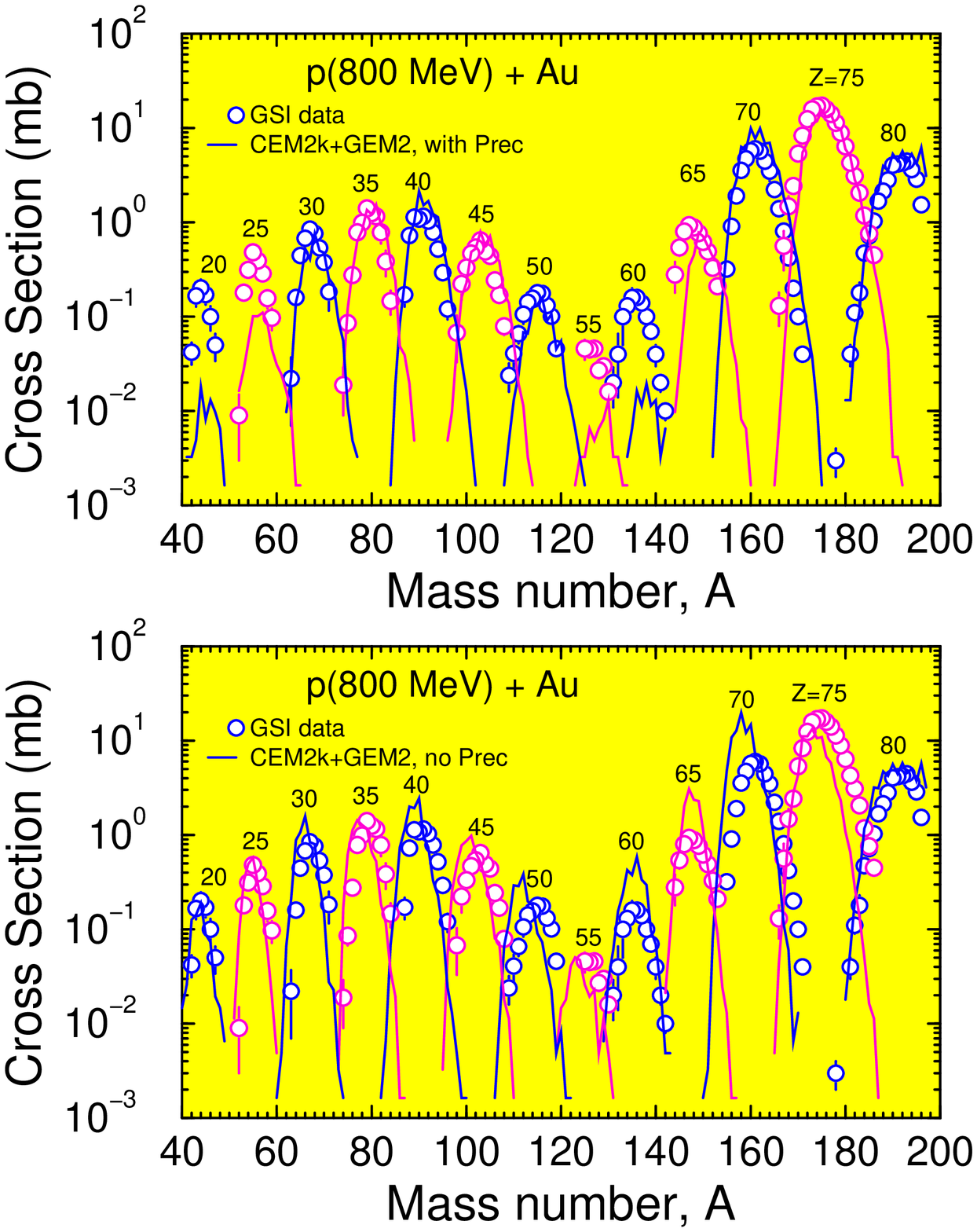}

\vspace*{-20mm}
{\noindent Figure 7. 
Experimental$^{9}$ mass distributions of the cross sections 
of twelve isotopes with the charge $Z$ from 20 to 80
compared with our CEM2k+GEM2 calculations. ``With Prec'' 
results are shown on the upper plot, while ``no Prec'' results 
are shown in the lower one.
}
\label{fig7}
\end{figure}
\end{center}
\vspace{-1.1cm}
The agreement (or disagreement) of our calculations with these data
is different from what we have for the integral
$A$ or $Z$ distributions in Figs.\ 5
and 6: We see that for the isotopes produced in the spallation
region (not too far from the target) and for fission fragments in
the region with the maximum yield, the version ``with Prec'' agree
much better with the data than the version ``no Prec''.
Only for production of isotopes at the border between spallation
and fission and between fission and fragmentation does
the version ``with Prec'' underestimates the data, due to too narrow
$A$ and $Z$ distributions in the simulation of fission fragments, as
we discussed previously. The ``no Prec'' version agrees better with
the data in these transition regions but are in worse
agreement for isotopes both in the spallation region and in the
middle of the fission region. We conclude that if a model agrees well
with some $A$ or $Z$ distributions it does not necessarily mean that 
it also describes well production of separate isotopes.
In other words, integral $A$ and $Z$ distributions
are not sensitive enough to develop and test such models, 
a practice which has often used in the literature.

It is more difficult for any model to describe correctly the
energy dependence for the production cross sections
of different isotopes, {\it i.e.,} excitation functions.
We calculated using both the ``with Prec'' and ``no Prec'' versions
of CEM2k+GEM2 all the excitation functions for the same reaction, p + Au,
for proton energies from 10 MeV to 3 GeV and compared our results with
all available data from our compilation referred to here as T-16 Library
(``T16 Lib'')$^{59}$. Only several typical examples from our comparison
are shown below.
 Figs.\ 8 and 9 show two examples of excitation functions for the production
of several isotopes in the spallation region. One can see a not too good 
but still reasonable agreement of both calculations with the data. 
Nevertheless, the version ``with Prec'' reproduces these experimental 
excitation functions better that the version ``no Prec''. Similar results 
were obtained for excitation functions of many other isotopes in the 
spallation region.

\begin{center}
\begin{figure} 
\vspace*{-13mm}
\hspace*{-16.mm}
\includegraphics[angle=-0,width=10.7cm]{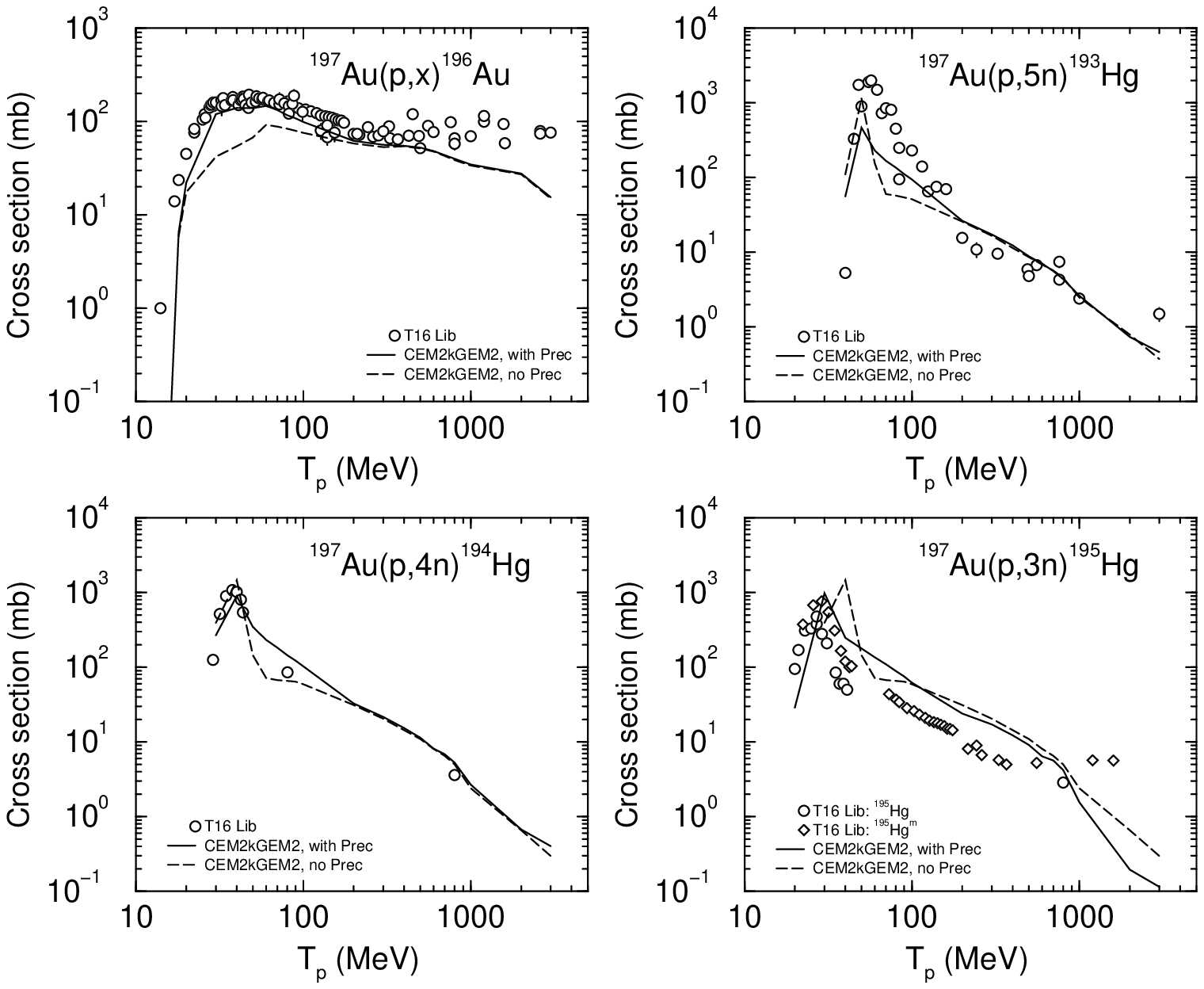}

\vspace*{-70mm}
{\noindent Figure 8. 
Excitation functions for the production of $^{196}$Au, $^{193}$Hg,
$^{194}$Hg, and $^{195}$Hg from p+$^{197}$Au. Results by CEM2k+GEM2
``with Prec'' are shown by solid lines and ``no Prec'' by dashed lines.
Experimental data (symbols) are from our LANL compilation 
(T16 Lib)$^{59}$ and are available from the authors upon request.
}
\label{fig8}
\end{figure}
\end{center}

\begin{center}
\begin{figure} 
\vspace*{-30mm}
\hspace*{-16.mm}
\includegraphics[angle=-0,width=10.7cm]{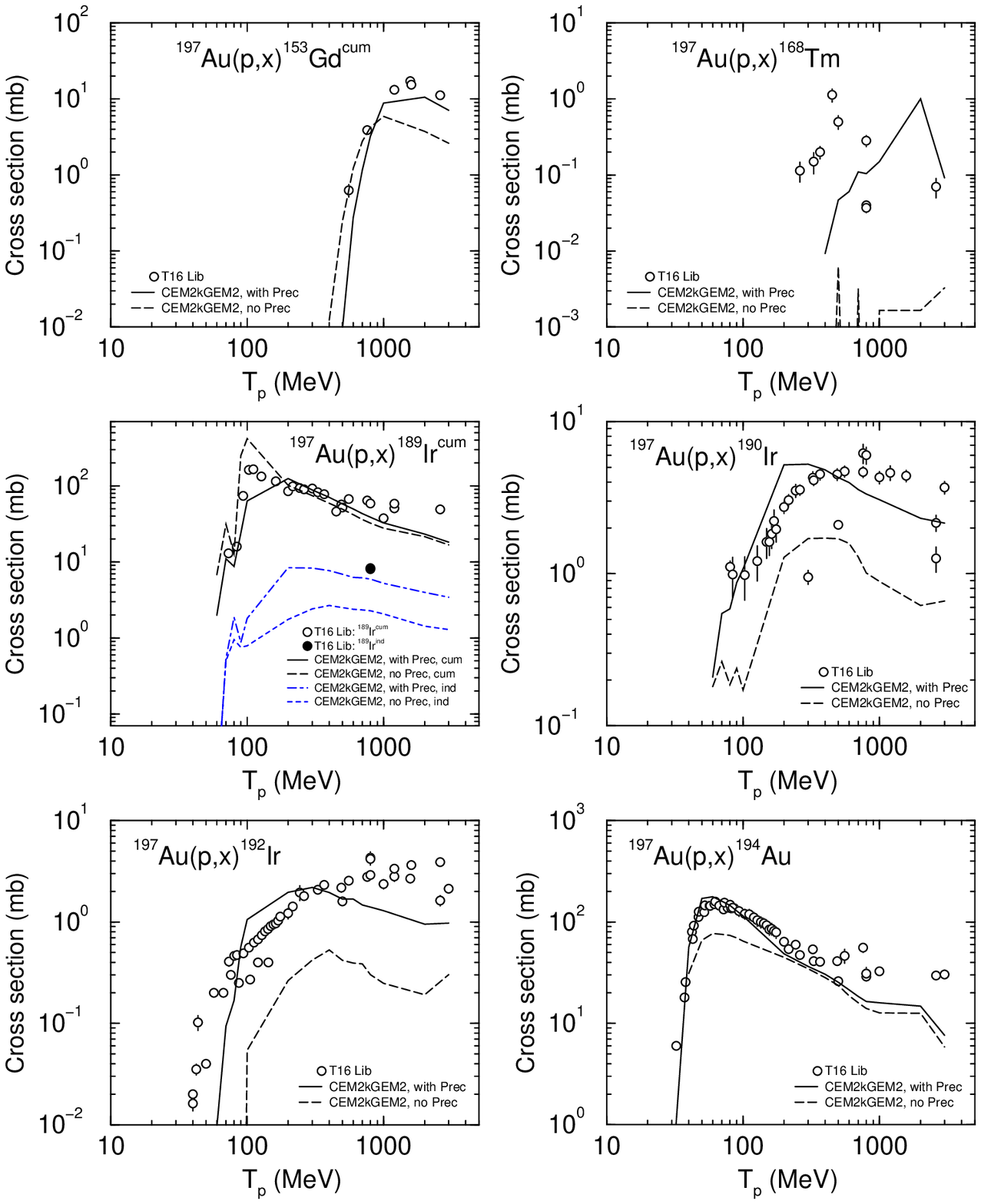}

\vspace*{-37mm}
{\noindent Figure 9. 
The same as Fig.\ 8 but for the production of
$^{153}$Gd$^{cum}$, $^{168}$Tm, $^{189}$Ir$^{cum}$, $^{190}$Ir, 
$^{192}$Ir, and $^{194}$Au. Note that for $^{153}$Gd, only the
cumulative yield is shown, where measured and calculated
cross sections contain contributions not only from a direct production
of $^{153}$Gd (``independent yield''), but also from all its 
decay-chain precursors.  For $^{189}$Ir, both cumulative and independent 
cross sections are shown.
}
\label{fig9}
\end{figure}
\end{center}
\vspace*{-17.0mm}

Figs.\ 10 and 11 show two examples of excitation functions for the production 
of fission fragments. We see that merging CEM2k with GEM2 allows us to
reasonably describe yields of fission fragments, while in the old standard
CEM2k we do not have any fission fragments and are not able to describe
such reactions at all. We see that as shown in Figs.\ 5 and 7 for
a single proton energy of 800 MeV, the ``with Prec'' version
agrees better with the data in the whole energy region 
for the production of most of the fission fragments.
Only on the border between fission and fragmentation regions ($^{54}$Mn
and $^{60}$Co in Fig.\ 10) does the ``no Prec'' version agree much better with
the data than the ``with Prec'' version; the reason for this we have
already discussed.

\begin{center}
\begin{figure} 
\vspace*{-13mm}
\hspace*{-16.mm}
\includegraphics[angle=-0,width=10.7cm]{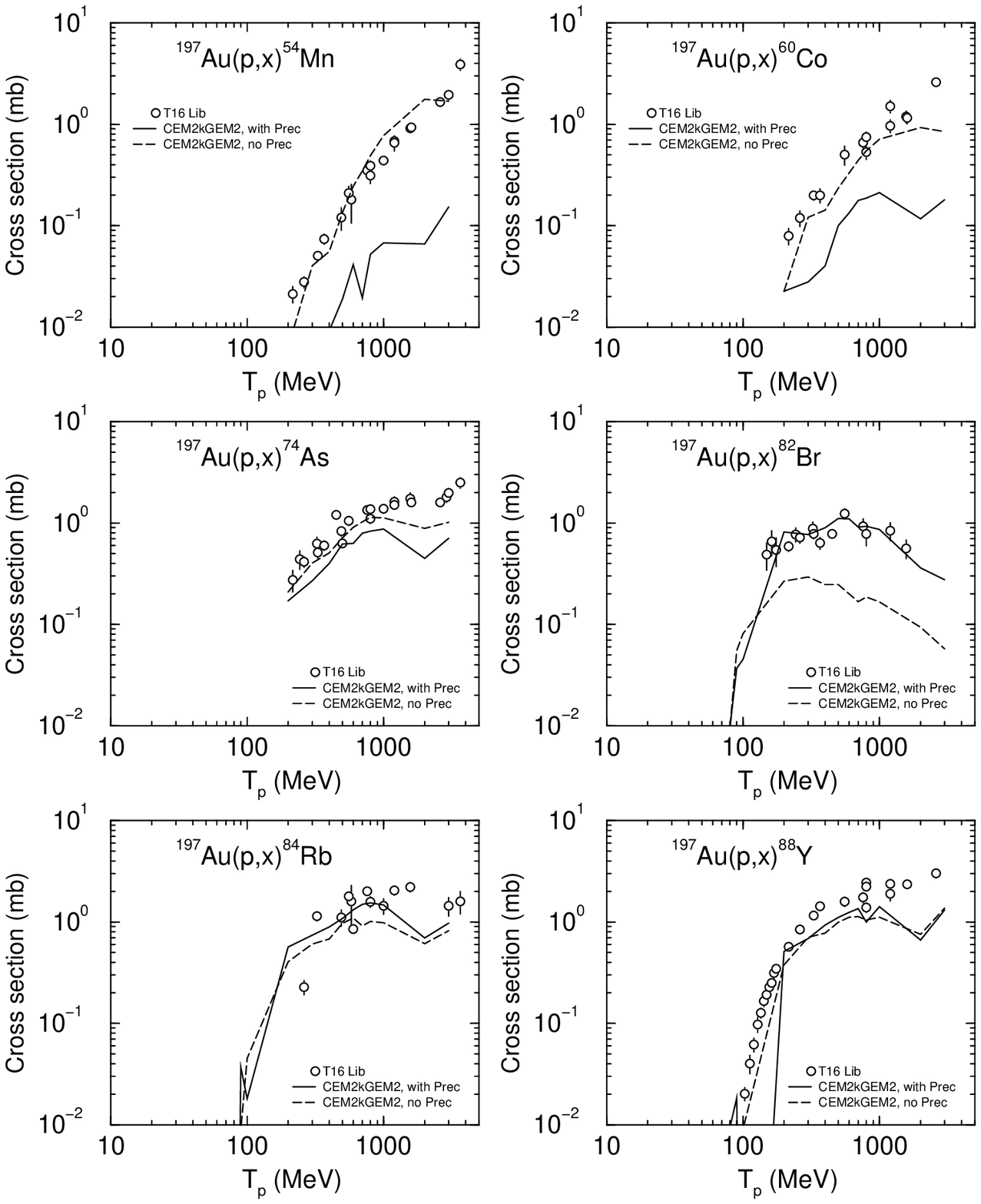}

\vspace*{-37mm}
{\noindent Figure 10. 
The same as Fig.\ 8 but for the production of
$^{54}$Mn, $^{60}$Co, $^{74}$As, $^{82}$Br, $^{84}$Rb, and $^{88}$Y. 
Note that while some minor contribution to the production
of these isotopes from deep spallation processes may be 
present, all of them are produced mainly via fission.
}
\label{fig10}
\end{figure}
\end{center}

\begin{center}
\begin{figure} 
\vspace*{-13mm}
\hspace*{-16.mm}
\includegraphics[angle=-0,width=10.7cm]{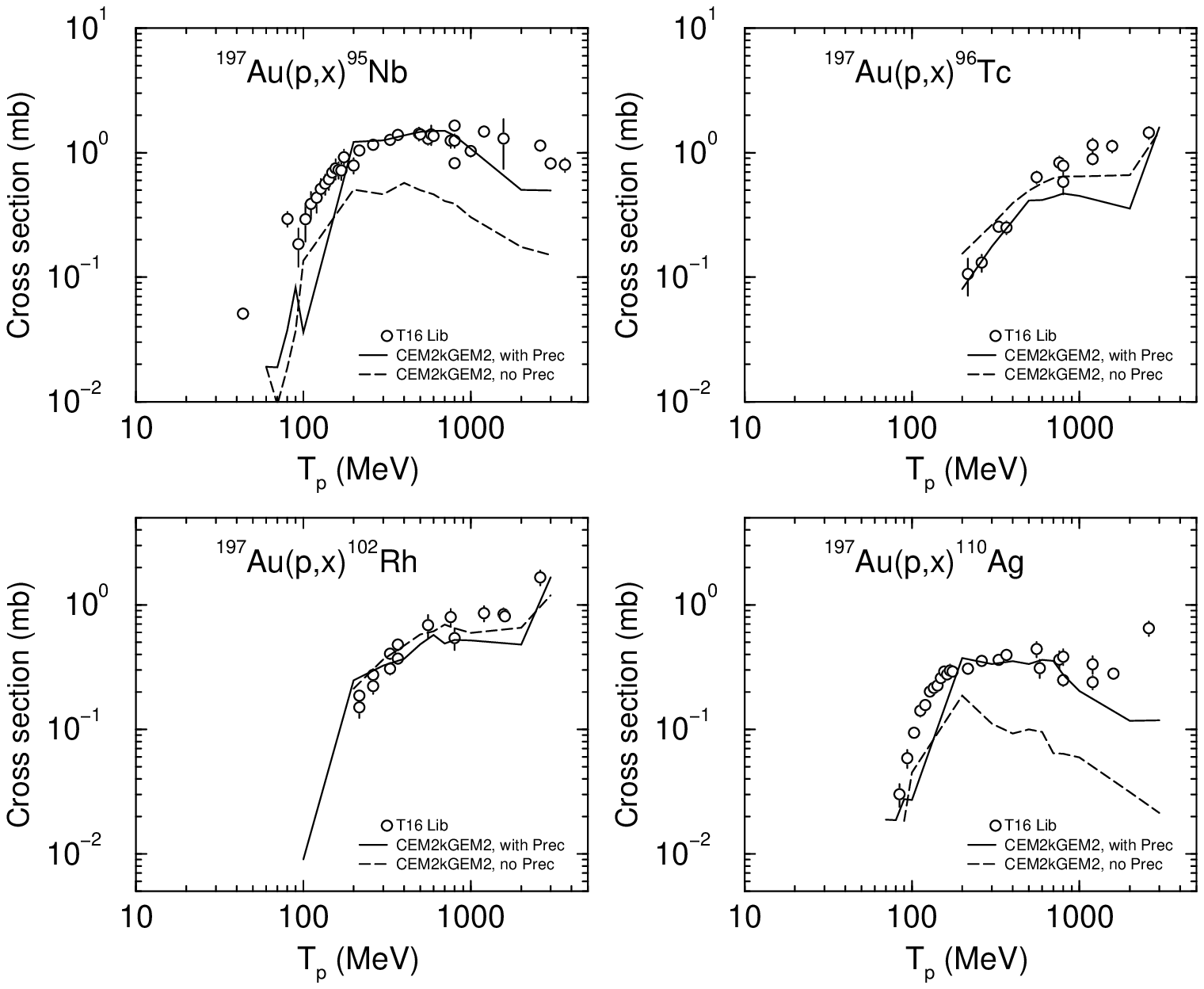}

\vspace*{-70mm}
{\noindent Figure 11. 
The same as in Fig.\ 8 but for the production of
$^{95}$Nb, $^{96}$Tc, $^{102}$Rh, and $^{110}$Ag. 
}
\label{fig11}
\end{figure}
\end{center}
\vspace*{-17mm}

In Fig.\ 12 we show examples of excitation functions for the production
of light fragments, in the fragmentation region, that are produced in 
CEM2k+GEM2 only via evaporation (the contribution to the yield of these
isotopes from fission or deep spallation is negligible).
We see that with the ``no Prec'' version, GEM2 reproduces correctly the
yields of light fragments $^{7}$Be and $^{18}$F, and not so well the
excitation functions for heavier fragments $^{22}$Na and $^{24}$Ne.
With increasing mass of the fragment, the calculations progressively
underestimate their yields ({\it e.g.}, for $^{28}$Mg, the calculated
excitations function is more than an order of magnitude below the data).
The version ``with Prec'' strongly underestimates the yields 
of all these fragments,
and this is again not surprising, as Furihata developed her model
and fitted all parameters without taking into account preequilibrium
processes. Undeniably, the parameters determining the yields of evaporated
fragments in GEM (inverse cross sections and Coulomb barriers)
could be adjusted to get a good agreement with the data
for the yields of light fragments with the version ``with Prec''.
This is not an aim of our present work and we will not 
do this here. Even if we were to
do this, we expect in advance to get similar results as we got for
the ``no Prec'' version: It would be possible to
describe correctly the yields of light fragments but not of heavy
fragments like $^{24}$Na and $^{28}$Mg. To describe such heavy fragments 
the model would need to be improved further, by considering 
other mechanisms for heavy fragment production in addition to the
evaporation process taken into account by GEM2.

\begin{center}
\begin{figure} 
\vspace*{-13mm}
\hspace*{-16.mm}
\includegraphics[angle=-0,width=10.7cm]{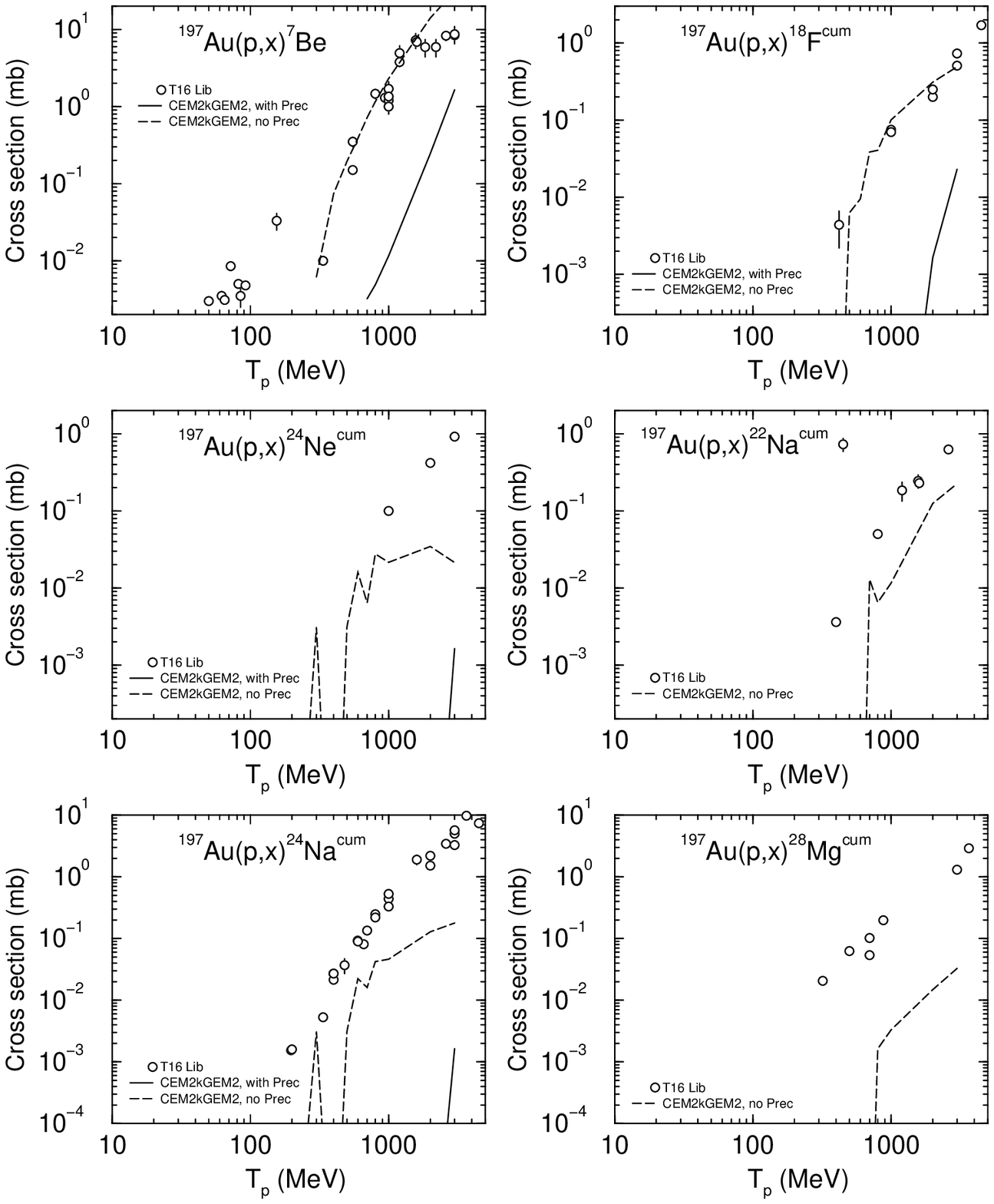}

\vspace*{-37mm}
{\noindent Figure 12. 
The same as Fig.\ 8 but for the production of
$^{7}$Be, $^{18}$F$^{cum}$, $^{24}$Ne$^{cum}$, 
$^{22}$Na$^{cum}$, $^{24}$Na$^{cum}$, and $^{28}$Mg$^{cum}$.
These fragments are produced in CEM2k+GEM2 only via evaporation.
}
\label{fig12}
\end{figure}
\end{center}

\begin{center}
\begin{figure} 
\vspace*{-20mm}
\hspace*{-16.mm}
\includegraphics[angle=-0,width=10.7cm]{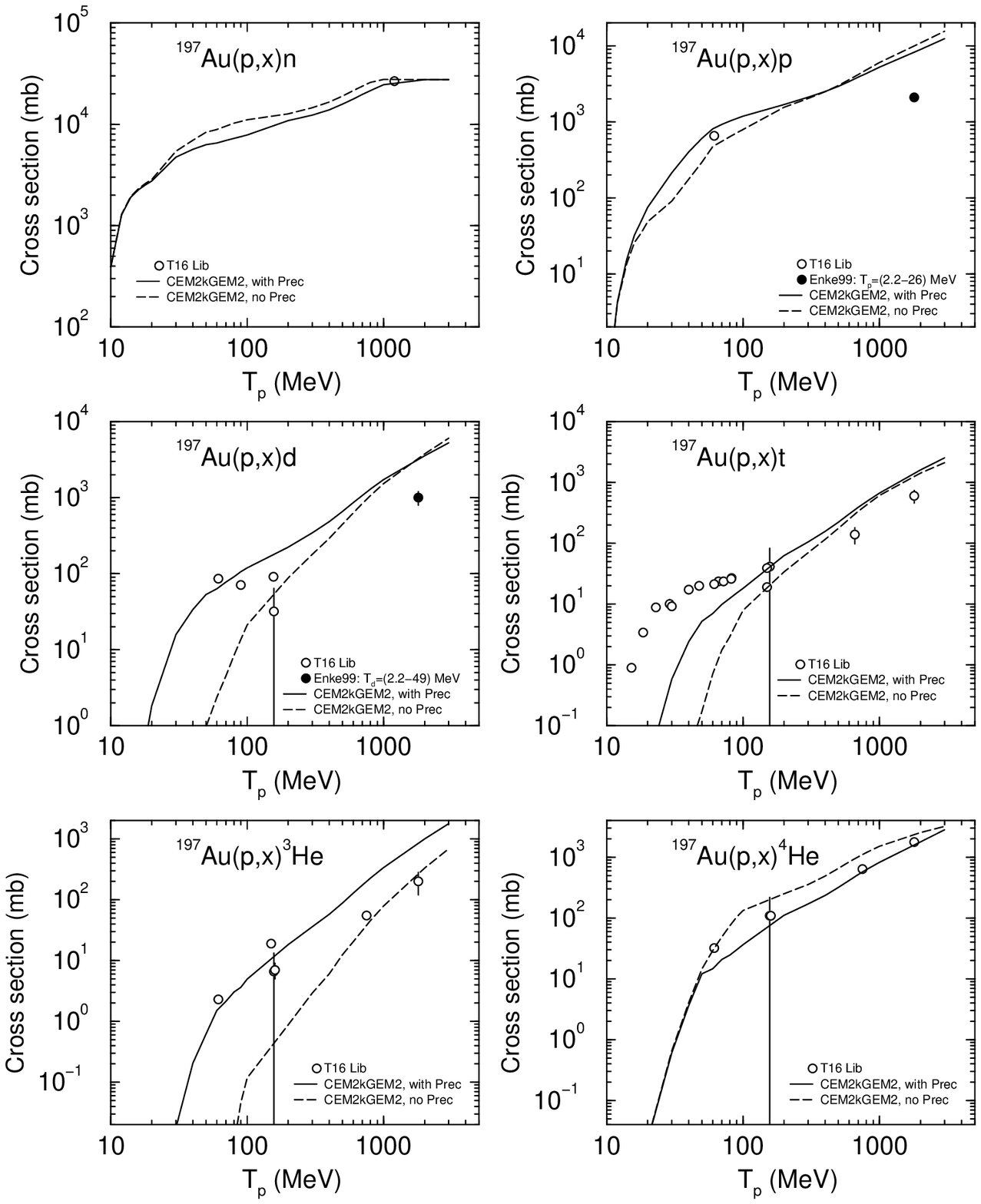}

\vspace*{-37mm}

{\noindent Figure 13. 
The same as Fig.\ 8 but for the production of n, p, d, t, 
$^{3}$He, and $^{4}$He.  The complex particles 
are produced in CEM2k+GEM2 via evaporation and preequilibrium emission;
n and p are also produced during the cascade stage.
}
\label{fig13}
\end{figure}
\end{center}



Finally, Fig.\ 13 shows the excitation functions for emission of nucleons 
and complex particles up to $\alpha$ for this reaction. 
Note that the data for these excitation functions are not so extensive 
and precise as we have for heavier
products: many data points were obtained by integration (plus extrapolation)
of the spectra of  particles measured only at several angles and only 
for a limited range of energy. But even from a comparison with
these sparce and imprecise data we see that the ``with Prec'' version
describes these excitation function much better than the ``no Prec''
version. This is an expected result as the high Coulomb
barriers for heavy nuclear targets oppose evaporation of low energy
complex particles and the main contribution to their yields comes
from preequilibrium emission from highly excited pre-compound nuclei.

Besides yields of reaction products, reliable models should also describe
their spectra. Data on particle spectra are important for
shielding calculations and other applications, in addition to the
scientifc interest in understanding nuclear reactions. For the reaction
discussed here, p + Au, there are measurements by Bertrand 
and Peelle for proton and complex-particle spectra at incident proton
energies of 29, 39, and 62 MeV$^{60}$.
We compare calculated angle-integrated energy spectra for p, d, t, $^3$He, 
and $^4$He with the data at 62 MeV in Fig.\ 14. 

\begin{center}
\begin{figure} 
\vspace*{-5mm}
\hspace*{-6.mm}
\includegraphics[angle=-0,width=92mm]{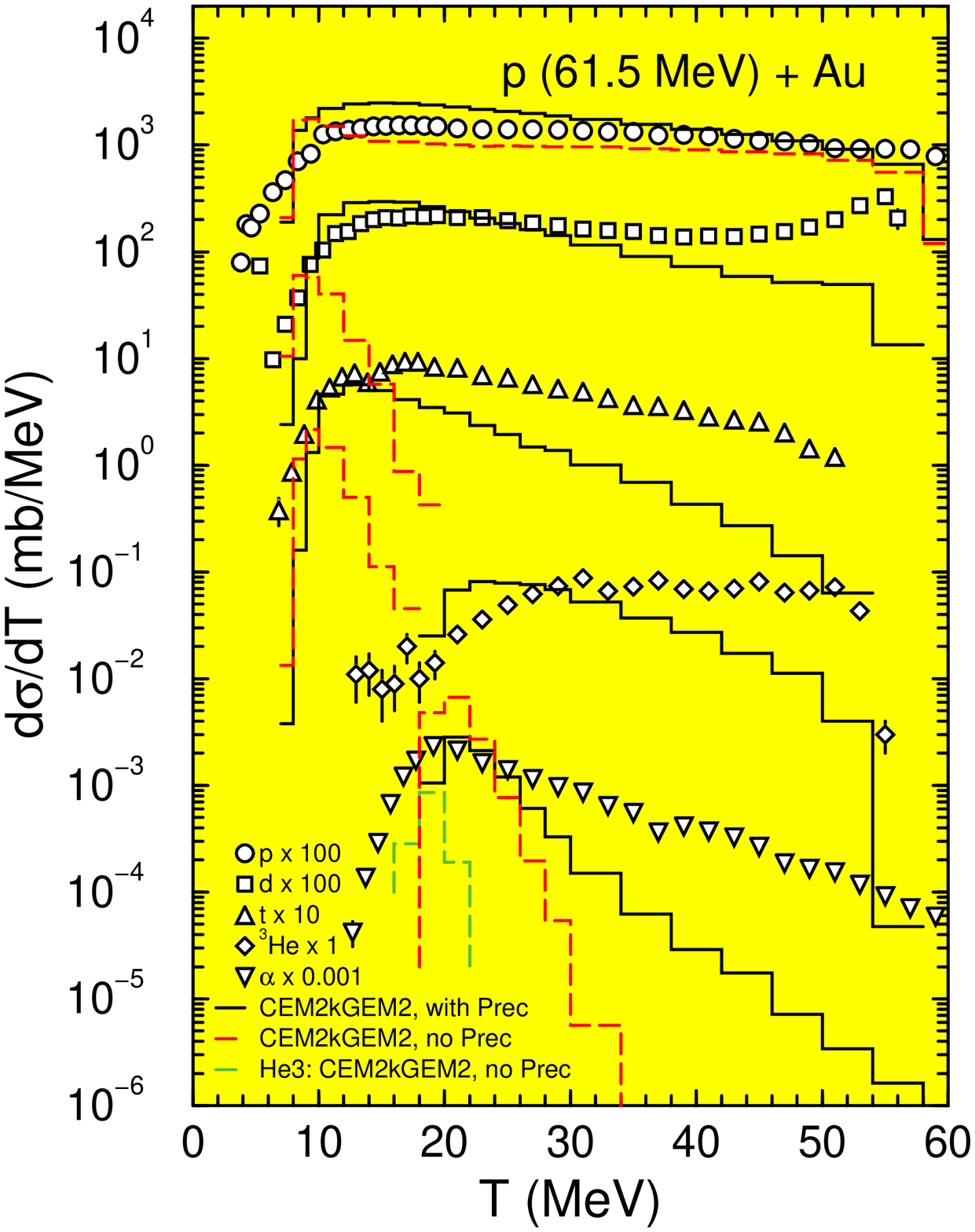}

\vspace*{-7mm}
{\noindent Figure 14. Angle-integrated energy spectra of p, d, t, 
$^{3}$He, and $^{4}$He from 61.5 MeV protons on $^{197}$Au.
Calculations from the merged CEM2k+GEM2 code
``with Prec'' are shown by solid black histograms and ``no Prec'', by
dashed red  (and green, for $^{3}$He) histograms. Experimental
data (symbols) are from Bertrand and Peelle$^{60}$. \\
}
\label{fig14}
\end{figure}
\end{center}

\vspace*{-10mm}
We see that the ``with Prec'' version (black solid histograms)
describes
a good part of the complex particle emission spectra, though the tails
of calculated spectra are significantly below the data (we discuss
in the next section an attempt to improve the description of complex 
particle spectra in CEM2k). The ``no Prec'' version 
(dashed red and green histograms) fails completely to
describe these spectra, as it contains contribution to complex 
particle emission only from evaporation, and evaporation spectra
do not extend significantly above particle energies of 20 MeV.
Even in the low-energy evaporation region, the evaporation component
for the $^3$He spectra (shown by a green dashed histogram in Fig.\ 14)
is more than an order of magnitude below the data$^{60}$.
Let us mention that this is a problem not only of GEM2 but also for all 
other similar models where preequilibrium emission is not modeled. What is
more, some of the evaporation models used in the literature, like the
GSI evaporation model by Schmidt {\it et al.}$^{61}$, which is
used in conjunction with the Liege INC by Cugnon {\it et al.}$^{15}$ ,
evaporate only n, p, and $^4$He, and do not include
evaporation of d, t, and $^3$He providing for them no yield.
 
For completeness sake, we show here also an example of results 
from a calculation with the merged
CEM2k+GEM2 code of a reaction on an actinide, p(190 MeV) + $^{232}$Th.
This reaction was recently measured by Duijvestijn {\it et al.}$^{62,63}$,
and at a similar energy of 200 MeV, by Titarenko {\it et al.}$^{64}$
To get for actinides a proper fission cross section, we need to adjust
in GEM2 the parameters $C(Z)$ (or, also $A_0(Z)$) in Eq.\ (19), as they
were fitted by Atchison to work the best with Bertini's INC and we have
in CEM2k our own INC. As mentioned above, for actinides, Eq.\ (13) is not
used in GEM2 and $a_f$ is not used in any calculations, 
therefore we do not need to adjust $a_f/a_n$,
for fissioning nuclei with $Z > 88$. We found that to get
with CEM2k+GEM2 a fission cross section in agreement with the
data for our reaction we need to use
$C(Z)^{CEM}/C(Z)^{RAL}= 8.50$ for the ``with Prec'' version and
$C(Z)^{CEM}/C(Z)^{RAL}= 1.77$ for ``no Prec''
(while we use $a_f^{CEM} / a_f^{RAL} = 1.0848$ which we fitted for the 
reaction p(800 MeV) + Au, ``with Prec'').
This is the only parameter we fitted for this reaction.
Nevertheless, we should mention that for reactions on actinides at 
intermediate or high energies, the parameter
$a_f^{CEM} / a_f^{RAL}$ should also be fitted along with
$C(Z)^{CEM}/C(Z)^{RAL}$. In some simulated events several protons
can be emitted at the cascade and preequilibrium stages of the reaction,
as well as at the evaporation stage, before the compound
nucleus actually fissions, and the charge of the fissioning nucleus can 
have $Z \le 88$, even when the initial
charge of the target has $Z > 88$. At the same time, for $Z \leq 88$,
due to charge exchange reactions, the charge of the fissioning nucleus may
exceed 88, so that we would need to fit as well $C(Z)^{CEM}/C(Z)^{RAL}$.
This is a peculiarity of treating the fission 
probability $P_f$ differently for the elements above and below $Z=89$ 
in the Atchison model.

Fig.\ 15 shows mass distributions of products from 
p(190 MeV) + $^{232}$Th calculated with both versions of CEM2k+GEM2
compared to the available experimental data$^{62,64}$.

\begin{center}
\begin{figure} 
\vspace*{-5mm}
\hspace*{-10.mm}
\includegraphics[angle=-90,width=96mm]{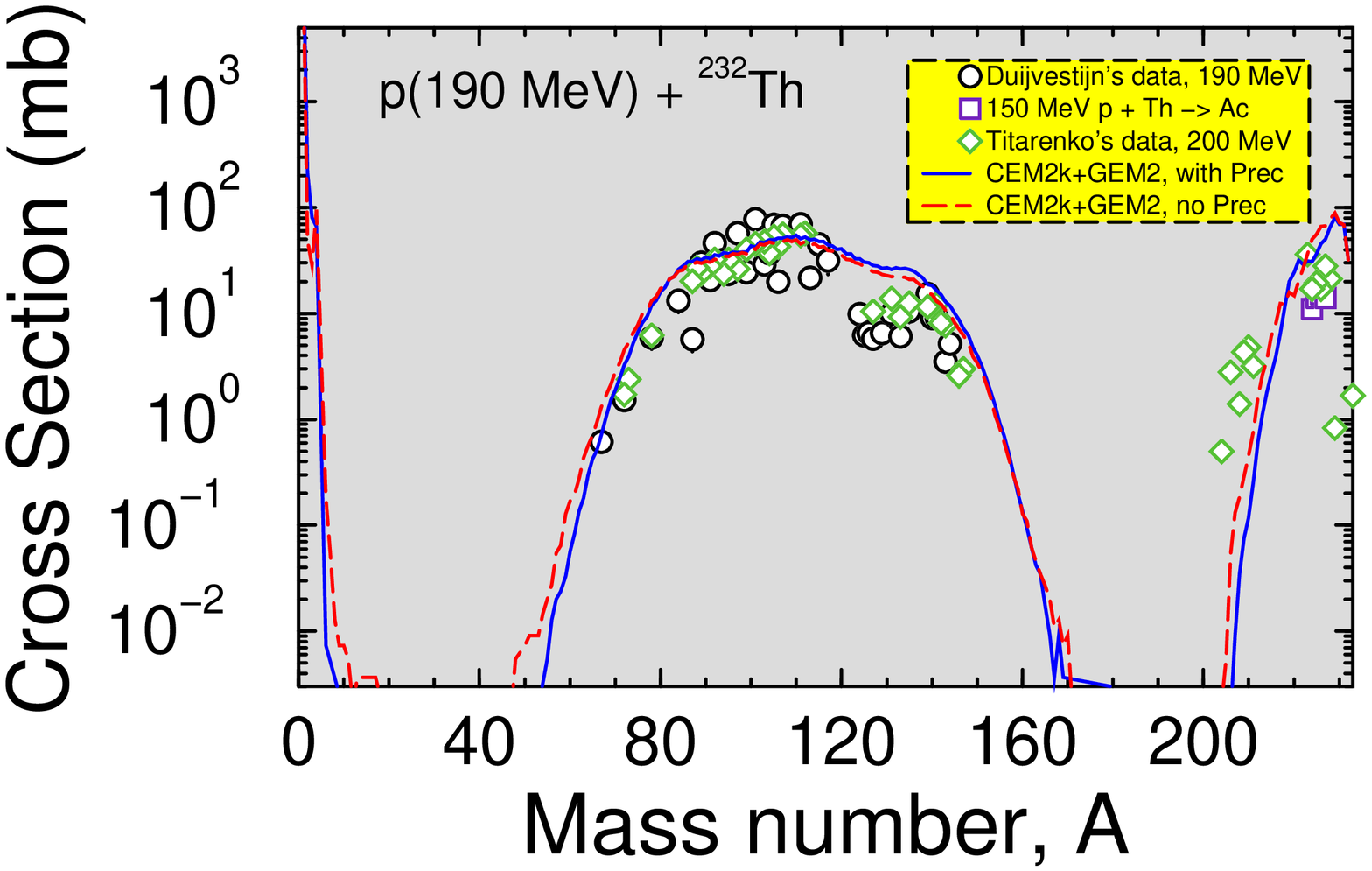}

\vspace*{3mm}
{\noindent Figure 15. 
Mass distribution of nuclides from p (190 MeV) + $^{232}$Th.
The 190 MeV data are from$^{62}$; the 200 MeV data are by 
Titarenko {\it et al.}$^{64}$; data for several Ac isotopes
from p (150 MeV) + $^{232}$Th are from the T-16 compilation$^{59}$.
Our ``with Prec'' results are shown by the solid blue line and
the ``no Prec'' results are shown by the dashed red line.
}
\label{fig15}
\end{figure}
\end{center}
\vspace*{-9mm}

We need to mention that these data are not as good for testing and 
developing models as are the GSI data measured in inverse kinematics
for the p + Au reaction discussed above: All the data shown in Fig.\ 15
were obtained by the $\gamma$-spectrometry method. Only some of the
produced isotopes were measured, and most of the data were measured
for the cumulative yields. To get the ``experimental'' A-distribution, we
summed for each $A$ the available data taking care to not sum the individual
cross sections already included in some cumulative yields;
but the resulting A-distribution is still not complete, as many 
isotopes were not measured. This means that some theoretical values 
can be above the experimental data (where some isotopes were not 
measured) without necessarily implying disagreement between calculations 
and measurements. What is more, many of Duijvestijn's data
tabulated in$^{62}$ differ significantly
from the same data tabulated in an earlier publication$^{63}$ for
some isotopes: {\it e.g.}, for $^{72}$Zn the
difference is a factor of 8.03 (!), possibly due to a misprint
in$^{63}$. We chose for our comparison Duijvestijn's data tabulated 
in$^{62}$ as
more reliable than the earlier tabulation published in$^{63}$.
(We thank Dr.\ Duijvestijn for clarifying this point.)

One can see that after the parameter $C(Z)$ is adjusted, both the
``with Prec'' and ``no Prec'' versions of CEM2k+GEM2
describe equally well the mass distribution of the
products from this reaction; therefore it is not possible
to choose between either of the versions from a comparison with these
not very informative data. It is more useful to compare calculations
with the yields of individually measured isotopes.
In Fig.\ 16, we compare our calculations with all cross sections measured
by Duijvestijn$^{62}$ for each nuclide separately, where we can compare our
results with the data (we do not include in our comparison the nuclides
measured only either in their isomer or ground states, as our model does not
provide such information: CEM2k+GEM2 provides only yields for the sum
of isotope production cross sections both in their ground and excited states). 
We see that on the whole, the ``with Prec'' version agrees better
with most of the individually measured cross sections than the ``no
Prec'' version and for many  of the measured isotopes the disagreement 
is less than a factor of two. Nevertheless, for several isotopes
like $^{72}$Ga, $^{96}$Tc, and $^{124}$I, we see some big disagreements.
For comparison, we also show in Fig.\ 16 calculations by the phenomenological
code YIELDX of Silberberg, Tsao, and Barghouty$^{23}$ and with the 
phenomenological code CYF by Wahl$^{65}$ often used in applications.
We see that both these phenomenological systematics completely fail
to describe the production of all isotopes from this reaction, 
indicating that we cannot rely on phenomenological systematics and 
must develop reliable models to be used in applications.
We note that the agreement of our CEM2k+GEM2 results with the measured
product yields is different for different reactions. So, in Ref.$^{56}$,
for the reaction p (100 MeV) + $^{238}$U, we got an almost perfect
agreement of calculations with the data: for this particular reaction,
both ``with Prec'' and ``no Prec'' results by CEM2k+GEM2 almost
coincide with the available data (see details in Ref.$^{56}$). 

\begin{center}
\begin{figure} 
\vspace*{-5mm}
\hspace*{-25.mm}
\includegraphics[angle=-0,width=120mm]{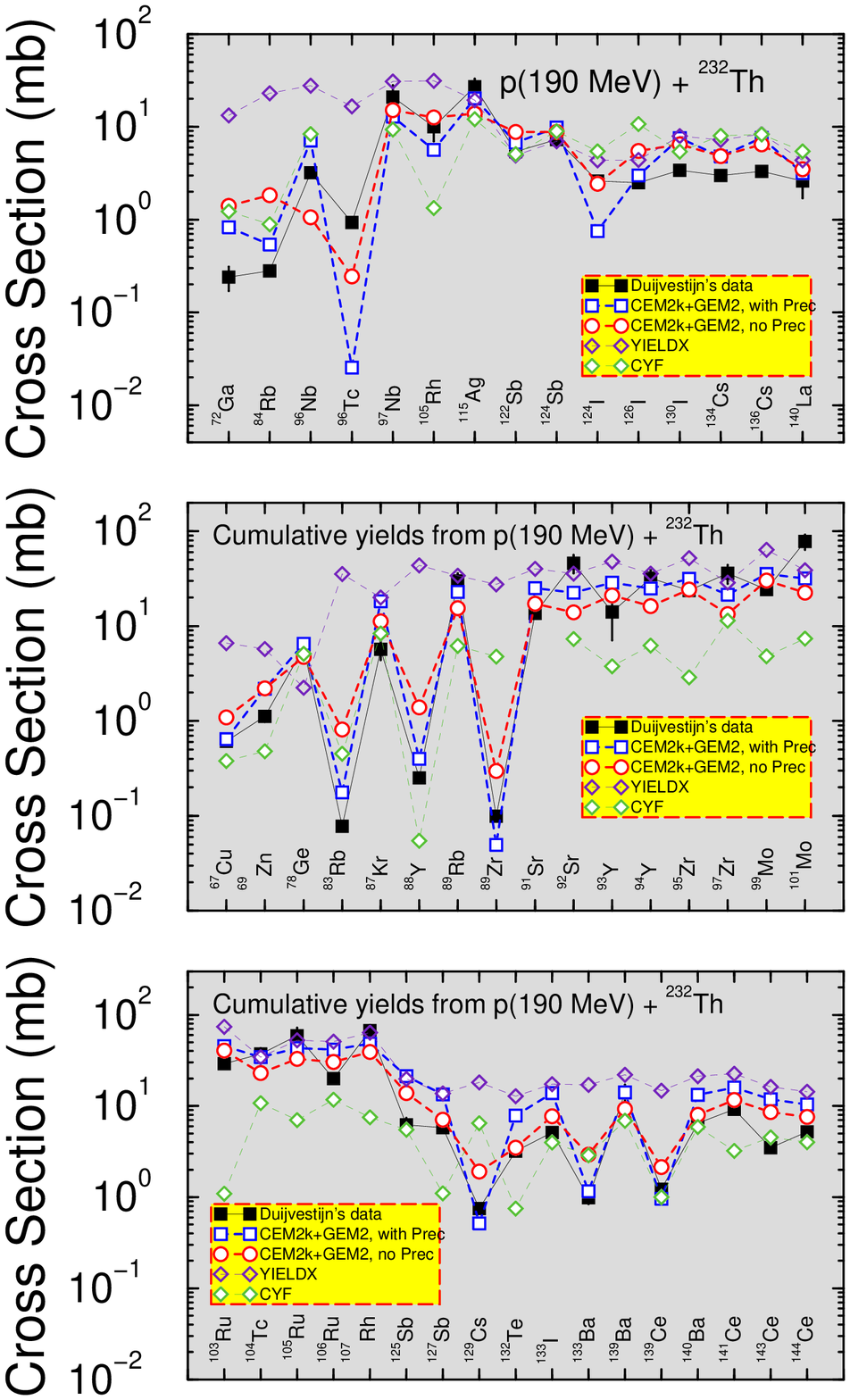}

\vspace*{-18mm}
{\noindent Figure\ 16. Detailed comparison between experimental$^{62}$ 
(black filled squares) and calculated cross sections of individual 
(upper plot) and cumulative (middle and bottom plots) reaction products.
Our CEM2k+GEM2 ``with Prec'' results are shown by connected blue squares
and the ``no Prec'' results are shown by connected red circles.
For comparison, predictions by the phenomenological systematics YIELDX
of Silberberg, Tsao, and Barghouty$^{23}$ are shown by 
connected indigo diamonds, and results calculated by Dr.\ W.\ B.\ Wilson with
the phenomenological code CYF by Wahl$^{65}$ are shown
with connected green diamonds. We thank Dr.\ Wilson for
providing us with results of his calculations with the CYF code
included in this figure.
}
\label{fig16}
\end{figure}
\end{center}
\vspace*{-11mm}

To summarize this section, merging CEM2k with GEM2 allows us to describe
reasonably well
many fission and fragmentation reactions in addition to the spallation
reactions already described well by CEM2k. Some reactions, like production
of fission fragments at the borders between fission and fragmentation 
or between fission and emission of heavy fragments
like Na and Mg are poorly described by CEM2k+GEM2 in its current
version. This disagreement does not discourage us;
the results of the present work suggest that some
of the fission and evaporation parameters of GEM2 can be adjusted
to get a much better description of all reactions.
This lends credibility to such an approach. 
There is one more drawback of this approach to be mentioned: considering
evaporation of up to 66 particles in GEM becomes extremely time consuming 
when calculating reactions with heavy targets at high incident energies. 
But even this disadvantage may be mitigated by the performance of modern 
computers.
We have nevertheless some more serious doubts about the current version
of GEM2 related to its lack of self-consistency, 
{\it e.g.}:

1) using different, not physically related parameterizations for inverse
cross sections and Coulomb barriers for different particles and fragments;

2) using different level density parameters for the same compound nuclei
when calculating evaporation and estimating fission probability from the
widths of neutron evaporation and fission;

3) different, and purely phenomenological treatments of
fission for pre-actinide and actinide nuclei;

4) not taking into account at all the angular momentum of compound and
fissioning nuclei;

5) rough estimations for the fission barriers and level density parameters, 
{\it etc.}

This means that an approach like GEM2 can in principle be used to describe
fission and evaporation of particles and fragments heavier that $^4$He
after the INC and preequilibrium parts of CEM2k and other models,
but it should be considerably improved striving first to progressively
incorporate better physics, and only after that looking on agreement with
the data.

The results of the present work and from$^{56}$ show that
on the whole, the merged CEM2k+GEM2 code agrees better 
with most of tested experimental data when we take 
into account the preequilibrium emission of particles, than 
when we neglect completely preequilibrium processes.
But there is still an open question to be solved here; we have had some
indications for many years that CEM accounts for too many preequilibrium
particles, at least at energies above the pion-production
threshold. To solve this problem, as a ``zero-step'' approximation, in
our original CEM2k version$^{10}$ we neglected in the exciton model
the transitions that decrease or do not change the number
of excitons $\Delta_n = -2$ and $\Delta_n = 0$, shortening in this way the
preequilibrium stage of reactions. We do not like this
approach as it is arbitrary, ``ad hoc'', even though this
``never come back'' approximation is used in some popular codes$^{14, 66}$.
In the present work we removed this arbitrary
condition in CEM2k and the ``with Prec'' version takes into account all
the preequilibrium transitions $\Delta_n = +2$, 0, and -2, making
the preequilibrium stage of a reaction longer and increasing the
number of emitted preequilibrium particles. The results of the present
work indicate to us once again that we need to take into account the
preequilibrium stage in reactions, but we need less 
particle emission than we currently calculate at this stage.
In$^{56}$, we explore an alternative way to shorten the preequilibrium
particle emission in CEM, based on Ref.\ $^{67}$, but this work is 
incomplete so we do not include these results here.\\ 

\noindent {\bf IV. EXTENSION OF CEM2K}\\
In this section we present an alternative way of describing fission-fragment, 
complex-particle and light-fragment emission with CEM2k,
by extending and developing further our model itself, without 
using independent evaporation models developed by other authors.

One of the unsolved problems in all versions of the CEM code
is using the Dostrovsky {\it et. al.}$^{39}$ approximations
for the inverse cross sections (Eq.\ (2) with the parameters shown
in Tab.\ 3) both for preequilibrium and evaporation.
As already mentioned, the Dostrovsky {\it et. al.}$^{39}$ formula
is simple and allows one to calculate the integral in Eq.\ (1)
analytically. This is the main reason we have kept Dostrovsky's formula
in all our previous versions of the CEM code, even though we know it is 
not reliable enough. We develop and incorporate
into CEM2k our own approximation for inverse cross sections. As a first
step, we collected experimental data on absorption or
fusion cross sections for all target nuclei and for all particles and
fragments for which we were able to find data, {\it i.e.}, just the 
``inverse'' cross sections when we deal with particle emission. Then, 
we collected from the literature
theoretical estimations and systematics for the inverse cross sections and
compared these systematics with the data to find which systematics
better describe the data. An example from this study is shown in Fig.\ 17,
where we compare available data on inverse cross sections for
n, p, d, t, $^3$He, and $^4$He on $^{27}$Al with the approximation from 
Ref.$^{39}$, NASA systematics by Tripathi, Cucinota, and Wilson$^{33}$,
a parameterization by Kalbach$^{34}$, an approximation by 
Tang, Srinivasan, and Azziz$^{69}$,
and the results calculated with the phenomenological code CROSEC by
Barashenkov and Polanski$^{70}$. Similar figures have been plotted for
other target nuclei for which we were able to find experimental
data. One can see a big disagreement between Dostrovsky 
{\it et al.}'s$^{39}$ approximation and the data.

Probably, the reason why Dostrovsky {\it et. al.}'s$^{39}$ 
formula works quite well in many evaporation models to describe
emission of neutrons and protons from many reactions is because
it overestimates both and in similar ways 
the neutron and proton inverse cross sections
(correspondingly, their width $\Gamma_n$ and $\Gamma_p$)
at energies above about 20 MeV. The
emission of n and p are the main channels of most evaporation processes,
and emission of particles are simulated in evaporation codes
using only the ratios $\Gamma_j / \sum_j \Gamma_j$ but not the
absolute values of particle widths, $\Gamma_j$.
Dostrovsky formula underestimates significantly the
inverse cross sections for the complex particles, and this is one of the 
reasons why all evaporation models that use Dostrovsky's
formula underestimate the yields of complex particles. An additional
problem introduced by the overestimation of the nucleon emission probabilities
occurs when fission is considered.  A realistic value of the fission
barrier and the saddle-point level density will give too small a
fission probability when compared to a too-large nucleon evaporation
probability.  This then leads to the empirical determination of unphysically
large values of the level density parameter, {\it etc.}

From Fig.\ 17 and similar figures for other nuclei which we have made, 
we determined
that the overall best agreement with experimental data is achieved 
with the NASA systematics by Tripathi, Cucinota, and Wilson$^{33}$; 
therefore we chose it as the basis to calculate inverse cross sections 
in CEM2k.
Nevertheless, this systematics does not reproduce correctly the inverse
cross sections for neutrons at energies below about 10 MeV. We
address this problem the following way: We calculate with the NASA 
systematics$^{33}$ the inverse cross sections for all charged particles
and for neutrons with energies above $T_n^{max}$, where a maximum in the
neutron inverse cross section is predicted by$^{33}$, and with the systematics
by Kalbach$^{34}$, for neutrons with energies below $T_n^{max}$. We
tabulated the values of $T_n^{max}$ for each nucleus and renormalize
the neutron inverse cross sections calculated with the systematics
by Kalbach$^{34}$ so that they coincide at $T_n^{max}$ with the NASA 
values$^{33}$, providing a continuity of the neutron inverse cross 
sections at this energy.

A detailed description of the NASA and Kalbach systematics may be found
in$^{33,34}$ and references therein. For completeness sake, we outline here
only their basic ideas and formulas.

The NASA systematics$^{33}$ is a universal parameterization
for any systems of colliding nuclei, therefore can be used to
calculate inverse cross sections not only for nucleons and complex
particles but also for heavier fragments. It uses the following form 
for the reaction cross sections (inverse cross sections, in our case):
\beq
\sigma_R = \pi r_0^2 ( A_P^{1/3} + A_T^{1/3} + \delta_E)^2
(1 - V_C/E_{cm}) ,
\eeq
where $A_P$ and $A_T$ are the projectile and target mass numbers, 
respectively, $r_0 = 1.1$ fm is energy independent, $E_{cm}$ is 
the total center-of-mass kinetic energy in MeV, $\delta_E$ is
an energy dependent parameter described below, and
$V_C$ [MeV] is the energy-dependent Coulomb barrier defined as:
\beq
V_C = 1.44 Z_P Z_T /R .
\eeq
Here,
\beq
R = r_P + r_T + 1.2 (A_P^{1/3} + A_T^{1/3} ) / E_{cm}^{1/3} ,
\eeq
with ($i = P$, $T$)
\beq
r_i = 1.29 (r_i)_{rms} ,
\eeq
with the root-mean-square radius, $(r_i)_{rms}$, obtained directly from 
experimental data. 

There is an energy dependence in the reaction cross section at
intermediate and higher energies mainly due to two 
effects---transparency and Pauli blocking. This is taken into account in
$\delta_E$ which is given by
\beq
\delta_E = 1.85S + 0.16 {S \over {E_{cm}^{1/3}} } - C_E + 0.91
{ (A_T - 2 Z_T) Z_P \over {A_T A_P} } ,
\eeq

where $S$ is the mass asymmetry term and is given by
\beq
S = A_P^{1/3} A_T^{1/3} / (A_P^{1/3} + A_T^{1/3} )
\eeq
and is related to the volume overlap of the collision system.
The last term on the right hand side of Eq.\ (32) accounts for
the isotopic dependence of the reaction cross section. 

The term $C_E$ is related to the transparency and Pauli blocking
and is given by
\bea
C_E = &D& (1 - \exp (-E/40)) - 0.292 \exp (-E/792)  \nonumber \\
&\times& \cos (0.229 E^{0.453} ) .
\eea
Here $D$ is related to the density dependence of the colliding 
system scaled with respect to the density of the C+C system,
{\it i.e.}:
\beq
D = 1.75 (\rho_{A_P} + \rho_{A_T} ) /(\rho_{A_C} + \rho_{A_C}) .
\eeq
The density of a nucleus is calculated in the sharp-surfaced
sphere model and for a nucleus of mass $A_i$ is given by
\beq
\rho_{A_i} = A_i / {4\over3}\pi r_i^3 ,
\eeq
where the radius of the nucleus $r_i$ is defined in Eq.\ (31).
The physics related to the constant $D$ is to simulate the 
modifications of the reaction cross sections due to Pauli blocking.
This effect helps present a universal picture of the reaction cross 
section$^{33}$.

At lower energies (below several tens of MeV) where the overlap of 
interacting nuclei is small (and where the Coulomb interaction 
modifies the reaction cross section significantly) the modifications of
the cross sections due to Pauli blocking are small, and gradually play 
an increasing role as the energy increases, since this leads to higher 
densities where Pauli blocking gets increasingly important. For proton-nucleus
interactions, where there is not much compression effect,
a single constant value of $D = 2.05$ gives good results for all
proton-nucleus collisions. For alpha-nucleus collisions,
where there is a little compression, the best value of $D$ is given by$^{33}$
\bea
D = 2.77 &-& 8.0 \times 10^{-3} A_T + 1.8 \times 10^{-5} A_T^2 \nonumber \\
&-& 0.8 / ( 1 + \exp (250 - E) /75 ) .
\eea
For lithium nuclei because of the ''halos'', compression is less and hence
the Pauli blocking effect is less important and a reduced value of $D/3$ gives
better results for the reaction cross sections at the intermediate and
higher energies. 

Note that for proton-nucleus collisions this method of calculating the
Coulomb energy underestimates its value for the very light closed shell 
nuclei of alpha and carbon, and these should be increased by a factor of
27 and 3.5 respectively for a better fit$^{33}$. 

Kalbach's approximation for inverse cross sections$^{34}$
is based on optical-model reaction cross sections using an empirical
parameterization of Chatterjee, Murthy, and Gupta$^{73}$ fitted
for different particles using different optical potentials,
with some additional modifications and described in$^{34}$.
For instance, the reaction cross section $\sigma_R$ (inverse cross 
section, in our case) for neutrons (used here) is written as
\beq
\sigma_R = \lambda \eps + \mu + \nu / \eps ,
\eeq
where $\lambda$, $\mu$, $\nu$ are mass-dependent parameters and $\eps$
is the neutron laboratory energy in MeV. As mentioned in$^{73}$,
the addition of the linear term $\lambda \eps$ in Eq.\ (38)
greatly improves the fit as compared with the Dostrovsky {\it et al.}
approximation$^{39}$. 
Separating out the energy dependence as in Eq.\ (38), the dependence
of $\lambda$, $\mu$, $\nu$ on target mass number $A$, was obtained
empirically as$^{73}$:
\bea
\lambda &=& \lambda_0 A^{-1/3} + \lambda_1 ,  \nonumber \\
\mu &=& \mu_0 A^{1/3} + \mu_1 A^{2/3} , \\
\nu &=& \nu_0 A^{4/3} + \nu_1 A^{2/3} + \nu_2 .   \nonumber 
\eea
\onecolumn

\begin{center}
\begin{figure}[h!] 

\vspace*{-20mm}
\hspace*{-18.mm}
\includegraphics[angle=-0,width=204mm]{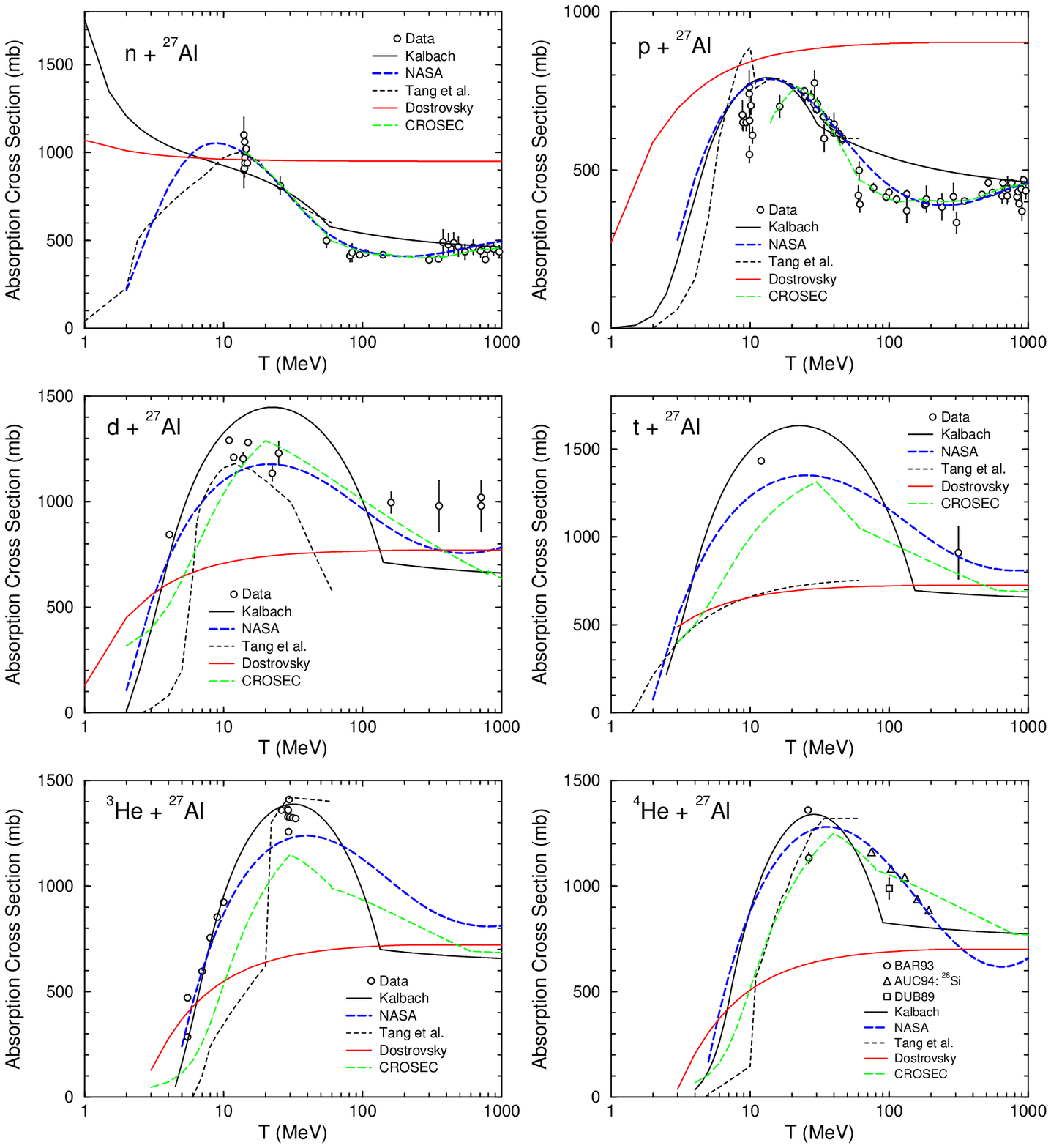}

\vspace*{-50mm}
{\noindent Figure 17. 
Comparison of experimental data on absorption cross sections
for n, p, d, t, $^3$He, and $^4$He on $^{27}$Al with
the Dostrovsky {\it et. al.}$^{39}$ approximation, the NASA 
systematics by Tripathi, Cucinota, and Wilson$^{33}$,
a parameterization by Kalbach$^{34}$, an approximation by 
Tang,  Srinivasan, and Azziz$^{69}$,
and results calculated with the phenomenological code CROSEC by
Barashenkov and Polanski$^{70}$. Most of the data points (circles,
either without a special label or labeled as ``BAR93'') 
shown here are from the
compilation by Barashenkov$^{35}$; several data points are by 
Auce {\it et al.}$^{71}$ (triangles labeled as ``AUC94''),
and a data point for $^4$He is measured by Dubar {\it et al.}$^{72}$
(the square labeled as ``DUB89'').
}
\label{fig17}

\vspace*{-10mm}
\end{figure}
\end{center}

\twocolumn

The seven parameters $\lambda_0$, $\lambda_1$,
$\mu_0$, $\mu_1$, $\nu_0$,  $\nu_1$, and $\nu_2$
in Eq.\ (39) were fitted and tabulated by  Chatterjee, Murthy, and Gupta$^{73}$
for different optical potentials (for different incident
particles from n up to $\alpha$), with a priority to the potential by
Mani {\it et al.}$^{74}$ for neutrons. In 1982, the original$^{73}$ neutron
parameterization was modified by S. K. Gupta to include an effective
barrier of 2.4 MeV (see details in$^{34}$), and is now used in the
Kalbach approximation for neutron inverse cross sections$^{34}$ with
the following values (different from the ones of Ref.$^{73}$)
for the parameters in Eq.\ (38):
$\lambda_0 = 12.10$, 
$\lambda_1 = -11.27$,
$\mu_0 = 234.1$, 
$\mu_1 = 38.26$, 
$\nu_0 = 1.55$,  
$\nu_1 = -106.1$, and 
$\nu_2 = 1280.8$.
In addition, Kalbach has changed somehow arbitrarily$^{34}$ the neutron
``barrier'' from 2.4 MeV to 0.5 MeV to raise the neutron emission cross 
section for emission energies below about 1.5 MeV.
Also, after the choice of the neutron potential of Mani {\it et al.}$^{74}$
was made in 1963, new data became available, and the earlier comparisons with 
measured non-elastic cross sections have been extended to include
additional targets. This led to a normalization factor for the
parameterized optical model cross section having the form$^{34}$:
\beq
R_n(A) =  \cases{0.7 + 0.3 A/40&for  $A < 40$ ,\cr
                 1.0&for $ A \geq 40$ .\cr}
\eeq
Thus, the current version of the 
Kalbach approximation for neutron inverse cross sections used
here reads like $\sigma_{inv} = \sigma_R  \times R_n(A)$ with
$\sigma_R$ and $R_n(A)$ defined by Eqs.\ (38--40) and the seven parameters 
listed above.

An example of inverse cross sections calculated with our method 
compared to experimental data for
n, p, d, t, $^3$He, and $^4$He on $^{12}$C, $^{40}$Ca, $^{56}$Fe,
and  $^{238}$U is shown in Fig.\ 18.

For comparison, calculations using the NASA systematics by Tripathi, 
Cucinota, and Wilson$^{33}$, the Kalbach parameterization$^{34}$,
and the Dostrovsky  {\it et. al.}$^{39}$
approximation are shown as well.
One can see that our hybrid approach combining the NASA systematics
by Tripathi, Cucinota, and Wilson$^{33}$ and the
Kalbach parameterization$^{34}$ reproduces
well the measured inverse cross sections  for
n, p, d, t, $^3$He, and $^4$He and agrees much better with the data
than the approximation by Dostrovsky  {\it et. al.}$^{39}$
used earlier in all versions of our CEM code.
Similar results are obtained for other nuclear targets
for which we found data.

We incorporated this hybrid approximation for $\sigma_{inv}$
in our CEM2k code and use it to calculate inverse cross sections for
all particles to be evaporated from a compound nucleus or
emitted at the preequilibrium stage of a reaction.
The widths for particle emission, $\Gamma_j$, are calculated by
integrating Eq.\ (1) (and a similar equation for preequilibrium
particles; see details in$^{68}$) numerically.

To be able to describe production of light fragments heavier than $^4$He,
as a first step, we extended the evaporation process including
emission of up to 66 different particles, the same considered by Furihata
in GEM2 and listed in Tab.\ 1. Following Furihata, we consider
evaporation of fragments both in the ground and 
excited states, and calculate the widths for the
emission of excited fragments and their energy in the same
manner as realized in GEM2 by Furihata and described above in Sec. III.A.

We enlarged the number of fragments that may be emitted at the
preequilibrium stage, considering the possibility of emission 
of up to 29 different preequilibrium particles
and fragments (the first 29 from Tab.\ 1, those having $Z \leq 6$).

As we use now in CEM2k the
Tripathi, Cucinota, and Wilson systematics$^{33}$ for
charged-particle inverse cross sections that employ the energy dependent
approximation for the Coulomb barriers defined by Eqs.\ (29--31),
we replaced the old routine that calculates Coulomb barriers
according to 
Dostrovsky  {\it et. al.}$^{39}$ used in all previous versions of CEM
code with a new routine that calculates Coulomb barriers using
Eqs.\ (29--31). We did this for self-consistency of the model:
we use the same approximation for Coulomb barriers while calculating
inverse cross sections and in all other parts of the code where 
Coulomb barriers are used.

In this extended version of the CEM2k, we consider also the possibility
of ``creating'' high-energy d, t, $^3$He, and $^4$He by final state
interactions among emitted cascade nucleons outside 
of the target, 
using the coalescence model as implemented by Gudima {\it et al.}$^{75}$
In contrast to most other coalescence models for heavy-ion induced 
reactions, where complex-particle spectra are estimated simply by
convolving the measured or calculated inclusive spectra of nucleons
with corresponding fitted coefficients (see, {\it e.g.}$^{76}$,
and references therein), CEM2k uses in its simulation of
particle coalescence real information about all emitted cascade nucleons
and does not use integrated spectra. CEM2k assumes that
all the cascade nucleons having differences in their momenta 
smaller than $p_c$ and a correct isotopic content form an appropriate
composite particle. This means that the formation probability for,
{\it e.g.}, a deuteron is
\bea
W_d(\vec p,t) = \int \int d \vec p_p  d \vec p_n 
\rho^C(\vec p_p,t) \rho^C(\vec p_n,t) \nonumber \\
\times \delta(\vec p_p + \vec p_n - \vec p)
\Theta(p_c - |\vec p\ ^{c.m.} - \vec n^{c.m.}|) ,
\eea
where the particle density in momentum space is related to the
one-particle distribution function $f$ by
\begin{equation}
\rho^C(\vec p,t) = \int d \vec r f^C (\vec r, \vec p,t) .
\end{equation}
Here, the
superscript $C$ shows that only cascade nucleons are taken into
account for the coalescence process. 
\onecolumn

\begin{center}
\begin{figure}[h!] 

\vspace*{-40mm}
\hspace*{-18.mm}
\includegraphics[angle=-0,width=204mm]{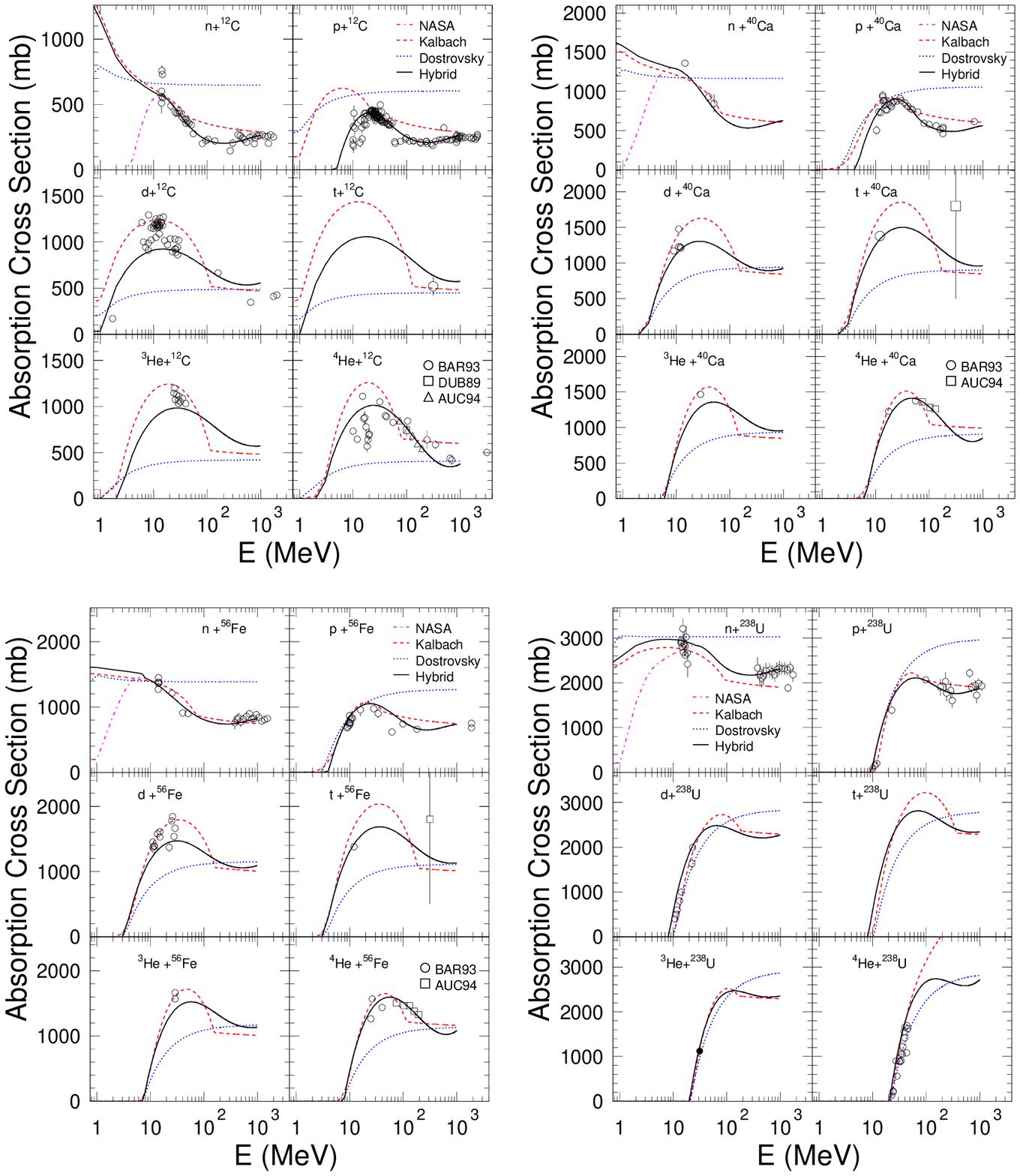}

\vspace*{-50mm}
{\noindent Figure 18. Inverse cross sections for
n, p, d, t, $^3$He, and $^4$He on $^{12}$C, $^{40}$Ca, $^{56}$Fe,
and  $^{238}$U calculated with our routine HYBRID (black solid lines)
compared with available experimental data (symbols) from the compilation
by Barashenkov$^{35}$ (circles,
either without a special label or labeled as ``BAR93''), 
by Auce {\it et al.}$^{71}$ (labeled as ``AUC94''),
and by Dubar {\it et al.}$^{72}$ (labeled as ``DUB89'').
For comparison, calculations using the NASA systematics by Tripathi, 
Cucinota, and Wilson$^{33}$ are shown by magenta dot-dashed lines,
results from  the Kalbach parameterization$^{34}$ are shown with red dashed
lines, and predictions by the Dostrovsky  {\it et. al.}$^{39}$
approximation are shown by blue dotted lines.
}
\label{fig18}

\vspace*{-10mm}
\end{figure}
\end{center}

\twocolumn
The coalescence radii $p_c$
were fitted for each composite particle in Ref.$^{75}$
to describe available data for the reaction Ne+U at 1.04 GeV/nucleon,
but the fitted values turned out to be quite universal and were
subsequently found
to satisfactorily describe high-energy  complex-particle production
for a variety of reactions induced both by protons and nuclei
at incident energies up to about 200 GeV/nucleon.  These parameters
used in the present version of CEM2k are:
\bea
p_c({\mbox d}) &=& 90 \mbox{ MeV/c,} \nonumber \\
p_c({\mbox t}) &=&  p_c(^3{\mbox He}) = 108 \mbox{ MeV/c,} \\
p_c(^4{\mbox He}) &=& 115 \mbox{ MeV/c.}
\nonumber\eea
If several cascade nucleons are chosen to coalesce into composite
particles, they are removed from the status of nucleons and
do not contribute further to such nucleon characteristics as spectra,
multiplicities, {\it etc.} 

To describe fission-fragment production from heavy fissioning compound
nuclei, we incorporated into the present version of CEM2k the 
thermodynamical fission model by Stepanov$^{77}$
with its own parameterizations for mass and charge widths, 
level-density parameters, fission barriers, {\it etc.} The already
extended length of the current paper does not allow us to 
present here details of the thermodynamical fission model by Stepanov$^{77}$
and sample results obtained with it. We plan to publish our fission 
results in a separate paper; for the moment, we only note  that
this model statistical by nature and coupled with CEM2k it provides 
similar results to those we got with the modified RAL 
model from GEM2, with some better agreement with the
data for pre-actinides and some worse agreement for actinides.

For brevity, we call the extended version of CEM2k described here
as CEM2k2f (shorthand for ``second fission model") and show below several 
results on complex-particle and fragment
emission from several reactions, referring to them as from CEM2k2f.

Fig.\ 19 shows examples of angle-integrated energy spectra of
p, d, t, $^3$He, and $^4$He from 62 MeV proton interactions with
Al, $^{56}$Fe, Y, and $^{120}$Sn calculated with CEM2k2f (red
histograms) compared
with experimental data by Bertrand and Peelle$^{60}$ (symbols). 
For comparison,
results obtained with previous versions of CEM are shown with
black histograms, namely:
calculations with CEM95$^2$ for $^{56}$Fe, Y, and $^{120}$Sn, and
with CEM97$^{3}$ for Al.
We still have some problems in a correct description of the high-energy
tails of complex-particle spectra (that is probably related to
direct processes of complex particle production like knock-out and
pick-up  which are not included in CEM2k2f),
although the improved CEM2k2f provides a much
better agreement with the data than the older versions CEM95 and CEM97. 

The new inverse cross sections and Coulomb barriers implemented
here provide a correct and better description of the evaporation parts 
of particle spectra, but require an adjustment of the ``condensation''
probability $\gamma_j$ at the preequilibrium stage of reactions
(see details in$^{1,68}$). CEM assumes$^1$ that 
during the preequilibrium part of a reaction $p_j$ excited particles
(excitons) are able to condense with probability $\gamma_j$
forming a complex particle which can be emitted during the 
preequilibrium state. In our first publication on CEM$^1$, $\gamma_j$
was roughly estimated as 
\beq
\gamma_j \simeq p^3_j (p_j /A)^{p_j - 1},
\eeq
($A$ is the mass number of the pre-compound nucleus before
emitting the particle $j$), and this estimation was used in all the
succeeding versions of CEM. Generally, $\gamma_j$ is a parameter
of the preequilibrium model, and in the literature it
is taken from fitting the
theoretical preequilibrium spectra to the experimental ones, which gives 
rise to an additional, as compared to (44), dependence of the
factor $\gamma_j$ on $p_j$, mass-number $A$, and excitation energy
(see, {\it e.g.} Ref.$^{78}$). Our calculations show that one may
obtain a reasonable description with CEM2k2f of all complex
particles up to $^4$He from many reactions
if we use values for $\gamma_j$ estimated using
Eq.\ (44) divided by 3 for d and t, and by 4 for $^3$He. 
The results shown in Fig.\ 19 are obtained with this estimation
for $\gamma_j$. It is clear that if we fit $\gamma_j$ for each particle and
each reaction as done in Ref.$^{78}$, one may obtain a much better
agreement with the data, but such exercises are outside the aim of our
work. Our calculations of several reactions show that 
if we were to fit $\gamma_j$ in CEM2k+GEM2,
we could also obtain a much better description of complex
particle spectra than the one shown in Fig.\ 14, but such a fit is as well
beyond the aim of the present work.

All spectra of particles from Al, $^{56}$Fe, and $^{120}$Sn shown on
Fig.\ 19 are calculated without introducing a
coalescence mechanism for complex-particle production. This is
because for proton-induced reactions at these incident energies
the multiplicity of fast nucleons emitted at the cascade stage of
reactions is so small that the role of coalescence of
complex particles from such nucleons is negligible. 
As an example, in Fig.\ 19  we show only for Y (by blue histograms) 
calculated 
spectra that contain contributions from coalescence production of
complex particles. One can see that indeed the contribution from
the coalescence mechanism is very small for this reaction
and complex particle spectra
calculated with and without this mechanism almost coincide. 

Another example of the role of different
mechanisms of particle production from these reactions is shown
in Table 6, where we present the mean total multiplicities of 
complex particles from Y and $^{120}$Sn calculated by CEM2k2f 
(last column) together with their contributions
from preequilibrium, evaporation, and coalescence reaction mechanisms
(third to fifth columns).\\

\begin{center}
Table 6. Role of different CEM2k2f reaction mechanisms in complex particle 
production (multiplicities) from p (62 MeV) + Y and $^{120}$Sn\\

\vspace{2mm}
\begin{tabular}{|c|c|c|c|c|c|}
\hline \hline
Nucleus & Ejectile & Prec & Evap & Coales & Total  \\
\hline \hline
$^{89}$Y        & d & 2.96e-2 & 1.41e-3 & 1.83e-3 & 3.28e-2 \\
                & t & 5.34e-3 & 1.13e-4 & 5.66e-6 & 5.46e-3 \\
                & $^3$He & 1.92e-3 & 1.13e-5 & 2.83e-6 & 1.94e-3 \\
                & $^4$He & 4.19e-2 & 1.56e-2 & --      & 5.75e-2 \\
\hline \hline
$^{120}$Sn        & d & 3.17e-2 & 6.51e-4 & 1.75e-3 & 3.41e-2 \\
                & t & 8.58e-3 & 1.57e-4 &   --    & 8.74e-3 \\
                & $^3$He & 7.59e-4 &   --    & --      & 7.59e-4 \\
                & $^4$He & 3.15e-2 & 6.19e-3 & --      & 3.77e-2 \\
\hline \hline
\end{tabular}
\end{center}
The percentage of the coalescence mechanism to the total complex
particle yields is only 5.6\% for d from Y and 5.1\% for d from $^{120}$Sn,
while for t and $^3$He from Y is only of the order of 0.1\% and below
the statistical errors of these reactions as simulated by CEM2k2f for 
$^4$He from Y and t, $^3$He, and $^4$He from $^{120}$Sn.
Nevertheless, our other studies show that at high incident 
energies$^{79,80}$ and for heavy-ion
induced reactions$^{75}$, the coalescence mechanism of complex-particle 
production is important and often is the main mechanism of high-energy
complex particle production; therefore we incorporate this mode
into CEM2k2f. 

It is informative to observe from Table 6 that due to high Coulomb
barriers for complex particles from medium and heavy nuclei, the 
main contribution to
production of such particles comes from preequilibrium 
emission, while the evaporation mode is less important for such reactions.

Fig.\ 20 shows examples of $^3$He and $^4$He double-differential
spectra from interaction of 300 and 480 MeV protons with Ag
calculated with CEM2k2f compared with experimental data by
Green and Korteling$^{81}$. One can see that CEM2k2f reproduces
reasonably these data though it overestimates
most of the spectra, especially the high-energy tails of them.
Probably, by fitting $\gamma_j$ for this reactions as discussed
above one may obtain a much better agreement with the measurement,
but we didn't do any additional fitting of $\gamma_j$ for these
reactions. We also see some problems with a correct description
of angular distributions of $^3$He and $^4$He from these
reactions and we plan to address this point in our further work.
These results are preliminary and though we did not get a very good
agreement with the data, they are encouraging to us as they help us
to clarify ways of further improving CEM2k.

From Fig.\ 20 we conclude that we probably get with CEM2k2f 
too much preequilibrium particle emission from these reactions.
Such a conclusion may also be drawn from the results presented in Table 7,
where we present a comparison of the total total yields 
of several particles
and light fragments from 480 MeV p + Ag calculated with CEM2k2f
with experimental data by Green, Korteling, and Jackson$^{82}$.
The results presented in Table 7 are preliminary, without any
additional fitting of any parameters of CEM2k2f, nevertheless
we see a reasonable agreement with the measurement not only for
production of $^3$He and $^4$He, but also for heavier fragments.
We calculated this reaction with CEM2k2f both including
preequilibrium-particle emission (third column in Table 7) and
ignoring it (last column in Table 7). We can see that for many of
the fragment-production cross sections ($^8$Li, $^9$Li, $^7$Be, 
$^9$Be, and $^{11}$Be shown in the table) the experimental
yields are higher than the ones calculated without 
preequilibrium emission, and lower than those
with preequilibrium emission. Though these results are preliminary
and will change when we develop further our model, they may
serve as a rough  indication that in the present version of CEM2k2f we
have too much preequilibrium particle emission, just as we got
with CEM2k+GEM2 discussed in Sec. III. We recall
again that in$^{56}$ we address this point both in our
CEM2k$^{10}$ and LAQGSM$^{79}$ codes merged with GEM2$^{37,38}$ 
using an approach based on Ref.$^{67}$, and the results obtained
there are promising.

\vspace{5mm}
Table 7.
Comparison of the measured$^{82}$ total production cross sections (mb)
for several ejectiles from 480 MeV p + Ag
with our preliminary CEM2k2f results  with and without 
preequilibrium emission 

\begin{center}
\vspace{2mm}
\begin{tabular}{|c|c|c|c|c|}
\hline \hline
Ejectile & Exp. data & CEM2k2f            & CEM2k2f          \\
         &           &          with Prec &          no Prec \\
\hline \hline
n  & --   & 3.97e+3 & 4.50e+3 \\
p  & --   & 9.30e+2 & 9.94e+2 \\
d  & --   & 2.85e+2 & 3.51e+2 \\
t  & --   & 1.27e+2 & 1.35e+2 \\
$^3$He  & 3.42e+1 & 3.98e+1 & 3.02e+1 \\
$^4$He  & 4.44e+2 & 7.68e+2 & 1.11e+3 \\
$^6$He  &   --    & 6.60e+0 & 3.61e+0 \\
$^8$He  &   --    & 1.31e-1 & 1.61e-2 \\
$^6$Li  & 3.67e+1 & 9.91e-2 & 9.28e+0 \\
$^7$Li  & 4.20e+1 & 3.59e-2 & 5.27e+0 \\
$^8$Li  & 4.48e-1 & 1.24e-3 & 5.31e-1 \\
$^9$Li  & 6.98e-2 & 1.08e-4 & 8.37e-2 \\
$^7$Be  & 8.36e-1 & 5.85e-3 & 1.20e+0 \\
$^9$Be  & 9.22e-1 & 1.06e-2 & 1.73e+0 \\
$^{10}$Be  & 4.16e-1 & 6.23e-4 & 1.52e-1 \\
$^{11}$Be  & 1.29e-2 & 2.24e-5 & 1.66e-2 \\
$^{12}$Be  &    --   & 7.65e-7 & 1.02e-3 \\
$^{8}$B   &  --     & 5.98e-5 & 3.46e-2 \\
$^{10}$B   & 4.85e-1 & 1.35e-3 & 4.22e-1 \\
\hline \hline
\end{tabular}
\end{center}

\onecolumn
\begin{center}
\begin{figure}[h!] 
\vspace*{-50mm}
\hspace*{-18.mm}
\includegraphics[angle=-0,width=198mm]{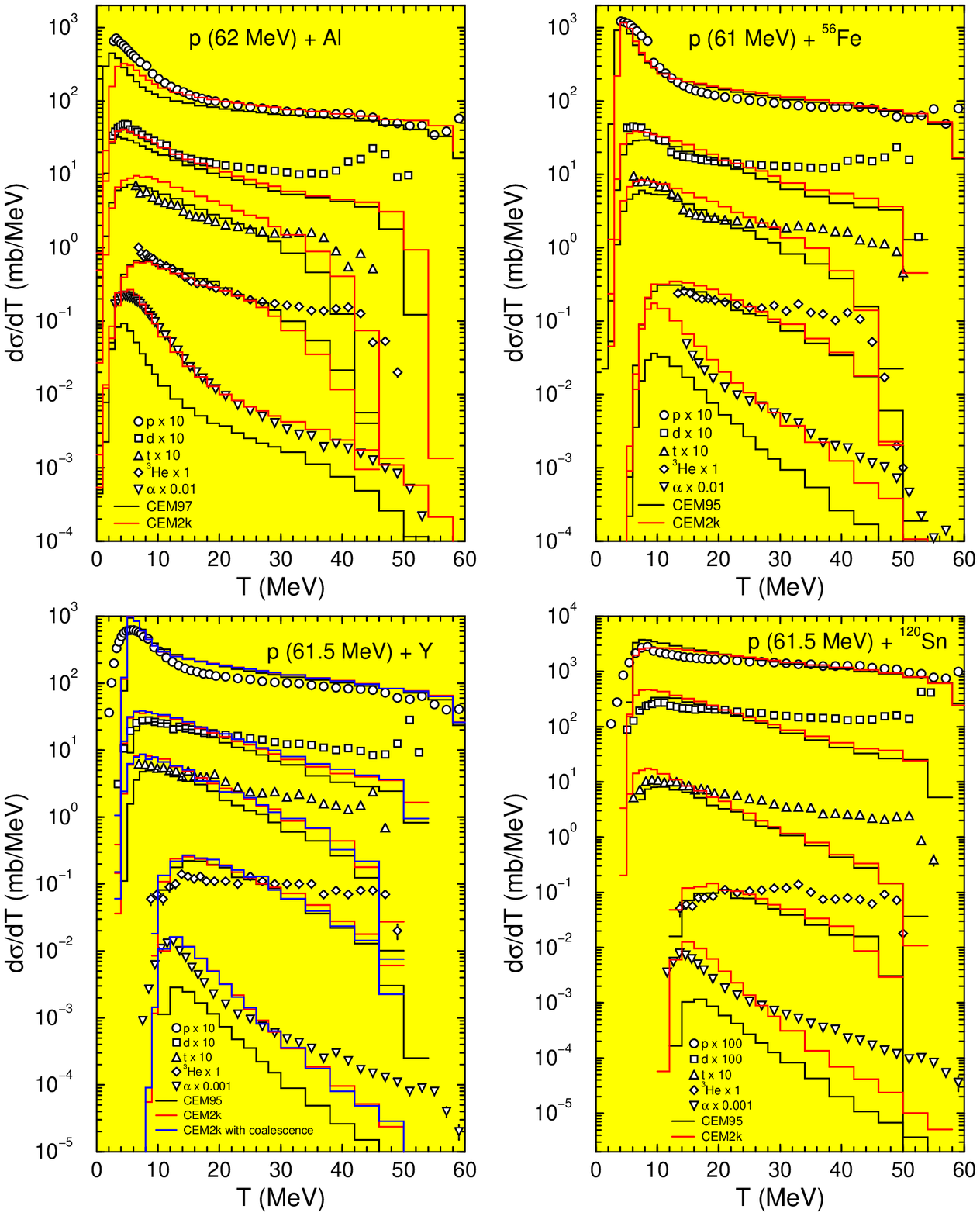}
\vspace*{-30mm}

{\noindent Figure 19. Angle-integrated energy spectra of  p, d, t, 
$^{3}$He, and $^{4}$He from 62 MeV protons on Al, 
$^{56}$Fe, Y, and $^{120}$Sn.  Calculations from the CEM2k2f
code are shown by red histograms and from CEM95$^{2}$ 
(for $^{56}$Fe, Y, and $^{120}$Sn)
and CEM97$^{3}$ (for Al) are shown by black histograms.
For Y, the blue histograms show CEM2k2f results taking into account
contributions from coalescence of complex particles from
fast cascade nucleons, as described in the text.
Experimental data (symbols) are by Bertrand and Peelle$^{60}$. 
}
\label{fig19}
\end{figure}
\end{center}

\begin{center}
\begin{figure}[h!] 
\vspace*{-48mm}
\hspace*{-18.mm}
\includegraphics[angle=-0,width=198mm]{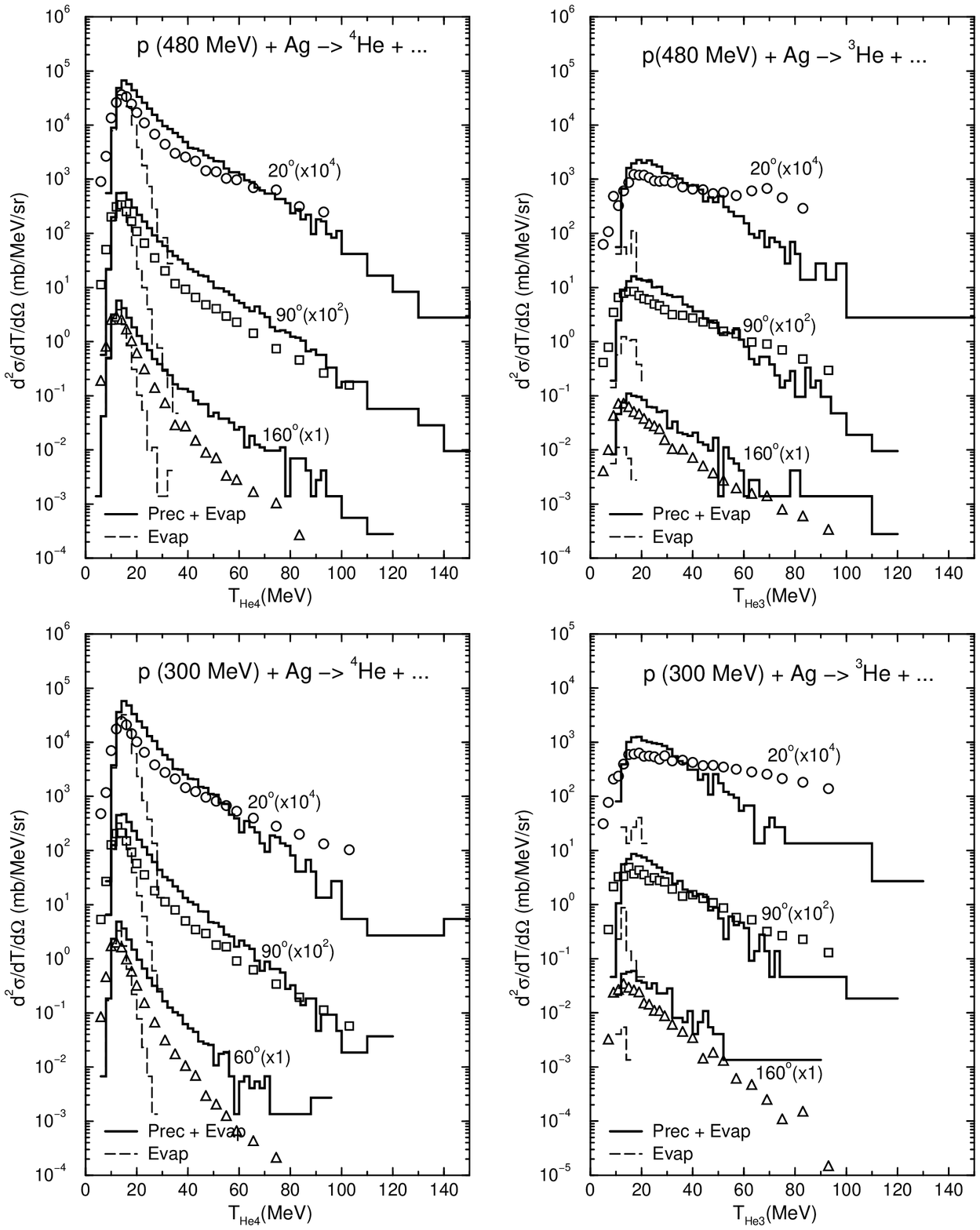}
\vspace*{-30mm}

{\noindent Figure 20. Double-differential spectra of  
$^{4}$He and $^{3}$He from 480 and 300 MeV protons on Ag. 
Calculations from the CEM2k2f code are shown by solid histograms
and contribution from only the evaporation mode,
by dashed histograms.
Experimental data (symbols) are by Green and Korteling$^{81}$. 
}
\label{fig20}
\vspace*{-10mm}
\end{figure}
\end{center}

\twocolumn

Finally, we have investigated the well known
code GEMINI by Charity$^{83}$ as an alternative way to describe
production of various fragments by merging GEMINI with both our
CEM2k and LAQGSM. The preliminary results we find
for spallation, fission, and 
fragmentation products from several reactions
are promising and we will present our results from this
study in a separate paper.

\vspace{0.4cm}
\noindent V. FUTURE WORK\\
The two modifications of the CEM2k model presented here allow us
to describe satisfactorily
many fission and fragmentation reactions in addition to the
spallation reactions which are already quite well described by CEM2k.

Our results presented in Sec. III
show that the Generalized Evaporation Model
code GEM2 by Furihata is a useful tool and when merged with CEM2k
(or LAQGSM) it allows us to reproduce a large variety of
fission and fragmentation measurements, though some tuning of
several GEM2 parameters will be required.
We may choose a model similar to GEM2 in our codes
to describe fission, fragmentation, and evaporation yields, but it must 
be significantly extended and improved in order to properly describe 
complex particles, light and heavy fragments 
not produced in fission, and more importantly,
to become self-consistent and better grounded from a physical 
point of view.

The latest version of CEM2k described in Sec. IV, CEM2k2f,
contains new inverse cross
sections and Coulomb barriers, new routines to calculate
widths of emitted particles and to simulate their kinetic energies using
arbitrary approximations for the inverse cross sections and Coulomb barriers,
allows evaporation of up to 66 different particles
both in the ground and excited states and
emission of up to 29 different preequilibrium particles, takes into account
coalescence of complex particles from
fast nucleons emitted at the cascade stage, and describes
much better complex-particle spectra and yields than its precursors.

Our work is not finished yet. For example, at the evaporation
and preequilibrium stages we still need to make an ``ad hoc'' decision
about how many different particles should be included into consideration
so that we have production of light fragments heavier than $^4$He but
maintain the running time of the code at a reasonable level.
Our new algorithms and
routines do not impose any limitations to the energy dependence
of the particle emission probabilities, and we plan to try to improve
the present description of the level densities and particle emission widths
with a hope to describe better the tails of complex-particle spectra.

We consider also implementation into CEM2k a model of fragmentation,
to be able to describe emission of fragments like Na, Mg, and heavier,
and possibly, incorporation of the Fermi break-up model to replace
the preequilibrium and evaporation models when the residual excited
nucleus after the cascade stage of a reaction is very light.

Our work on fission in CEM2k is also not finished. For instance, 
we are not satisfied with the situation that in both the improved
versions of the CEM2k discussed here we still have an additional 
input parameter to describe fission cross sections: either $C(Z)$,
in the GEM2/RAL approach for nuclei with $Z > 88$, and/or $a_f/a_n$, 
in the same model for lighter fissioning nuclei with $Z \leq 88$ and 
when using the model by Stepanov.
In addition, to get with CEM2k a better description of fission-fragment
distributions, the approximations used in both GEM/RAL and Stepanov's 
fission models to simulate $A$, $Z$, and kinetic-energy distributions
for fission fragments should be further adjusted in our code.

We emphasize that it is not sufficient to analyze only $A$ and 
$Z$ distributions 
of the product yields when evaluating the type of model discussed here, 
as is often done in the literature, but that it is essential to 
study all the separate isotope yields as well as the spectra of light 
particles and fragments. This should be done for different target nuclei,
different incident particles, and at different energies.

\vspace{0.4cm}
\noindent ACKNOWLEDGMENTS\\
We thank Dr.\ Shiori Furihata for several useful discussions, for 
sending us her
Generalized Evaporation Model code GEM, and providing us further
with its updated version, GEM2. We thank Dr. Nikita Stepanov for
providing us with his thermodynamical fission model implemented
in the ITEP code INUCL and useful explanations on his code.
We thank Prof.\ Barashenkov and Dr.\ Polanski for providing us with their
phenomenological code CROSEC and Dr.\ Tsao for providing our colleagues
with his and Dr.\ Silberberg's phenomenological
code YIELDX we used here and in our other works.
We are grateful to Drs.\ M.\ B.\ Chadwick, 
R.\ E.\ Prael, D.\ D.\ Strottman, L.\ S.\ Waters, and W.\ B.\ Wilson
for useful discussions, interest, and support.

This study was supported by the U.\ S.\ Department of Energy and by the
Moldovan-U.\ S.\ Bilateral Grants Program, CRDF Project MP2-3025.

\vspace{0.5cm}
\noindent REFERENCES\\
1. K.\ K.\ Gudima, S.\ G.\ Mashnik, and V.\ D.\ Toneev,
``Cascade-Exciton Model of Nuclear Reactions,"
{\it Nucl.\ Phys.\ }{\bf A401}, 329--361 (1983).\\

\vspace*{-0.2cm}
\noindent
2. S.\ G.\ Mashnik, ``User Manual for the Code CEM95," JINR, Dubna (1995);
OECD NEA Data Bank, Paris, France (1995); \\
http://www.nea.fr/abs/html/iaea1247.html;
RSIC-PSR-357, Oak Ridge, 1995.\\

\vspace*{-0.2cm}
\noindent
3. S.\ G.\ Mashnik and A.\ J.\ Sierk,
``Improved Cascade-Exciton Model of Nuclear Reactions,"
{\it Proc.\ Fourth Workshop on Simulating Accelerator Radiation Environments
(SARE4)}, Knoxville, TN, USA, September 14--16, 1998, pp.\ 29--51,
T.\ A.\ Gabriel, Ed., ORNL, Oak Ridge (1999). \\

\vspace*{-0.2cm}
\noindent
4. A.\ J.\ Sierk and S.\ G.\ Mashnik,
``Modeling Fission in the Cascade-Exciton Model,"
{\it Proc.\ Fourth Workshop on Simulating Accelerator Radiation Environments
(SARE4)}, Knoxville, TN, USA, September 14--16, 1998, pp.\ 53--67,
T.\ A.\ Gabriel, Ed., ORNL, Oak Ridge (1999).\\

\vspace*{-0.2cm}
\noindent
5. {\it MCNPX$^{TM}$ User's Manual, Version 2.1.5}, edited by Laurie S.\ Waters,
Los Alamos National Laboratory Report LA-UR-99-6058, Los Alamos (1999).\\

\vspace*{-0.2cm}
\noindent
6. W.\ Wlazlo {\it et el.}, 
``Cross Sections of Spallation Residues Produced in 
1$A$ GeV $^{208}$Pb on Proton Reactions,"
{\it Phys.\ Rev.\ Lett.\ }{\bf 84}, 5736--5739 (2000).\\

\vspace*{-0.2cm}
\noindent
7. T.\ Enqvist {\it et al.}, 
``Isotopic Yields and Kinematic Energies of Primary Residues in 
1$A$ GeV $^{208}$Pb + p Reactions,"
{\it Nucl.\ Phys.\ }{\bf A686}, 481--524 (2001).\\

\vspace*{-0.2cm}
\noindent
8. J.\ Taieb {\it et al.}, 
``Measurement of $^{238}$U Spallation Product Cross Sections at 
1 GeV per Nucleon,"\\
http://www-wnt.gsi.de/kschmidt/publica.htm\#Conferences.\\

\vspace*{-0.2cm}
\noindent
9. F.\ Rejmund {\it et al.}, 
``Measurement of Isotopic Cross Sections of Spallation Residues 
in 800 $A$ MeV $^{197}$Au + p Collisions,"
{\it Nucl.\ Phys.\ }{\bf A683}, 540--565 (2001);\\
J.\ Benlliure {\it et al.},  
``Isotopic Production Cross Sections of Fission Residues 
in $^{197}$Au-on-proton Collisions at 800 $A$ MeV,"
{\it Nucl.\ Phys.\ }{\bf A683},  513--539 (2001).\\

\vspace*{-0.2cm}
\noindent
10. S.\ G.\ Mashnik and A.\ J.\ Sierk,
``CEM2k---Recent Developments in CEM,"
{\it Proc.\ 
AccApp00, November 13--15, 2000, Washington, DC, USA}, pp.\ 328--341,
American Nuclear Society, La Grange Park, IL, USA, (2001);
http://xxx.lanl.gov/ps/nucl-th/0011064.\\

\vspace*{-0.2cm}
\noindent
11. Stepan\ G.\ Mashnik and Arnold\ J.\ Sierk,
``Recent Developments of the Cascade-Exciton Model of Nuclear Reactions,"
{\it Proc.\ 
Int.\ Conf.\ on Nuclear Data for Science and Technology (ND2001),
Oct.\ 7--12, 2001, Tsukuba, Japan};
http://lib-www.lanl.gov/la-pubs/0081526.pdf, to be published in
{\it Journal of Nuclear Science and Technology (2002)}.\\

\vspace*{-0.2cm}
\noindent
12. F.\ Atchison,
``Spallation and Fission in Heavy Metal Nuclei under Medium
Energy Proton Bombardment," 
in {\it Proc.\ Meeting on Targets for Neutron Beam Spallation Source,
Julich, June 11--12, 1979}, pp.\ 17--46,
G.\ S.\ Bauer, Ed.,
Jul-Conf-34, Kernforschungsanlage Julich GmbH, Germany (1980).\\

\vspace*{-0.2cm}
\noindent
13. F.\ Atchison,
``A Treatment of Fission for HETC,''
in {\it Intermediate Energy Nuclear Data: Models and Codes}, pp.\ 199--218,
Proc.\ of a Specialists's Meeting, May 30--June 1, 1994, 
Issy-Les-Moulineaux, France, OECD, Paris, France (1994).\\ 

\vspace*{-0.2cm}
\noindent
14. R.\ E.\ Prael and H.\ Lichtenstein, 
``User Guide to LCS: The LAHET Code System," 
LANL Report No.\ LA-UR-89-3014, Los Alamos (1989); \\
http://www-xdiv.lanl.gov/XTM/lcs/lahet-doc.html.\\

\vspace*{-0.2cm}
\noindent
15. J.\ Cugnon, C.\ Volant, and S.\ Vuillier,
``Improved Intranuclear Cascade Model for Nucleon-Nucleus Interactions,
{\it Nucl.\ Phys.\ }{\bf A620},  475--509, (1997).\\

\vspace*{-0.2cm}
\noindent
16. S.\ G.\ Mashnik, R.\ J.\ Peterson, A.\ J.\ Sierk, and M.\ R.\ Braunstein,
``Pion-Induced Transport of $\pi$ Mesons in Nuclei,''
{\it Phys.\ Rev.\ C} {\bf 61}, 034601 (2000).\\ 

\vspace*{-0.2cm}
\noindent
17. A.\ V.\ Prokofiev, S.\ G.\ Mashnik, and A.\ J.\ Sierk,
``Cascade-Exciton Model Analysis of Nucleon-Induced Fission Cross 
Sections of Lead and Bismuth at Energies from 45 to 500 MeV,''
{\it Nucl.\ Sci.\ Eng.\ }{\bf 131}, 78--95, (1999).\\

\vspace*{-0.2cm}
\noindent
18. G.\ Audi and A.\ H.\ Wapstra,
``The 1995 Update to the Atomic Mass Evaluation,"
{\it Nucl.\ Phys.\ }{\bf A595}, 409--480 (1995).\\ 

\vspace*{-0.2cm}
\noindent
19. P.\ M\"oller, J.\ R.\ Nix, W.\ D.\ Myers, and W.\ J.\ Swiatecki,
``Nuclear Ground-States Masses and Deformations,"
{\it Atomic Data and Nuclear Data Tables}, {\bf 59}, 185--383 (1995).\\

\vspace*{-0.2cm}
\noindent
20. P.\ M\"oller, J.\ R.\ Nix, and K.-L.\ Kratz,
``Nuclear Properties for Astrophysical and Radioactive-Ion-Beam Application,"
{\it Atomic Data and Nuclear Data Tables} {\bf 66}, 131--345 (1997).\\

\vspace*{-0.2cm}
\noindent
21. A.\ V.\ Ignatyuk {\it et al.}, 
``Fission of Pre-Actinide Nuclei. Excitation Functions for the
($\alpha,f$) Reactions,"
{\it Yad.\ Fiz.\ {\bf 21}}, 1185--1205 (1975)
[{\it Sov.\ J.\ Nucl.\ Phys.\ }{\bf 21}, 612--621 (1975)].\\

\vspace*{-0.2cm}
\noindent
22. A.\ S.\ Iljinov {\it et al.}, 
``Decay Width and Lifetimes of Excited Nuclei,"
{\it Nucl.\ Phys.\ }{\bf A543}, 517--557 (1992).\\

\vspace*{-0.2cm}
\noindent
23. R.\ Silberberg, C.\ H.\ Tsao, and A.\ F.\ Barghouty,
``Updated Partial Cross Sections of Proton-Nucleus Reactions,"
{\it Astrophys.\ J.\ }{\bf 501}, Part 1, 911--819 (1998).\\

\vspace*{-0.2cm}
\noindent
24. J.-J.\ Gaimard and K.-H.\ Schmidt,
``A Reexamination of the Abrasion-Ablation Model for the Description
 of the Nuclear Fragmentation Reaction,''
{\it Nucl.\ Phys.\ }{\bf A531}, 709--745 (1991).\\

\vspace*{-0.2cm}
\noindent
25. A.\ V.\ Ignatyuk, N.\ T.\ Kulagin, V.\ P.\ Lunev, and K.-H.\ Schmidt,
``Analysis of Spallation Residues within the Intranuclear Cascade
Model,'' 
{\it Proc.\ XV Workshop on Phys.\ of Nucl.\ Fission, Oct.\ 3--6, 2000,
Obninsk, Russia};
http://www-wnt.gsi.de/kschmidt/publica.htm.\\

\vspace*{-0.2cm}
\noindent
26. S.\ G.\ Mashnik {\it et al.}, 
``Benchmarking Ten Codes Against the Recent GSI Measurements of the 
Nuclide Yields from $^{208}$Pb, $^{197}$Au, and $^{238}$U + p Reactions
at 1 GeV/nucleon,''
{\it Proc.\ 
Int.\ Conf.\ on Nuclear Data for Science and Technology (ND2001),
Oct.\ 7--12, 2001, Tsukuba, Japan};
http://lib-www.lanl.gov/la-pubs/0081527.pdf, to be published in
{\it Journal of Nuclear Science and Technology (2002)}.\\

\vspace*{-0.2cm}
\noindent
27. T.\ W.\ Armstrong and K.\ C.\ Chandler,
``HETC A High Energy Transport Code,''
{\it Nucl.\ Sci.\ Eng.\ }{\bf 49}, 110--111 (1972).\\

\vspace*{-0.2cm}
\noindent
28. 
V.\ S.\ Barashenkov {\it et al.}, 
``CASCADE Program Complex for Monte Carlo Simulations of Nuclear 
Processes Initiated by High-Energy Particles and Nuclei in Gaseous
and Condensed Matter,''
JINR Report R2-85-173, Dubna, 1985;
V.\ S.\ Barashenkov, F.\ G.\ Zheregi, and Zh.\ Zh.\ Musul'manbekov,
``The Cascade Mechanism of Inelastic Interactions of High-Energy Nuclei,''
{\it Yad.\ Fiz.\ }{\bf 39}, 1133--1134 (1984) 
[{\it Sov.\ J.\ Nucl.\ Phys.\ }{\bf 39}, 715--716 (1984)];
V.\ S.\ Barashenkov, B.\ F.\ Kostenko, and A.\ M.\ Zadorogny,
``Time-Dependent Intranuclear Cascade Model,''
{\it Nucl.\ Phys.\ }{\bf A338}, 413--420 (1980).\\

\vspace*{-0.2cm}
\noindent
29.
V.\ S.\ Barashenkov, A.\ Yu.\ Konobeev, Yu.\ A.\ Korovin, and V.\ N.\ Sosnin,
``CASCADE/INPE Code System,"
{\it Atomnaya Energiya }{\bf 87}, 283--286 (1999)
[{\it Atomic Energy }{\bf 87}, 742--744 (1999)].\\

\vspace*{-0.2cm}
\noindent
30.
G.\ A.\ Lobov, N.\ V.\ Stepanov, A.\ A.\ Sibirtsev, and Yu.\ V.\ Trebukhovskii,
``Statistical Simulation of Hadron and Light-Nuclei Interactions
with Nuclei. Intranuclear Cascade Model.''
Institute for Theoretical and Experimental Physics (ITEP)
Preprint No.\ ITEP-91, Moscow, 1983\\

\vspace*{-0.2cm}
\noindent
31.
Yu.\ E.\ Titarenko {\it et al.}, 
``Cross Sections for Nuclide Production in 1 GeV Proton-Irradiated Pb",
{\it Phys.\ Rev.\ C }{\bf 65}, 064610 (2002).\\

\vspace*{-0.2cm}
\noindent
32. K.\ Ishibashi {\it et al.},
``Measurement of Neutron-Production Double-Differential
Cross Sections for Nuclear Spallation Reaction Induced by 0.8,
1.5 and 3.0 GeV Protons,"
{\it J.\ Nucl.\ Sci.\ Techn.\ }{\bf 34}, 529--537 (1997).\\

\vspace*{-0.2cm}
\noindent
33. 
R.\ K.\ Tripathi, F.\ A.\ Cucinotta, and J.\ W.\ Wilson,
``Accurate Universal Parameterization of Absorption Cross Sections,'' 
{\it Nucl.\ Instr.\ Meth.\ Phys.\ Res.\ }{\bf B117}, 347--349 (1996).\\

\vspace*{-0.2cm}
\noindent
34.
C.\ Kalbach,
``Towards a Global Exciton Model; Lessons at 14 MeV,''
{\it J.\ Phys.\ G: Nucl.\ Part.\ Phys.\ }{\bf 24}, 847--866 (1998).\\

\vspace*{-0.2cm}
\noindent
35.
V.\ S.\ Barashenkov,
{\it Cross Sections of Interactions of Particles and Nuclei with
Nuclei},
JINR, Dubna, Russia (1993);
tabulated data are available from the NEA/OECD Data Bank Web site at\\
http://www.nea.fr/html/dbdata/bara.html.\\

\vspace*{-0.2cm}
\noindent
36.
M.\ Blann,
``New Precompound Decay Model,''
{\it Phys.\ Rev.\ C} {\bf 54}, 1341--1349 (1996);
M.\ Blann and M.\ B.\ Chadwick,
``New Precompound Decay Model: Angular Distributions,''
{\it Phys.\ Rev.\ C }{\bf 57}, 233--243 (1998).\\

\vspace*{-0.2cm}
\noindent
37.
S. Furihata,
``Statistical Analysis of Light Fragment Production from Medium Energy
Proton-Induced Reactions,''
{\it Nucl.\ Instr.\ Meth.\ in Phys.\ Research }{\bf B171}, 252--258 (2000);
``The Gem Code---The Generalized Evaporation Model and the Fission Model,''
{\it Proc.\ of Monte Carlo 2000 Conference, October 23--26, 2000, Lisbon,
Portugal}, to be published by Springer Verlag;
``The Gem Code Version 2 Users Manual,''
Mitsubishi Research Institute, Inc., Tokyo, Japan (November 8, 2001).\\

\vspace*{-0.2cm}
\noindent
38.
Shiori Furihata {\it et al.},
``The Gem Code---A simulation Program for the Evaporation and Fission
Process of an Excited Nucleus,''
JAERI-Data/Code 2001-015, JAERI, Tokai-mura, Naka-gam, Ibaraki-ken, 
Japan (2001).\\
 
\vspace*{-0.2cm}
\noindent
39.
I.\ Dostrovsky, Z.\ Frankel, and G.\ Friedlander,
``Monte Carlo Calculations of Nuclear Evaporation Processes.
III.\ Application to Low-Energy Reactions,"
{\it Phys.\ Rev.\ }{\bf 116}, 683--702 (1959).\\

\vspace*{-0.2cm}
\noindent
40.
V.\ F.\ Weisskopf and D.\ H.\ Ewing,
``On the Yield of Nuclear Reactions with Heavy Elements,"
{\it Phys.\ Rev.\ }{\bf 57}, 472--485 (1940).\\

\vspace*{-0.2cm}
\noindent
41.
P.\ E.\ Haustein,
``An Overview of the 1986--1987 Atomic Mass Predictions,''
{\it Atomic Data and Nuclear Data Tables }{\bf 39}, 185--393 (1988).\\

\vspace*{-0.2cm}
\noindent
42. A.\ G.\ W.\ Cameron,
``A Revised Semiempirical Atomic Mass Formula,"
{\it Can.\ J.\ Phys.\ }{\bf 35}, 1021--1032 (1957).\\

\vspace*{-0.2cm}
\noindent
43.
T.\ Matsuse, A.\ Arima, and S.\ M.\ Lee,
``Critical Distance in Fusion Reactions,''
{\it Phys.\ Rev.\ C }{\bf 26}, 2338--2341 (1982).\\

\vspace*{-0.2cm}
\noindent
44.
Shiori Furihata and Takashi Nakamura,
``Calculation of Nuclide Productions from Proton Induced Reactions on
Heavy Targets with INC/GEM,''
{\it Proc.\ 
Int.\ Conf.\ on Nuclear Data for Science and Technology (ND2001),
Oct.\ 7--12, 2001, Tsukuba, Japan},
 to be published in
{\it Journal of Nuclear Science and Technology (2002)}.\\

\vspace*{-0.2cm}
\noindent
45.
A.\ S.\ Botvina {\it et al.},
``Statistical Simulation of the Break-up of Highly Excited Nuclei,''
{\it Nucl Phys.\ }{\bf A475}, 663--686 (1987)\\.

\vspace*{-0.2cm}
\noindent
46.
A.\ Gilbert and A.\ G.\ W.\ Cameron,
``A Composite Nuclear-Level Density Formula with Shell Corrections,''
{\it Can.\ J.\ Phys.\ }{\bf 43}, 1446--1496 (1965).\\

\vspace*{-0.2cm}
\noindent
47.
J.\ L.\ Cook, H.\ Ferguson, and A.\ R.\ del Musgrove,
``Nuclear Level Densities in Intermediate and Heavy Nuclei,''
{\it Australian Journal of Physics }{\bf 20}, 477--487 (1967).\\

\vspace*{-0.2cm}
\noindent
48.
William A.\ Friedman and William G.\ Lynch,
``Statistical Formalism for Particle Emission,''
{\it Phys.\ Rev.\ C }{\bf 28}, 16--23 (1983).\\

\vspace*{-0.2cm}
\noindent
49.
The Evaluated Nuclear Structure Data File (ENSDF) maintained by the 
National Nuclear Data Center (NNDC), Brookhaven National Laboratory,\\
http://www.nndc.bnl.gov/.\\

\vspace*{-0.2cm}
\noindent
50.
Robert Vandenbosch and John R.\ Huizenga,
{\it Nuclear Fission}, Academic Press, New York (1973).\\ 

\vspace*{-0.2cm}
\noindent
51.
E.\ F.\ Neuzil and A.\ W.\ Fairhall,
``Fission Product Yields in Helium Ion-Induced Fission
of Au$^{197}$, Pb$^{204}$, and Pb$^{206}$ Targets,''
{Phys.\ Rev.\ }{\bf 129}, 2705--2710 (1963).\\

\vspace*{-0.2cm}
\noindent
52.
A.\ Ya.\ Rusanov, M.\ G.\ Itkis, and V.\ N.\ Okolovich,
``Features of Mass Distributions of Hot Rotating Nuclei,''
{\it Yad.\ Fiz.\ }{\bf 60}, 773--803 (1997)
[{\it Phys.\ At.\ Nucl.\ }{\bf 60}, 683--712 (1997)];
M.\ G.\ Itkis {\it et al.},
``Fission of Excited Nuclei with $Z^2/A = $20--33:
Mass-Energy Distributions of Fragments, Angular Momentum,
and Liquid-Drop Model,''
{\it Yad.\ Fiz.\ }{\bf 58}, 2140--2165 (1995)
[{\it Phys.\ At.\ Nucl.\ }{\bf 58}, 2026--2051 (1995)].\\

\vspace*{-0.2cm}
\noindent
53.
W.\ D.\ Myers and W.\ J.\ Swiatecki,
``Thomas-Fermi Fission Barriers,''
{\it Phys.\ Rev.\ C }{\bf 60}, 014606 (1999).\\

\vspace*{-0.2cm}
\noindent
54.
M.\ G.\ Itkis {\it et al.},
``Experimental Study of the Mass and Energy Distributions
of Fragments from Fission,''
{\it Ya.\ Fiz.\ }{\bf 52}, 23--35 (1990)
[{\it Sov.\ J.\ Nucl.\ Phys.\ }{\bf 52}, 15--22 (1990)].\\

\vspace*{-0.2cm}
\noindent
55.
F.\ Atchison,
``A Revised Calculational Model for Fission,''
Paul Scherrer Insititut Report No. 98-12,
Villigen PSI (1998).\\

\vspace*{-0.2cm}
\noindent
56.
S.\ G.\ Mashnik, K.\ K.\ Gudima, and A.\ J.\ Sierk,
``Merging the CEM2k and LAQGSM Codes with GEM2 to Describe Fission 
and Light-Fragment Production,''
{\it Proc.\ 6th Intl.\ Workshop on Shielding Aspects of Accelerators,
Targets and Irradiated Facilities (SATIF-6), April 10--12, 2002,
SLAC, Menlo Park, CA 94025, USA};
http://www.nea.fr/html/science/satif/satif6t2.html.\\

\vspace*{-0.2cm}
\noindent
57.
H.\ W.\ Bertini,
``Low-Energy Intranuclear Cascade Calculation,"
{\it Phys.\ Rev.\ }{\bf 131}, 1801--1871 (1963);
``Intranuclear Cascade Calculation of the Secondary Nucleon Spectra from
Nucleon-Nucleus Interactions in the Energy Range 340 to 2900 MeV and
Comparison with Experiment,"
{\it Phys.\ Rev.\ }{\bf 188}, 1711--1730 (1969).\\

\vspace*{-0.2cm}
\noindent
58.
Y.\ Yariv and Z.\ Frankel,
``Intranuclear Cascade Calculation of High-Energy Heavy-Ion Interactions,"
{\it Phys.\ Rev.\ C }{\bf 20}, 2227--2243 (1979); 
``Inclusive Cascade Calculation of High Energy Heavy Ion Collisions:
Effect of Interactions between Cascade Particles,"
{\it Phys.\ Rev.\ C }{\bf 24}, 488--494 (1981).\\ 

\vspace*{-0.2cm}
\noindent
59.
 S.\ G.\ Mashnik, A.\ J.\ Sierk, K.\ A.\ Van Riper, and W.\ B.\ Wilson,
``Production and Validation of Isotope Production Cross Section
Libraries for Neutrons and Protons to 1.7 GeV,"
{\it Proc.\ Fourth Workshop on Simulating Accelerator Radiation Environments
(SARE4)}, Knoxville, TN, USA, September 14--16, 1998, pp.\ 29--51,
T.\ A.\ Gabriel, Ed., ORNL, Oak Ridge (1999);
Eprint: {\bf nucl-th/9812071}; our compilation (T-16 Lib) is permanently
updated as new data become available to us. \\

\vspace*{-0.2cm}
\noindent
60.
F.\ E.\ Bertrand and R.\ W.\ Peelle,
"Complete Hydrogen and Helium Particle Spectra from 30 to 60 MeV
Proton Bombardment on Nuclei with A = 12 to 209 and Comparison with the
Intranuclear Cascade Model",
{\it Phys.\ Rev.\ C }{\bf 8}, 1045--1064 (1973).\\

\vspace*{-0.2cm}
\noindent
61.
A.\ R.\ Junghans {\it et al.},
``Projectile-Fragment Yields as a Probe for the Collective Enhancement
in the Nuclear Level Density,''
{\it Nucl.\ Phys.\ }{\bf A629}, 635--655 (1998).\\

\vspace*{-0.2cm}
\noindent
62. 
Marieke Duijvestijn,
{\it Nucleon-Induced Fission at Intermediate Energies},''
PhD thesis, NRG Petten, The Netherlands (2000).\\

\vspace*{-0.2cm}
\noindent
63.
M.\ C.\ Duijvestijn {\it et al.},
``Proton-Induced Fission at 190 MeV of $^{nat}$W, $^{197}$Au,  
$^{nat}$Pb, $^{208}$Pb, and $^{232}$Th,''
{\it Phys.\ Rev.\ C }{\bf 59}, 776--788 (1999).\\

\vspace*{-0.2cm}
\noindent
64.
Yury E.\ Titarenko {\it et al.}, 
``Experimental and Theoretical  Study of the  Yields of Radionuclides
Produced in 232-Th Thin Targets Irradiated by 100 and 800 MeV Protons",
{\it Proc.\ 3rd Int.\ Conf.\ on Accelerator Driven Transmutation Technologies
and Applications (ADTTA'99), Praha (Pruhonice), 7--11 June 1999, 
Czech Republic},
Paper \# P-C24 on the ADTTA'99 Web page 
and Proceedings CD; 
Numerical values of measured cross sections are tabulated in
Yu.\ E.\ Titarenko,
{\it Experimental and Theoretical Study of the Yields of Residual Product
Nuclei Produced in Thin Targets Irradiated by 100--2600 MeV Protons},''
Final Project Technical Report of ISTC 839B-99, ITEP, Moscow (2001).\\

\vspace*{-0.2cm}
\noindent
65.
Arthur C.\ Wahl,
``Systematics of Fission-Product Yields,''
Los Alamos National Laboratory Report LA-UR-01-5944,
Los Alamos (2001).\\

\vspace*{-0.2cm}
\noindent
66.
A.\ Ferrari,
``Physics and Modeling of Hadronic Interactions,'' Tutorials,
Int.\ Conf.\ on Advanced Monte Carlo for Radiation Physics,
Particle Transport Simulation and Application (MC 2000),
23--26 October, 2000, Lisbon, Portugal; more references 
and many details on FLUKA may be found at the Web page
http://pcfluka.mi.infn.it/.\\

\vspace*{-0.2cm}
\noindent
67.
M.\ Veselsky,
``Production Mechanism of Hot Nuclei in Violent Collisions in the
Fermi Energy Domain,''
E-print: nucl-th/0107062 v2 (8 October 2001);
M.\ Veselsky {\it et al.}
``Production of Fast Evaporation Residues by the Reaction $^{20}$Ne +
$^{208}$Pb at Projectile Energies of 8.6, 11.4 and 14.9 A.MeV,''
{\it Z.\ Phys.\ }{\bf A356}, 4003--410 (1997).\\

\vspace*{-0.2cm}
\noindent
68.
S.\ G.\ Mashnik, A.\ J.\ Sierk, O.\ Bersillon, and T.\ Gabriel, 
``Cascade-Exciton Model Analysis of Proton Reactions from
10 MeV to 5 GeV",
Los Alamos National Laboratory Report LA-UR-97-2905 (1997); \\
http://t2.lanl.lanl.gov/publications/publications.html.\\

\vspace*{-0.2cm}
\noindent
69.
H.\ H.\ K.\ Tang, G.\ R.\ Srinivasan, and N.\ Azziz,
``Cascade Statistical Model for Nucleon-Induced Reactions
on Light Nuclei in the Energy Range 50 MeV--1 {GeV}",
{\it Phys.\ Rev.\ C }{\bf 42}, 1598--1622 (1990).\\

\vspace*{-0.2cm}
\noindent
70.
V.\ S.\ Barashenkov and A.\ Polanski, 
``Electronic Guide for Nuclear Cross-Sections",
Joint Institute for Nuclear Research
Report JINR E2-94-417, 
Dubna, Russia (1994). \\      

\vspace*{-0.2cm}
\noindent
71.
A.\ Auce {\it et al.},
``Reaction Cross Section for 75--190 MeV Alpha Particles on Targets from
$^{12}$C to $^{208}$Pb,''
{\it Phys.\ Rev.\ C }{\bf 50}, 871--879 (1994).\\

\vspace*{-0.2cm}
\noindent
72.
L.\ V.\ Dubar, D.\ Sh.\ Eleukenov, L.\ I.\ Slyusarenko, and N.\ P.\ Yurkuts,
``Parameterization of the Total Cross Sections of Reactions in the 
Intermediate Energy Region,''
{\it Yad.\ Fiz.\ }{\bf 49}, 1239--1242 (1989)
[{\it Sov.\ J.\ Nucl.\ Phys.\ }{\bf 49}, 771--773 (1989)].\\

\vspace*{-0.2cm}
\noindent
73.
A.\ Chatterjee, K.\ H.\ N.\ Murthy, and S.\ K.\ Gupta,
``Optical Reaction Cross-Section for Light Projectiles,''
{\it Pramana }{\bf 16}, 391--402 (1981).\\

\vspace*{-0.2cm}
\noindent
74.
G.\ S.\ Mani, M.\ A.\ Melkanoff, and A.\ Zucker,
Centre d'Etudes Nucleaire de Saclay Report CEA 2380 (1963).\\

\vspace*{-0.2cm}
\noindent
75.
K.\ K.\ Gudima, G.\ R\"opke, H.\ Schulz, and V.\ D.\ Toneev,
``The Coalescence Model and Pauli Quenching in High-Energy Heavy-Ion
Collisions,"
Joint Institute for Nuclear Research Report JINR-E2-83-101, Dubna (1983);
H.\ Schulz, G.\ R\"opke, K.\ K.\ Gudima, and V.\ D.\ Toneev,
``The Coalescence phenomenon and the Pauli Quenching in High-Energy Heavy-
Ion Collisions,"
{\it Phys.\ Lett.\ }{\bf B124} 458--460 (1983);
V.\ D.\ Toneev and K.\ K.\ Gudima,
``Particle Emission in Light and Heavy-Ion Reactions,"
{\it Nucl.\ Phys.\ }{\bf A400} 173c--190c (1983).\\

\vspace*{-0.2cm}
\noindent
76.
Joseph I.\ Kapusta,
``Mechanisms for Deuteron Production in Relativistic Nuclear Collisions,"
{\it Phys.\ Rev.\ C }{\bf 21}, 1301--1310 (1980).\\

\vspace*{-0.2cm}
\noindent
77.
N.\ V.\ Stepanov,
``Statistical Simulation of Excited Nuclei Fission. 
1. Formulation of the Model'', ITEP Preprint 81 (1987),
Institute for Theoretical and Experimental Physics,
Moscow, USSR (1987);
N.\ V.\ Stepanov,
``Statistical Simulation of Excited Nuclei Fission. 
2. Calculations and Comparison with Experiment,''
ITEP Preprint 55-88 (1988),
Institute for Theoretical and Experimental Physics,
Moscow, USSR (1988).\\

\vspace*{-0.2cm}
\noindent
78.
J.\ R.\ Wu and C.\ C.\ Chang,
``Complex-Particle Emission in the Pre-Equilibrium Exciton Model,''
{\it Phys.\ Rev.\ C }{\bf 17}, 1540--1549 (1978).\\

\vspace*{-0.2cm}
\noindent
79.
Konstantin K.\ Gudima, Stepan G.\ Mashnik, and Arnold J.\ Sierk,
``User Manual for the Code LAQGSM,''
Los Alamos National Laboratory Report LA-UR-01-6804, Los Alamos (2001);
http://lib-www.lanl.gov/la-pubs/00818645.pdf.\\

\vspace*{-0.2cm}
\noindent
80.
S.\ G.\ Mashnik {\it et al.},
`` Benchmarking Codes for Proton Radiography Applications,''
{\it Proc.\ 6th Intl.\ Workshop on Shielding Aspects
of Accelerators, Targets and Irradiation Facilities (SATIF-6),
April 10--12, 2002, Stanford Linear Accelerator Center, CA 94025, USA};
{http://www.nea.fr/html/science/satif/3-03.html}
LANL Report LA-UR-02-0609, Los Alamos (2002).\\

\vspace*{-0.2cm}
\noindent
81.
Ray E.\ L.\ Green and Ralph G.\ Korteling,
``Implications for Statistical Theories of Fragmentation from Measurements
of $Ag(p,^3He)$ and $Ag(p,^4He)$ at Intermediate Proton Energies,''
{\it Phys.\ Rev.\ C }{\bf 18}, 311--316   (1978);
tabulated values are available from the EXFOR database at the Web page:
http://www.nea.fr/html/dbdata/x4/.\\

\vspace*{-0.2cm}
\noindent
82
Ray E.\ L.\ Green, Ralph G.\ Korteling, and K.\ Peter Jackson,
``Inclusive Production of Isotopically Resolved Li through Mg Fragments by
480 MeV p + Ag Reactions,''
{\it Phys.\ Rev.\ C }{\bf 29}, 1806--1824   (1984).\\

\vspace*{-0.2cm}
\noindent
83
R.\ J.\ Charity {\it et al.}, 
``Systematics of Complex Fragment Emission in 
Niobium-induced Reactions'',
{\it Nucl.\ Phys.\ }{\bf A483}, 371--405 (1988);
R.\ J.\ Charity,
``$N-Z$ Distributions of Secondary Fragments and the
Evaporation Attractor Line'', 
{\it Phys.\ Rev.\ C }{\bf 58}, 1073--1077 (1998);
R.\ J.\ Charity {\it et al.},
``Emission of Unstable Clusters from Yb Compound Nuclei,''
{\it Phys.\ Rev.\ C }{\bf 63}, 024611 (2001);
http://wunmr.wustl.edu/~rc/.

\end{sf}
\end{document}